\documentclass[traditabstract]{aa} % for the abstract without structuration 
 \usepackage{longtable,lscape}

\usepackage{graphicx}
\usepackage{txfonts}
\def\gsim{~\rlap{$>$}{\lower 1.0ex\hbox{$\sim$}}}

\begin{document}

   \title{Nearby early-type galaxies with ionized gas.}

   \subtitle{IV. Origin and powering mechanism of the ionized gas.\fnmsep\thanks{Based on observations
   obtained at the European Southern Observatory, La Silla, Chile.}\fnmsep\thanks{Full Tables 2 and 3 
   are only available at the CDS via anonymous ftp to {\tt cdsarc.u-strasbg.fr}}}

   \author{F. Annibali\inst{1}, A. Bressan\inst{1,2,3}, R. Rampazzo\inst{1},
  W.W. Zeilinger\inst{4}, O. Vega\inst{3}, P. Panuzzo\inst{5}
          }

   \institute{INAF- Osservatorio Astronomico di Padova, 
              Vicolo dell'Osservatorio 5, I - 35122 Padova, ITALY\\
              \email{francesca.annibali@oapd.inaf.it alessandro.bressan@oapd.inaf.it roberto.rampazzo@oapd.inaf.it}
         \and
SISSA/ISAS, via Beirut 2-4, 34151 Trieste, ITALY\\
		\and
INAOE, Luis Enrique Erro 1,72840 Tonantzintla, Puebla, Mexico \\
\email{ovega@inaoep.mx}
         \and
Institut f\" ur Astronomie der Universit\" at  Wien, T\" urkenschanzstra$\ss$e 17, A-1180 Wien, Austria\\
\email{zeilinger@astro.univie.ac.at}
 		\and
CEA Saclay/Service dÕAstrophysique, Laboratoire AIM,CEA/DSM/IRFU-CNRS-Universit\'e Paris Diderot, 
F-91191 Gif-sur-Yvette C\'edex, France \\
\email{pasquale.panuzzo@cea.fr}
             }

   \date{Received ; Accepted}

% \abstract{}{}{}{}{} 
% 5 {} token are mandatory
 
  \abstract
   {
 {\sffamily\itshape Aims.} A significant fraction of early-type galaxies (ETGs) show emission lines in their optical spectra.
   We aim at understanding the powering mechanism and the origin of the ionized gas in ETGs, and its connection 
   with the host galaxy evolution.
%{

 {\sffamily\itshape Methods.} We analyzed intermediate-resolution optical spectra of a sample of 65 ETGs, mostly located 
in low density environments and biased toward the presence of ISM traces, for which we already derived in the previous papers of the series the stellar population properties.
To extract the emission lines from the galaxy spectra, we developed a new fitting procedure that properly 
subtracts the underlying stellar continuum, and that accounts for the uncertainties due to the age-metallicity degeneracy. %The procedure makes use of new SSP models based on the MILES 
%spectral library, which represents a substantial improvement over previous libraries used in population synthesis models.  
The emission line luminosities derived in annuli of increasing galacto-centric distance
were used to constrain the excitation mechanism and the metallicity of the ionized gas. 
 
 {\sffamily\itshape Results.} Optical emission lines are detected in 89\% of the sample. The detection fraction drops to 57\% if only 
   the galaxies with EW(H$\alpha$ $+$ [NII])$>$3 \AA \ are considered. The incidence and strength of 
   emission do not correlate either with the E/S0 classification, or with the fast/slow rotator classification. 
  Comparing the nuclear r$<$r$_e$/16 emission with the classical  [OIII]/H$\beta$ vs [NII]/H$\alpha$ diagnostic   diagram, the galaxy activity is so classified:
     72\% of the galaxies with emission are LINERs, 9\% are Seyferts, 12\% are Composite/Transition objects, 
     and 7\% are non-classified.
  Seyferts have young luminosity-weighted ages ($\lesssim$ 5 Gyr), and appear, on average, significantly 
  younger than LINERs and Composites. Excluding the Seyfert galaxies from our sample, we find that the spread in the ([OIII], H$\alpha$ or [NII]) emission strength increases with the galaxy central velocity dispersion $\sigma_c$, low-$\sigma_c$ galaxies having all weak emission lines, and high-$\sigma_c$ galaxies displaying both weak and strong emission lines. Furthermore, the [NII]/H$\alpha$ ratio tends to increase with  $\sigma_c$. 
 A spatial analysis of the emission line properties within the individual galaxies reveals that 
the [NII]/H$\alpha$ ratio decreases with increasing galacto-centric distance, indicating either a decrease of the nebular metallicity, or a progressive ``softening'' of the ionizing spectrum. 
 The average nebular oxygen abundance is slightly less than solar.  A comparison with the stellar metallicities derived in Paper~III shows that the gas oxygen abundance is $\approx$ 0.2 dex lower than that of stars.

{\sffamily\itshape Conclusions.}
The stronger nuclear (r$<r_e/16$) emission can be explained with photoionization by PAGB stars alone
 only for $\approx$ 22\% of the LINERs/Composite sample.
 On the other hand, we can not exclude an important role of PAGB star photoionization at larger radii.
 For the major fraction of the sample, the nuclear emission is consistent with excitation from a low-accretion rate AGN, fast shocks (200 -500 km/s) in a relatively gas poor environment (n $\lesssim100$ cm$^{-3}$) , or coexistence of the two. The derived [SII]6717/6731 ratios are consistent with the low gas densities required 
 by the shock models.
 % The natural explanation for  the observed trends of the [NII] intensity and of the [NII]/H$\alpha$ ratio with $\sigma_c$ is a scenario in which the shock mechanical energy comes from turbulent motions of clouds within the potential well of the galaxy.
%A scenario in which the shock mechanical energy comes from turbulent motions of clouds within the potential well of the galaxy is very appealing, since it naturally explains the observed trends of the [NII] intensity and of the [NII]/H$\alpha$ ratio with $\sigma_c$. 
The derived nebular metallicities suggest either an external origin of the gas, or an overestimate of the oxygen yields by SN models.
  }

   \keywords{Galaxies: elliptical and lenticular, cD -- Galaxies: ISM -- Galaxies: active--Galaxies:abundances.}
               
\authorrunning{Annibali et al.}

 \maketitle
%
%________________________________________________________________

\section{Introduction}

Early-type galaxies (ETGs) have long been considered to be inert stellar systems,
essentially devoid of gas and dust. However, this view has radically changed 
since a number of imaging and spectroscopy studies from both  
the ground and space have revealed the presence of a multiphase 
interstellar medium (ISM): 
a hot ($T \sim 10^6 - 10^7$ K), X-ray emitting halo (Forman \& Jones~\cite{fj85}; 
Fabbiano et al.~\cite{fab92}; O'Sullivan et al.~\cite{sul01}); a warm ($T \sim 10^4$ K)  
gas (often referred to as ''ionized gas``)  
(e.g. Phillips et al.~\cite{ph86}; Sadler \& Gerhard~\cite{sg85}; 
%van Dokkum \& Franx \cite{vdf95};  
see also Goudfrooij~\cite{gou99} for a review); 
and colder components detected in the mid and far infrared, CO and HI 
(see review of Sadler et al.~\cite{sad02}; 
Temi et al.~\cite{temi04}; Bressan et al.~\cite{bress06}; Sage \& Welch~\cite{sw06}; 
Morganti et al.~\cite{mo06}; Helmboldt~\cite{hel07}; Oosterloo et al.~\cite{oos07};
Panuzzo et al.~\cite{panu07}).
The hot gas component dominates the ISM in E/S0 galaxies, 
and only comparatively small quantities are detected in the warm and cold 
phases (Bregman et al.~\cite{bre92}). The derived masses of the warm ISM range from 
$\sim 10^3$ to $\sim 10^6 M_{\odot}$ (Phillips et al.~\cite{ph86}; 
Trinchieri \& di Serego Alighieri~\cite{tds91}; Macchetto et al.~\cite{mac96};
Goudfrooij et al.~\cite{gou94a,gou94b,gou94c}; Goudfrooij \& de Jong~\cite{gouj95}),
orders of magnitude below the X-ray emitting hot ISM and the stellar mass.
Ionized gas is detected in 40-80\% of early-type galaxies via its
 optical emission lines (Caldwell~\cite{cal84}; Phillips et al.~\cite{ph86};
 Macchetto et al.~\cite{mac96}; Sarzi et al.~\cite{sar06,sar09}; Yan et al.~\cite{yan06}; 
 Serra et al.~\cite{serra08}).
 The ionized gas is also usually morphologically associated 
 with dust (Kim~\cite{kim89}; Goudfrooij et al.~\cite{gou94c}).
 
Despite the number of studies, several issues remain still open. The 
first question is the origin of the ISM in ETGs. 
Narrow band imaging centered around H$\alpha + $[NII]$\lambda$6584
shows that the ionized gas presents a variety of morphologies, from regular,
disk-like structures, to filamentary structures 
(Demoulin-Ulrich et al.~\cite{du84}; Buson et al.~\cite{bus93}; Macchetto et al.~\cite{mac96}; 
Zeilinger et al.~\cite{zei96}; Martel et al.~\cite{mar04}; Sarzi et al.~\cite{sar06}).
Evidence for the acquisition of external gas comes from
kinematical studies showing that the angular momentum
of the gas is often decoupled from that of the stars
(Bertola et al.~\cite{ber92}; van Dokkum \& Franx~\cite{vdf95}; Caon et al.~\cite{caon00}), 
even if according to Sarzi et al.~(\cite{sar06}) the angular 
momenta are inconsistent with a purely external origin for the gas.
The fact that the ionized gas emission is always associated with dust 
(e.g. Tran et al.~\cite{tr01}) tends to exclude ``cooling flows'' as the origin for the warm gas in E/S0 
galaxies (Goudfrooij~\cite{gou99}). In the scenario proposed by
Sparks et al.~(\cite{spa89}) and de Jong et al.~(\cite{dej90}), the dusty filaments can arise from
the interaction of a small gas rich galaxy with the giant elliptical, or from a tidal
accretion event in which the gas and dust are stripped from a passing spiral
(see e.g. Domingue et al.~\cite{Domingue03}; Tal et al.~\cite{Tal09}).

The history of star formation and evolution leaves its chemical signature 
in the ISM of ETGs.
Historically, X-ray abundance measurements of the hot ISM 
have been problematic, as typified by the so-called 
Fe discrepancy (Arimoto et al.~\cite{ari97}).
More recently, {\it Chandra} and {\it XMM} measurements of X-ray bright 
galaxies at the center of groups have supported roughly
solar or even slightly supersolar abundances for the  ISM 
(e.g. Gastaldello \& Molendi~\cite{gm02}; Buote et al.~\cite{buo03}; 
Tamura et al.~\cite{tamu03}), and have claimed no evidence for very subsolar Fe abundances 
(Humphrey \& Buote~\cite{hb06}; Ji et al.~\cite{ji09}). 
These studies also report anomalously low oxygen abundances 
compared to iron and magnesium. This is in conflict with 
results from SN stellar yields if one assumes that the source of 
ISM is the material injected into interstellar space by evolved stars.
Up to now, the only study of metallicity in the warm ISM is that 
of Athey \& Bregman~(\cite{ab09}), who derived oxygen abundances for seven 
ETGs from optical emission lines.
The authors find an average solar metallicity, 
favoring an internal origin of the warm ISM. However, this study is limited
by the small sample. Determining the metallicity of the warm ISM 
is fundamental to create a link between the hot gas phase and the galaxy stellar population,
and to discriminate between an internal and an external origin of the interstellar matter 
in E/S0 galaxies.

The second still open issue concerns the ionizing source of the 
warm gas. Optical spectroscopic studies show that ETGs 
are typically classified as LINERs (Low-Ionization Nuclear Emission-line Regions, Heckman~\cite{heck80}) 
according to their emission line ratios (e.g. Phillips et al.~\cite{ph86}; Yan et al.~\cite{yan06}).
However, there is still strong debate about the ionization mechanism in LINERs.
At present, the most viable excitation mechanisms are: 
a low accretion-rate AGN (e.g., Ho~\cite{ho99b}; Kewley et al.~\cite{kew06}, Ho~\cite{ho09a}), 
photoionization by old post-asymptotic giant branch (PAGB) stars (e.g.  Trinchieri \& di Serego Alighieri~\cite{tds91}; 
Binette et al.~\cite{bin94}; Stasi{\'n}ska et al.~\cite{sta08}), 
fast shocks (Koski \& Osterbrock~\cite{ko76}; Heckman~\cite{heck80}; Dopita \& Sutherland~\cite{dosu95};
Allen et al.~\cite{allen08}). 
The AGN mechanism is strongly supported by the detection of broad emission lines 
in optical spectra (e.g. Ho et al.~\cite{ho97a}; Wang \& Wei~\cite{ww08}), 
by the properties of the X-ray emission 
(Terashima et al.~\cite{tera02}; Flohic et al.~\cite{flo06}; Gonzalez-Martin et al.~\cite{go09}),
by the presence of UV and X-ray variability (e.g Maoz et al.~\cite{maoz05}, Pian et al.~\cite{pian10}), 
and by the evidence that massive black holes (BHs) appear to be a generic component of galaxies with a bulge
(Magorrian et al.~\cite{mago98}; Ho~\cite{ho99a}; Kormendy~\cite{kor04}).
PAGB stars are good candidates because their radiation, much harder than that from young stars, 
is able to reproduce the observed emission-line ratios in LINERs (Trinchieri \& di Serego Alighieri~\cite{tds91}; Binette et al.~\cite{bin94}; Stasi{\'n}ska et al.~\cite{sta08}).
The most compelling evidence in support of the PAGB mechanism 
is the finding that the emission-line flux correlates very well with the host
galaxy stellar luminosity within the emission-line region (Macchetto et al.~\cite{mac96}),
and that the emission-line flux distribution follows that of the stellar continuum
(Sarzi et al.~\cite{sar06,sar09}). 
These correlations suggest that the source of ionizing photons is distributed 
in the same way as the stellar population.
LINERs emission has also been observed in
extranuclear regions associated with large-scale outflows and related 
shocks (L{\'{\i}}pari et al.~\cite{li04}), or regions shocked by radio jets
(Cecil et al.~\cite{cec00}).

Evidence is growing that more than one mechanism may be at work in LINERs.
In a recent study, Sarzi et al.~(\cite{sar09}) showed that the role of AGN activity 
is confined to the central regions, whereas  at larger radii the stellar and 
nebular fluxes follow each other, thus suggesting a stellar ionizing source.
Eracleous et al.~(\cite{era10}) showed that low accretion-rate AGNs do not
produce enough ionizing photons to explain the observed H$\alpha$ luminosities
within a 2'' $\times$ 4'' aperture, so that other mechanisms are needed.

With the aim of understanding the formation history of ETGs, 
and the connection between the stellar populations and the ionized gas, 
we have analyzed intermediate-resolution optical spectra
for a sample of 65 field E/S0 galaxies 
%biased toward the presence of ISM emission 
(Rampazzo et al.~\cite{ramp05}, hereafter Paper~I; 
 Annibali et al.~\cite{anni06}, hereafter Paper~II). 
The stellar population properties (i.e. ages, metallicities, and 
elements abundance ratios) were derived at different radii 
in Annibali et al.~(\cite{anni07}) (hereafter Paper~III) 
through the analysis of the Lick indices. 
In this paper we present the study of the optical emission lines in annuli of 
increasing galacto-centric distance. The purpose is to characterize 
the properties of the ionized gas, its origin, and the possible excitation 
mechanisms at different radii.

The paper is organized as it follows. In section~2 we provide a brief overview 
of the sample. In section~3 we describe the method used to extract the emission-line 
fluxes from the optical spectra. In section~4 we provide a spectral classification 
through standard diagnostic diagrams, and investigate possible correlations 
with the host galaxy properties. In section~5 we determine the oxygen metallicity.
 In section~6 we compare our data with the models. A summary and the conclusions are 
 given in section~7.

%----------------------------------------------Table 1------------------------------------------------------
\begin{table*}
\caption{Sample overview}
\label{sample}
\centering
{\tiny{
\begin{tabular}{lccccccccc}
\hline \hline
Galaxy Id  & RSA & RC3 & $\sigma_c$  & $r_e$  & $V_{hel}$    &   $\rho_{xyz}$ & Age & Z &  [$\alpha$/Fe] \\ 
                 &          &         &  km s$^{-1}$ & arcsec & km s$^{-1}$ & gal Mpc$^{-3}$  & Gyr &  &  \\ 
\hline
 NGC~128  & S02(8) pec      & S0 pec sp     &  183    & 17.3  &  4227 &        &    9.7   $\pm$    1.7   &    0.024   $\pm$   0.004   &   0.16    $\pm$   0.03    \\
 NGC~777  & E1                   & E1                 &  317    & 34.4  & 5040  &        &    5.4   $\pm$    2.1   &    0.045   $\pm$   0.020   &    0.28    $\pm$   0.10    \\
 NGC~1052 & E3/S0             & E4                 &  215   & 33.7   &1475  & 0.49 &   14.5  $\pm$    4.2   &    0.032   $\pm$   0.007   &    0.34    $\pm$     0.05    \\
 NGC~1209 & E6                  & E6:                &  240    &18.5  & 2619 & 0.13  &    4.8    $\pm$    0.9  &     0.051   $\pm$   0.012   &    0.14    $\pm$     0.02      \\
 NGC~1297 & S02/3(0)         & SAB0 pec:    &  115    &  28.4 & 1550  & 0.71 &   15.5   $\pm$   1.2   &    0.012   $\pm$    0.001   &    0.29    $\pm$     0.04    \\
 NGC~1366 & E7/S01(7)      & S0 sp             & 120    &  10.6 & 1310 & 0.16  &    5.9    $\pm$     1.   &     0.024   $\pm$    0.004   &    0.08    $\pm$     0.03     \\
 NGC~1380 & S03(7)/Sa      & SA0               &  240   &  20.3 & 1844 & 1.54   &    4.4    $\pm$   0.7   &     0.038   $\pm$    0.006   &    0.24    $\pm$     0.02   \\
 NGC~1389 & S01(5)/SB01  & SAB(s)0-:      & 139    & 15.0 & 986 & 1.50      &    4.5     $\pm$   0.6   &     0.032   $\pm$    0.005   &    0.08    $\pm$     0.02    \\
 NGC~1407 & E0/S01(0)      & E0                  &  286   & 70.3 & 1766 & 0.42   &    8.8     $\pm$   1.5   &      0.033   $\pm$    0.005   &    0.32    $\pm$     0.03    \\
 NGC~1426 & E4                  & E4                  &  162   & 25.0 & 1443 & 0.66   &    9.0     $\pm$    2.5   &     0.024   $\pm$    0.005    &   0.07    $\pm$     0.05     \\
 & & & & \\
 NGC~1453 & E0                  &    E2         & 289 &  25.0 & 3906 &  &    9.4         $\pm$       2.1   &      0.033   $\pm$      0.007     &    0.22    $\pm$     0.03      \\
 NGC~1521 & E3                  &    E3         & 235 & 25.5 & 4165 &   &    3.2         $\pm$       0.4   &      0.037   $\pm$      0.006     &    0.09    $\pm$     0.02       \\
 NGC~1533 & SB02(2)/SBa & SB0-          & 174  & 30.0 & 773 & 0.89  &   11.9   $\pm$       6.9   &      0.023   $\pm$      0.020     &    0.21    $\pm$     0.10        \\
 NGC~1553 & S01/2(5)pec  & SA(r)0        & 180 & 65.6 & 1280 & 0.97 &    4.8     $\pm$       0.7   &      0.031   $\pm$      0.004     &    0.10    $\pm$     0.02       \\
 NGC~1947 & S03(0) pec    & S0- pec     &  142 & 32.1 & 1100  & 0.24   &    5.9     $\pm$       0.8   &      0.023   $\pm$      0.003     &    0.05    $\pm$     0.02   \\
 NGC~2749 & E3                  & E3           &  248 & 33.7 & 4180 &  &   10.8     $\pm$       2.3   &      0.027   $\pm$      0.006     &    0.25    $\pm$     0.04    \\
 NGC~2911 & S0p or S03(2)& SA(s)0: pec  & 235  & 50.9 &  3131  & &    5.7     $\pm$       2.0   &      0.034   $\pm$      0.019     &    0.25    $\pm$     0.10      \\
 NGC~2962 & RSB02/Sa     & RSAB(rs)0+ &   & 23.3 & 2117 & 0.15  & & &  \\
 NGC~2974 & E4                  & E4           & 220 & 24.4 & 1890  & 0.26   &   13.9     $\pm$       3.6   &      0.021   $\pm$      0.005     &    0.23    $\pm$     0.06      \\
 NGC~3136 & E4                  & E:           & 230 & 36.9 & 1731  & 0.11 &    1.5     $\pm$       0.1   &      0.089   $\pm$      0.004     &    0.36    $\pm$     0.02     \\
 & & & & \\
 NGC~3258 & E1                  & E1          & 271 & 30.0 & 2778 & 0.72  &    4.5     $\pm$       0.8   &      0.047   $\pm$      0.013     &    0.21    $\pm$     0.03      \\
 NGC~3268 & E2                  & E2          & 227 & 36.1 & 2818 & 0.69  &    9.8     $\pm$       1.7   &      0.023   $\pm$      0.004     &    0.34    $\pm$     0.04    \\
 NGC~3489 & S03/Sa           & SAB(rs)+   & 129   & 20.3 & 693  & 0.39 &    1.7     $\pm$       0.1   &      0.034   $\pm$      0.004     &    0.05    $\pm$     0.02         \\
 NGC~3557 & E3                  & E3          & 265 & 30.0 & 3038 & 0.28 &    5.8     $\pm$       0.8   &      0.034   $\pm$      0.004     &    0.17    $\pm$     0.02     \\
 NGC~3607 & S03(3)            & SA(s)0:     & 220 & 43.4 &  934 & 0.34 &    3.1     $\pm$       0.5   &      0.047   $\pm$      0.012     &    0.24    $\pm$     0.03       \\
 NGC~3818 & E5                  & E5            & 191  & 22.2  & 1701 & 0.20   &    8.8     $\pm$       1.2   &      0.024   $\pm$      0.003     &    0.25    $\pm$     0.03    \\
 NGC~3962 & E1                  & E1          & 225 & 35.2 & 1822 & 0.32  &   10.0     $\pm$       1.2   &      0.024   $\pm$      0.003     &    0.22    $\pm$     0.03      \\
 NGC~4374 & E1                  & E1      & 282 & 50.9 & 1060 & 3.99  &    9.8     $\pm$       3.4   &      0.025   $\pm$      0.010     &    0.24    $\pm$     0.08     \\
 NGC~4552 & S01(0)            & E           &  264 & 29.3 &  322 & 2.97  &    6.0     $\pm$       1.4   &      0.043   $\pm$      0.012     &    0.21    $\pm$     0.03      \\
 NGC~4636 & E0/S01(6)      & E0-1        &  209 & 88.5 & 937 & 1.33  &   13.5     $\pm$       3.6   &      0.023   $\pm$      0.006     &    0.29    $\pm$     0.06   \\
 & & & & \\   
 NGC~4696 &(E3)            & E+1 pec  & 254  & 85.0 & 2958 & 0.00   &   16.0     $\pm$       4.5   &      0.014   $\pm$      0.004     &    0.30    $\pm$     0.10   \\
 NGC~4697 & E6              & E6      & 174  & 72.0 & 1241 & 0.60  &   10.0     $\pm$       1.4   &      0.016   $\pm$      0.002     &    0.14    $\pm$     0.04       \\
 NGC~5011 & E2              & E1-2    & 249 & 23.8 & 3104 & 0.27  &    7.2     $\pm$       1.9   &      0.025   $\pm$      0.008     &    0.25    $\pm$     0.06    \\
 NGC~5044 & E0              & E0      & 239 & 82.3 & 2704 & 0.38  &   14.2     $\pm$       10.   &      0.015   $\pm$      0.022     &    0.34    $\pm$     0.17     \\
 NGC~5077 & S01/2(4)     & E3+        & 260 & 22.8  &  2764 & 0.23  &   15.0     $\pm$       4.6   &      0.024   $\pm$      0.007     &    0.18    $\pm$     0.06       \\
 NGC~5090 & E2              & E2      & 269 & 62.4 &  3421 & &   10.0     $\pm$       1.7   &      0.028   $\pm$      0.005     &    0.26    $\pm$     0.04       \\
 NGC~5193 & S01(0)        & E pec  & 209 & 26.7 & 3711 &   &    6.8     $\pm$       1.1   &      0.018   $\pm$      0.002     &    0.26    $\pm$     0.04     \\
 NGC~5266 & S03(5) pec & SA0-: & 199 & 76.7 & 3074 & 0.35 &    7.4     $\pm$       1.4   &      0.019   $\pm$      0.003     &    0.15    $\pm$     0.05    \\
 NGC~5328 & E4              & E1:          & 303 & 22.2 & 4671 &  &   12.4     $\pm$       3.7   &      0.027   $\pm$      0.006     &    0.15    $\pm$     0.05     \\
 NGC~5363 & [S03(5)]      & I0:          &  199 & 36.1 & 1138 & 0.28 &   12.1     $\pm$       2.3   &      0.020   $\pm$      0.004     &    0.16    $\pm$     0.05    \\
  & & & & \\  
 NGC~5638 & E1                  & E1   &  168 & 28.0 & 1676 & 0.79   &    9.1     $\pm$       2.3   &      0.024   $\pm$      0.008     &    0.24    $\pm$     0.05       \\
 NGC~5812 & E0                  & E0   & 200 & 25.5 & 1930 & 0.19  &    8.5     $\pm$       2.1   &      0.027   $\pm$      0.008     &    0.22    $\pm$     0.05    \\
 NGC~5813 & E1                  & E1-2 & 239 & 57.2 & 1972 & 0.88   &   11.7     $\pm$       1.6   &      0.018   $\pm$      0.002     &    0.26    $\pm$     0.04      \\
 NGC~5831 & E4                  & E3   & 164 & 25.5 & 1656 & 0.83  &    8.8     $\pm$       3.5   &      0.016   $\pm$      0.011     &    0.21    $\pm$     0.09      \\
 NGC~5846 & S01(0)            & E0+         &250 & 62.7  & 1709 & 0.84  &    8.4     $\pm$       1.3   &      0.033   $\pm$      0.005     &    0.25    $\pm$     0.03     \\
 NGC~5898 & S02/3(0)         & E0           & 220  & 22.2 & 2267& 0.23  &    7.7     $\pm$       1.3   &      0.030   $\pm$      0.004     &    0.10    $\pm$     0.03      \\
 NGC~6721 & E1                  & E+:          & 262 & 21.7 &  4416 & &    5.0     $\pm$       0.8   &      0.040   $\pm$      0.007     &    0.24    $\pm$     0.02       \\
 NGC~6758 & E2 (merger)   & E+:  & 242 & 20.3 & 3404 &  &   16.0     $\pm$       2.5   &      0.016   $\pm$      0.002     &    0.32    $\pm$     0.05      \\
 NGC~6776 & E1 pec           & E+pec & 242 & 17.7 & 5480 & &  2.7     $\pm$       0.5   &      0.033   $\pm$      0.010     &    0.21    $\pm$     0.05   \\
 NGC~6868 & E3/S02/3(3)   & E2           & 277 & 33.7 & 2854 & 0.47  &    9.2     $\pm$       1.8   &      0.033   $\pm$      0.006     &    0.19    $\pm$     0.03      \\
  & & & & \\ 
NGC~6875 & S0/a(merger)    & SAB(s)0- pec: & & 11.7  &  3121 &  & & &    \\
NGC~6876 & E3                     & E3          & 230 & 43.0 & 3836 &  &    9.8     $\pm$       1.6   &      0.023   $\pm$      0.003     &    0.26    $\pm$     0.03        \\
NGC~6958 & R?S01(3)          & E+         & 223 & 19.8 & 2652 & 0.12  &    3.0     $\pm$       0.3   &      0.038   $\pm$      0.006     &    0.20    $\pm$     0.03      \\
NGC~7007 & S02/3/a             & SA0-:      & 125 & 15.4 & 2954 & 0.14   &    3.4     $\pm$       0.6   &      0.031   $\pm$      0.010     &    0.15    $\pm$     0.05       \\
NGC~7079  & SBa                  & SB(s)0     & 155 & 19.8 & 2670 & 0.19  &    6.7     $\pm$       1.1   &      0.016   $\pm$      0.003     &    0.21    $\pm$     0.05       \\
NGC~7097 & E4                     & E5          & 224 & 18.1 & 2404 & 0.26   &   10.5     $\pm$       2.4   &      0.024   $\pm$      0.005     &    0.30    $\pm$     0.05     \\
NGC~7135 & S01 pec            & SA0- pec   & 231 & 31.4 & 2718 &  0.32  &    2.2     $\pm$       0.4   &      0.047   $\pm$      0.010     &    0.46    $\pm$     0.04    \\
NGC~7192 & S02(0)               & E+:         & 257 & 28.6 & 2904 &  0.28   &    5.7     $\pm$       2.0   &      0.039   $\pm$      0.015     &    0.09    $\pm$     0.05    \\
NGC~7332 & S02/3(8)            & S0 pec sp & 136 & 14.7 &  1207 &  0.12 &    3.7     $\pm$       0.4   &      0.019   $\pm$      0.002     &    0.10    $\pm$     0.03       \\
NGC~7377 & S02/3/Sa pec    & SA(s)0+  & 145 & 36.9 & 3291 & &    4.8     $\pm$       0.6   &      0.020   $\pm$      0.002     &    0.10    $\pm$     0.03       \\
& & & & \\
IC~1459    & E4                      & E          & 311 & 34.4 & 1659 & 0.28  &    8.0     $\pm$       2.2   &      0.042   $\pm$      0.009     &    0.25    $\pm$     0.04     \\
IC~2006    & E1                      & E          & 122 & 28.6 & 1350 & 0.12   &    8.1     $\pm$       0.9   &      0.026    $\pm$     0.003     &    0.12     $\pm$    0.02    \\
IC~3370    & E2 pec               & E2+       & 202 & 38.6 & 2934 & 0.20   &    5.6     $\pm$       0.9   &      0.022   $\pm$      0.004     &    0.17    $\pm$     0.04    \\
IC~4296    & E0                      & E          & 340 & 41.4 & 3762  & &    5.2     $\pm$       1.0   &      0.044   $\pm$      0.008     &    0.25    $\pm$     0.02      \\
IC~5063   & S03(3)pec/Sa    & SA(s)0+: & 160 & 26.7 & 3402 & & & &  \\
\hline
\end{tabular}
}}
\end{table*}

%-------------------------end Table 1-----------------------------------------------------

%macro spettroxpaper/xpaper.sm , command fits2
%----------------------    Figure 1--------------------------------------------------
 \begin{figure*}
   \centering
  \includegraphics[width=17cm]{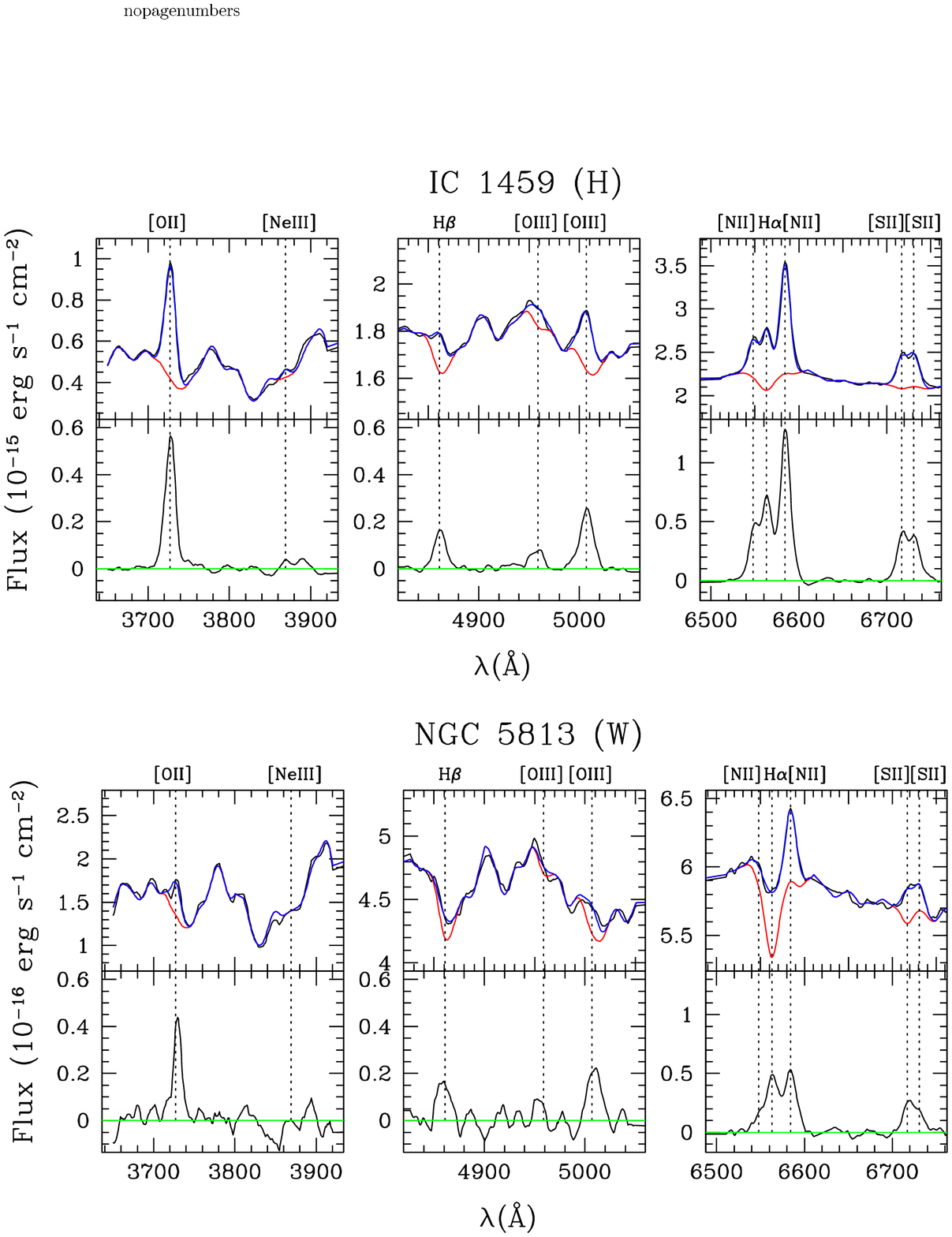}
  \caption{Central (r $\leq$r$_e$/16) spectra for a high emission galaxy (IC~1459) 
  and for a weak emission galaxy (NGC 5813) in the  
  wavelength ranges where the most important emission lines are measured. For each galaxy, 
  we show in the top panels the observed spectrum (black line), 
  the underlying stellar spectrum model (red), and the fit to the observed spectrum obtained 
  by adding emission lines (Gaussian curves) to the underlying stellar continuum (blue). 
 In the bottom panels we show the subtracted spectrum, where the emission lines emerge.
 Notice that in this figure the subtracted stellar continuum is an SSP of age, Z, and extinction equal
 to the average of all the solution with ${\rm \chi^2 < 2}$. 
 However, this is only for illustration purposes, and it is slightly different from the procedure described in Section~3.1.
 }
\label{spec}
\end{figure*}
%----------------------  end  Figure 1--------------------------------------------------

\section{Sample Overview}

The Rampazzo et al.~(\cite{ramp05}) + Annibali et al.~(\cite{anni06}) sample 
(hereafter R05$+$A06 sample) was selected from a compilation of ETGs showing 
ISM traces in at least one of the following bands:
IRAS 100 $\mu$m, X-ray, radio, HI and CO (Roberts et al.~\cite{rob91}).
All galaxies belong to the {\it Revised Shapley Ames Catalog of 
Bright Galaxies} (RSA) (Sandage \& Tammann~\cite{RSA})  and have 
a redshift lower than $\sim$ 5500 km s$^{-1}$.
Because of the selection criteria, the sample is biased toward the presence 
of emission lines.
Table~1 summarizes the main characteristics of the sample. 
Column (1) gives the galaxy identification name; 
Col. (2) and (3) provide the galaxy morphological classification
according to RSA (Sandage \& Tammann~\cite{RSA}) and 
RC3 (de Vaucouleurs et al.~\cite{RC3}), respectively: only in few cases 
do the two catalogues disagree in the distinction between E and S0 classes;
Col. (4) gives the central velocity dispersion from HYPERLEDA
(http://leda.univ-lyon1.fr//);
Col. (5) gives the effective radius $r_e$ from RC3;
Col. (6) gives the galaxy systemic velocity, V$_{hel}$, which is lower
than $\sim$5000 km~s$^{-1}$; Col. (7) provides the richness 
parameter $\rho_{xyz}$ (Tully~\cite{Tu88}): it represents the density of
galaxies brighter than -16 apparent B-mag in the vicinity of the entry,
in galaxies $\times$Mpc$^{-3}$. Cols (8), (9) and (10) list the age,
metallicity and [$\alpha$/Fe] ratios derived in Paper III.

The galaxies of the R05$+$A06 sample are mainly located in low
density environments. The local density of the galaxies varies from
$\rho_{xyz}$ $\approx$ 0.1 Mpc$^{-3}$, typical of very isolated galaxies, to
$\rho_{xyz}$ $\approx$ 4 Mpc$^{-3}$, characteristic of denser regions in the 
Virgo cluster. For comparison, in the Tully~(\cite{Tu88}) catalogue objects like NGC~1399
and NGC~1389, members of the Fornax cluster, have $\rho_{xyz}$=1.59
and 1.50 Mpc$^{-3}$, respectively.  Thus, the R05$+$A06 sample, even though biased towards low
density environments, contains a fraction  of galaxies in relatively
dense environments. The sample spans a wide range in central velocity dispersion, $\approx$
from 115 to 340 km~s$^{-1}$ (see Papers~I and II for details).

Following the RC3 classification, the sample is composed of $\sim$ 70\% ellipticals and 
$\sim$ 30\% lenticulars. However, if we classify the galaxies according to their amount of rotation 
(see Appendix A), we end up with $\sim$ 70\% fast rotators (F) and $\sim$ 30\% slow rotators (S). 

Intermediate resolution (FWHM $\approx$7.6~\AA\ at 5550~\AA) spectra in the 
(3700 - 7250) \AA \ wavelength range were acquired with the 
1.5 m ESO-La Silla telescope for the 65 galaxies of the sample.
The spectra were extracted for 7 apertures of increasing radius 
(1.5\arcsec, 2.5\arcsec, 10\arcsec,  r$_e$/10, r$_e$/8, r$_e$/4, and r$_e$/2), 
corrected for the galaxy ellipticity, and 4 adjacent annulii
(r $\leq$r$_e$/16, r$_{e}$/16 $<$ r $\leq$r$_e$/8, r$_{e}$/8$<$ r $\leq$r$_e$/4, and r$_{e}$/4 $<$ r $\leq$r$_e$/2).
The data reduction and the computation of the Lick indices are described 
in Papers~I and II. In Paper~III we derived ages, metallicities, 
and [$\alpha$/Fe] ratios by comparing the data with our new
Simple Stellar Population (SSP) models.
No stellar population parameters are derived for IC~5063, since 
line emission is too strong to allow a measurement of the Lick indices.
Summarizing the results of Paper~III, we derive a large age spread, with SSP-equivalent ages 
ranging from a few Gyrs to a Hubble time. 
The galaxies have metallicities and [$\alpha$/Fe] ratios above solar.
Both the total metallicity and the [$\alpha$/Fe] present a positive correlation with the central velocity
dispersion, indicating that the chemical enrichment was more efficient and the duration 
of the star formation shorter in more massive galaxies.
We also find that the youngest objects in our sample are all located in the 
lowest density environments ($\rho_{xyz}<0.4$ Mpc$^{-3}$).
We suggest that the young galaxies in the lowest  density environments underwent 
secondary episodes of star formation, which we call ``rejuvenation episodes''.
Within the individual galaxies, the stellar metallicity tends 
to decrease from the center outwards.
We derive an average metallicity gradient of  
$\Delta \log Z / \Delta \log (r/r_e) \sim -0.21$.

\section{Emission lines}

\subsection{Starlight Subtraction and Line Intensities}

There are a number of emission lines that we expect to detect 
within the wavelength range (3700 - 7250) \AA  \ sampled by our spectra:
[OII] $\lambda$3727, [NeIII] $\lambda$3869, H$\beta$ $\lambda$4861, 
[OIII] $\lambda\lambda$4959, 5007, [OI] $\lambda$6300, H$\alpha$ $\lambda$ 6563,
[NII] $\lambda\lambda$6548, 6584, [SII] $\lambda\lambda$6717, 6731.     

The emission line fluxes were measured in residual spectra 
obtained after subtracting the stellar population contribution from the 
observed galaxy spectra.
The starlight subtraction is a crucial step for the correct determination of the
emission line properties, in particular for H$\beta$ and H$\alpha$, which are 
superimposed to the Balmer absorption features of the 
underlying stellar population.
The  H$\beta$ emission, much fainter than the 
H$\alpha$ ($FH\beta$ $\sim$ 1/3 $FH\alpha$ in absence of extinction),
is the most difficult to measure.
With only few exceptions, we never see  in our spectra 
a true emission feature in H$\beta$, but rather 
an {\it infilling} of the stellar absorption line.
It follows that a reliable measurement of this 
line rests on the appropriate modeling of the 
underlying starlight contribution. 

%----------------------    Figure 2--------------------------------------------------
%MACRO: /Users/annibali/science/ellittiche/paperIV/plots_xpaper/fig.havshb.sm; command: vaipaper
%treshold to define no emission galaxies has been set to 5
%file to draw E(B-V) lines: myrun_err/EBV_hahb.txt
\begin{figure}
 \centering
  \includegraphics[width=8.5cm]{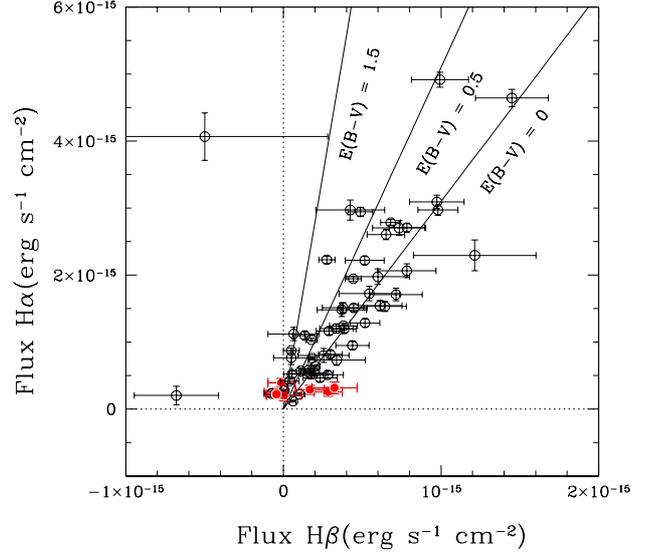}
\caption{H$\alpha$ versus H$\beta$ emission fluxes for the galaxy sample measured in a central r $\leq$r$_e$/16 region.
IC~5063, with derived emission fluxes falling outside the plot boundaries, is not included.
Full dots denote the seven ``no emission'' galaxies 
(NGC~5638, NGC~5831, NGC~1407, NGC~1366, NGC~1426, NGC~5812, and NGC~1389) in the sample.
The solid lines indicate regions of different reddening.}
\label{havshb}
\end{figure}
%----------------------  end  Figure 2--------------------------------------------------

% ----------------------------------- Figure 3 ------------------------------------------------------
%MACRO: /Users/annibali/science/ellittiche/paperIV/plots_xpaper/ebv.sm; command: go
%read file myrun_err/EWs.sample.clean.txt produced by ./myrun_err/tabellepaper.f90
%an cleaned for the 8 galaxies with no emission. also N2962 removed.
  \begin{figure}
   \centering
  \includegraphics[width=8.5cm]{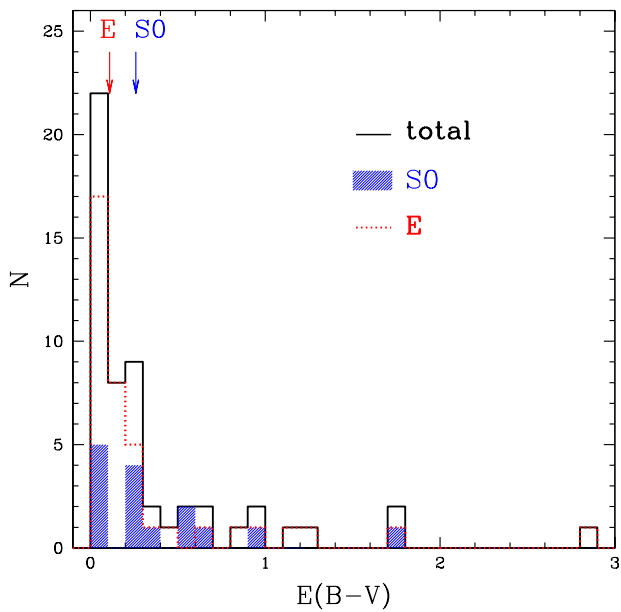}
  \caption{${\rm E(B-V)}$ distribution derived for our sample from the ${\rm F_{H\alpha}/F_{H\beta}}$ ratios.   
 The shaded histogram denotes S0 galaxies, while the dotted line is for E galaxies.
 Vertical arrows indicate the median values.}
\label{ebv}
\end{figure}
%------------------------------ end Figure 3 -------------------------------------------------------

To subtract the underlying stellar population from the galaxy spectra
we used new SSPs (Bressan, unpublished, see also 
Clemens et al.~\cite{Clemens09}, Chavez et al.~\cite{Chavez09}).
The SSPs were computed from the isochrones of Padova 94
(Bertelli et al.~\cite{Bertelli94}) with the revision of Bressan, Granato \& Silva~(\cite{Bressan98}),
including a new AGB mass-loss treatment calibrated on LMC cluster colors.
These SSPs are particularly suited for our analysis because they make use of  the MILES spectral library
(S{\'a}nchez-Bl{\'a}zquez et al.~\cite{sanb06}). MILES consists of  2.3 \AA \ FWHM 
optical spectra of $\approx$ 1000 stars spanning a large range in atmospheric parameters,
and represents a substantial improvement over previous libraries used in population synthesis models. 

For our study, we considered the galaxy spectra extracted in 4 concentric annulii 
(r $\leq$r$_e$/16, r$_{e}$/16 $<$ r $\leq$r$_e$/8, r$_{e}$/8$<$ r $\leq$r$_e$/4, and r$_{e}$/4 $<$ r $\leq$r$_e$/2),
as described in Section~2. 
The galaxy continua were fitted with the new SSPs, 
previously smoothed to match the instrumental resolution and the galaxy velocity dispersion,
through a $\chi^2$ criterion. 
In the fit, we considered selected spectral regions chosen to be particularly 
sensitive either to age or to metallicity.
For each SSP of given age and metallicity, 
and ${\rm E(B-V)}$ ranging from the Galactic foreground value to  ${\rm E(B-V)_{MAX}=0.5}$, 
the $\chi^2$ is:

 \begin{equation}
{\rm  \chi^2 =  \frac{1}{Nr}  \sum_{i=1}^{Nr}  \frac{1}{(N\lambda)_i} \sum_{j=1}^{(N\lambda)_i} \frac{(F_{obs}(\lambda_j) - F_{SSP}(\lambda_j) \times C_{norm,i})^2 }  { (\sigma (\lambda_j))^2} }   
\end{equation}

where ${\rm Nr}$ is the number of selected spectral regions, ${\rm (N\lambda)_i}$ is the number of points in the 
$i$-th region, ${\rm F_{obs}(\lambda_j)}$ is the observed galaxy flux at a given wavelength, 
${\rm F_{SSP}(\lambda_j) \times C_{norm,i}}$ is the SSP flux normalized to the observed flux
in the $i$-th region, ${\rm \sigma (\lambda_j)}$ is the error on the observed flux.
The ${\rm Nr}$ spectral regions have widths between $\approx 100$ \AA \ 
 and  $\approx 300$ \AA. The normalization constant ${\rm C_{norm,i}}$ 
is computed in a  $\approx 20$ \AA \ wide region adjacent to the $i$-th band.
This approach guarantees that the solution is mostly driven by the 
strengths of the absorption features, rather than 
by the global slope of the observed galaxy spectrum.
This minimizes the effect of extinction 
and of the uncertainties due to flux calibration on the solution. 
The spectral regions were selected in order to include both features particularly sensitive to age
(the relative strength of the Ca II H + K lines,  the 4000 \AA \ break ($D_{4000}$) , the H$\gamma$ and H$\delta$ lines), and features more sensitive  to metallicity (the Fe lines at $\lambda$ 4383, 4531, 5270, 5335, and the Mg absorption features around $\lambda$ 5175).
The region around  H$\beta$, where emission is expected, was obviously excluded from the fit. 
On the other hand, negligible emission is expected in the higher-order Balmer lines (H$\gamma$, H$\delta$ and H$\epsilon$, blended with the CaII H line at our resolution).
The residual spectrum around each line is derived 
by normalizing the SSP in two continuum bands adjacent to
the line of interest.

Because of the degeneracy between age, metallicity, and extinction, 
fits with similarly good  $\chi^2$ values 
can produce significantly different residual spectra, in 
particular around H$\beta$.
Thus, to obtain a statistically meaningful determination of the emission lines,
we considered all the N fits with ${\rm \chi^2 < 2}$, and computed the emission lines 
on the corresponding residual spectra.

 For each line, we adopted the average emission flux  ${\rm F_{\mu}}$ computed as:

 \begin{equation}
 {\rm 
\displaystyle   F_{\mu}=   \bigg( \sum_{\chi^2 < 2}  e^{-\chi^2} \times F_{Z,t,A_V}  \bigg)  \Bigg/  \bigg(   \sum_{\chi^2 < 2} e^{-\chi^2} \bigg)}
\end{equation}

where  ${\rm F_{Z,t,A_V}}$ is line flux obtained by subtracting the 
SSP of metallicity Z, age t, and reddening ${\rm A_V}$ to the observed galaxy spectrum,
and  ${\rm e^{-\chi^2}}$ is its weight.
The line flux  ${\rm F_{Z,t,A_V}}$ is determined by 
fitting the residual spectrum with Gaussian curves 
of variable width and intensity.
The emission line width is treated as a free parameter because the 
velocity dispersion of the gas can be significantly different from that of the stars.
More specifically, single Gaussians were used to fit 
all the lines, with the exceptions of [SII]$\lambda\lambda$6717, 31,
which was fitted with a sum of 2 Gaussians, and the 
 [NII]$\lambda\lambda 6548, 84$ $+$ H$\alpha$ complex, which was fitted with a sum of 3 Gaussians. 
 The [OII]$\lambda$3727 feature is actually a doublet ($\lambda\lambda 3726, 29$),
 but the two lines appear completely blended  in our spectra.
 
The error on the average emission flux is:

\begin{equation}
{\rm 
\displaystyle   \sigma_{\mu}^2=   \bigg( \sum_{\chi^2 < 2}  e^{-\chi^2} \times (F_{Z,t,A_V} - F_{\mu})^2  \bigg)  \Bigg/  \bigg(   \sum_{\chi^2 < 2} e^{-\chi^2} \bigg)
}
\end{equation}

This error takes into account the uncertainty due to the intrinsic degeneracy of the 
stellar populations. 
Because of the high signal-to-noise of our spectra,
this is actually the largest source of 
error in the measure of the emission fluxes.

 The use of a single SSP in the fitting procedure may lead to spurious effects  where recent star formation is present in some amounts.  We showed in Paper~III that some galaxies 
in our sample have very young luminosity weighted ages ($<$4 Gyr), possibly as a consequence 
of ``rejuvenation'' episodes. Therefore, we repeated the fitting procedure using, instead of one SSP, a combination 
of a young (t $<$ 8 Gyr) and an old (t$>$ 8 Gyr) SSP. This approach guarantees a more reliable spectral 
subtraction under the Balmer lines for the ``rejuvenated'' galaxies.
The two-SSP fitting procedure results in some differences for the galaxies with faint H$\beta$ emission.
This may be important when discriminating between LINERs and Seyfert
through the [OIII]/H$\beta$ ratio (see Section~4). On the other hand, we checked that there are no significant differences in 
the resulting emission lines for galaxies randomly selected among those without signs of recent star formation.

%------------------------ Figure 4 -----------------------------------
 
%MACRO: /Users/annibali/science/ellittiche/paperIV/plots_xpaper/ew.sm; command: go
%read file myrun_err/EWs.sample.txt produced by ./myrun_err/tabellepaper.f90
%it uses all galaxies, also those with no-emission
%new macro to do together both E/S0 and F/S is all_ew.sm
  \begin{figure*}
   \centering
  \includegraphics[width=19cm]{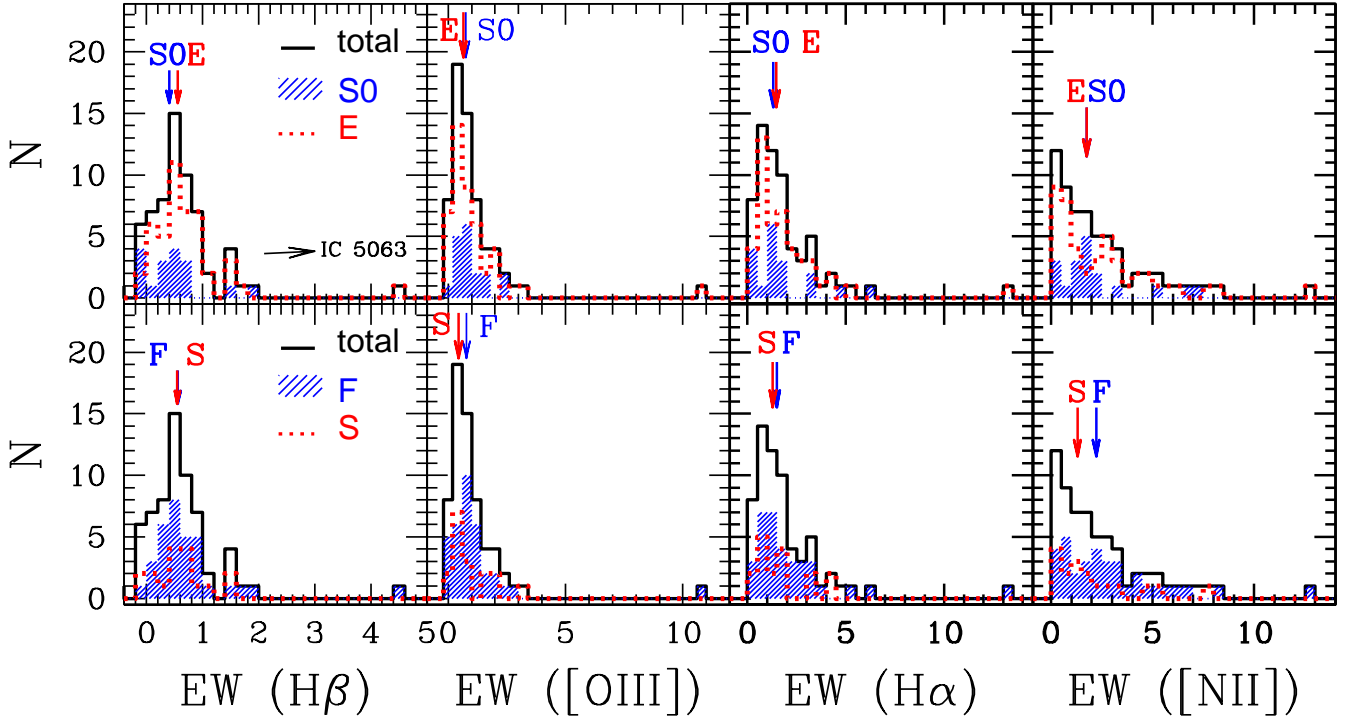}
  \caption{H$\beta$, [OIII]$\lambda$ 5007, H$\alpha$ and [NII] $\lambda$ 6584 EW distributions at 
  r $\leq$r$_e$/16 for the total galaxy sample, for the S0 and E subsamples,
  and for the fast (F) and slow (S) rotators subsamples.
   IC~5063, with EW(H$\beta$) $\sim$ 15,  EW([OIII]) $\sim$ 122, EW(H$\alpha$) $\sim$ 58,
  and EW([NII]) $\sim$ 38, falls outside the plot boundaries.
  The EWs are not corrected for extinction. The vertical arrows indicate the median values 
  for the E, S0, F, and S subsamples. }
\label{ew}
\end{figure*}
%------------------------ end Figure 4 ----------------------------------

In Fig.~\ref{spec} we show the stellar continuum subtraction procedure for two illustrative examples:
IC~1459, with relatively strong emission lines, and NGC~5813, with weak emission lines.
Notice that in the figure the subtracted stellar continuum is an SSP of age, metallicity, and extinction equal
to the average of all the N solution with ${\rm \chi^2 < 2}$. This is slightly different from our procedure,
where we have subtracted the individual solutions 
N times from the same galaxy spectrum, and have computed the final emissions as average of the 
N emission fluxes.
In Fig.~\ref{spec}, the emission features are easily visible in IC~1459. The errors on the emission line fluxes 
introduced by the uncertainties in the stellar continuum subtraction are thus relatively small.
In the case of NGC~5813, instead, the determination of the emissions is dominated by 
starlight subtraction effects, in particular for H$\beta$ and H$\alpha$. Consequently the errors 
on the emission lines are quite large.

We provide in Table~2 the emission line fluxes in units of 
${\rm 10^{-16}}$ erg s$^{-1}$ cm$^{-2}$ arcsec$^{-2}$ derived for the R05$+$A06 sample 
in 4 annuli of increasing galacto-centric distance 
(r $\leq$r$_e$/16, r$_{e}$/16 $<$ r $\leq$r$_e$/8, r$_{e}$/8$<$ r $\leq$r$_e$/4, and r$_{e}$/4 $<$ r $\leq$r$_e$/2).
In Table~3 we provide the line equivalent widths (EW), derived as 
${\rm F_{\mu}/F_{cont}}$, where ${\rm F_{cont}}$
is the flux computed in a continuum band adjacent to the line of interest.  
The high errors in H$\beta$ are due 
to its intrinsic faintness, accompanied by a strong
dependence of the H$\beta$ absorption line on age.
The values given in Tables~2 and 3 are not corrected for reddening.
The extinction corrected emissions can be easily derived from the ${\rm E(B-V)}$ values given 
in Col.~3 of Table~2, and computed as described in the next paragraph.

%------------------------- Figure 5 -------------------------------------------
%MACRO: /Users/annibali/science/ellittiche/paperIV/plots_xpaper/ew.grad.sm; command: vaiall
%read file myrun_err/lista_clean_gr and NGC#_myemiss.dat.
%EWs not corrected for emission.
%The two black empty points at lg2 are !C 5063 (which decreases at larger radii) and NGC 6721
%The levels are at 1, 0.68,0.4688,0.316, the sigma are 0, 0.24, 0.2479, 0.286.
%the macro to produce the levels is plots_xpaper/ew.grad.sm, command stat.

 \begin{figure}[!h]
   \centering
  \includegraphics[width=8.5cm]{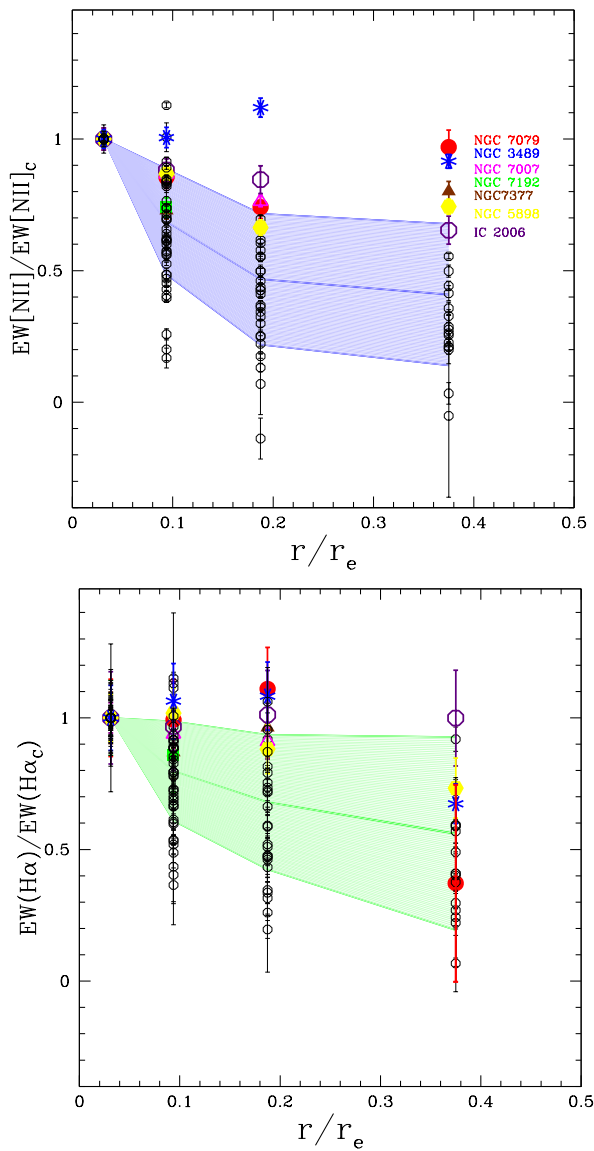}
  
  \caption{Emission EW derived in annuli of increasing galacto-centric distance 
  (r $\leq$r$_e$/16, r$_{e}$/16 $<$ r $\leq$r$_e$/8, r$_{e}$/8$<$ r $\leq$r$_e$/4, and r$_{e}$/4 $<$ r $\leq$r$_e$/2) over the central value (r $\leq$r$_e$/16)  as a function of ${\rm r/r_e}$. 
  The top and bottom panels refer to the [NII] 6584 \AA \ and H$\alpha$ lines, respectively.
   The shaded regions indicates the average trends and 
   the $\pm$ 1 $\sigma$ levels. Only galaxies classified as H or W in Table~4 were included.
  Galaxies with flat or increasing emission trends were labeled.}
\label{ewgrad}
\end{figure}
%%------------------------ end Figure 5 -----------------------------------

\subsection{Extinction}

The extinction in the lines was derived through the relative observed strengths of the Balmer lines.
The intrinsic flux ratio ${\rm (FH\alpha / FH\beta)_0}$   
is  $\approx$ 2.85 for HII regions and  $\approx$ 3.1
for AGN-like objects (Osterbrock~\cite{Osterbrock89}). 
Values larger than these are due to absorption
from an intervening medium.

More specifically, given an observed  ${\rm FH\alpha / FH\beta}$ ratio,
the reddening is computed as: 

\begin{equation}
{\rm 
\displaystyle  E(B-V) = \frac{  \log_{10} \left[ (FH\alpha / FH\beta )/ (FH\alpha / FH\beta)_0 \right] }{ 0.4 \times R_V \times (A_{H\beta}/A_V - A_{H\alpha}/A_V)      }
}
\label{eqebv}
\end{equation}

where ${\rm A_{\lambda}}$ is the magnitude attenuation at a given wavelength, 
and  ${\rm A_{H\beta}/A_V}$, ${\rm A_{H\alpha}/A_V}$, and  ${\rm R_V = A_V / E(B-V)}$, 
depend on the adopted extinction curve. 
We adopted the Cardelli et al. (1989) extinction law,  where ${\rm R_V = 3.05}$.
We also assumed ${\rm (FH\alpha / FH\beta)_0=3.1}$,
since all the galaxies in the R05$+$A06 sample display AGN - like emission-line 
ratios (see Section~4). The assumption of ${\rm (FH\alpha / FH\beta)_0=2.85 }$
does not change significantly our results.

In Figure~\ref{havshb}, we plot the H$\alpha$ versus H$\beta$
fluxes derived in a central r $\leq$r$_e$/16 region.
The majority of the galaxies have ${\rm E(B-V) < 0.3}$.
Some galaxies display instead very strong extinction, as high as
 ${\rm E(B-V) \approx 1.5}$ or even more.
 They have very faint H$\beta$  intensities,
 accompanied by relatively strong H$\alpha$ emissions. 
 Vice versa,  a few objects fall below the ${\rm E(B-V) =0}$ line, in the
 region of  ``negative reddening''. In these cases, the emission derived in 
H$\beta$ is incompatible (too large) with the derived H$\alpha$. 
An obvious explanation is that the underlying continuum was not 
subtracted properly. More specifically, the fact that we underestimate the 
H$\alpha$/H$\beta$ ratio suggests that we have subtracted populations 
that are systematically too young or too metal poor.
To verify this hypothesis, we simulated the situation in which a too-young population 
is subtracted from the galaxy spectrum in the following way.
We created a synthetic spectrum by adding H$\beta$ and H$\alpha$ emission lines 
(in the ratio ${\rm FH\alpha = 3.1 \   FH\beta}$) to an old SSP.
Then we re-derived the emission intensities by subtracting from the
synthetic spectrum increasingly younger SSPs. 
We obtained that the ${\rm FH\alpha / FH\beta}$ ratio 
decreases as the $\Delta$ age increases.
The effect is as much stronger as the initial emission
intensities are weaker. It is due to a combination of 
the  stronger dependence of the H$\beta$ absorption line on age,
 and of the intrinsic faintness of the H$\beta$ emission line 
 with respect to H$\alpha$. 
 
 Finally, there are some galaxies 
 whose H$\beta$ and H$\alpha$ emissions are both 
 consistent with zero within a 3 $\sigma$ error (NGC~1366, NGC~1389, NGC~1407, NGC~1426, 
 NGC~5638, NGC~5812, and NGC~5831).
 These galaxies show no emission in the other lines as well, and are classified as 
 {\it no emission} (N) galaxies.

The derived  ${\rm E(B-V)}$ values at different
galacto-centric distances are  given in Col.~3 of Table~2.
For the galaxies with ``negative reddening'', and for the galaxies with ${\rm E(B-V)}$ lower than
the Galaxy foreground extinction, the ${\rm E(B-V)}$ was set to 
the foreground value.
Very large  ${\rm E(B-V)}$ values are derived for 
NGC~777, NGC~1521, NGC~3136, NGC~3557, NGC~6776, NGC~7007, 
NGC~3489. Such strong extinction values, not observed in the 
continuum, suggest that the distribution of the dust is patchy. 
Notice however that the errors in the derived ${\rm E(B-V)}$ values are quite large, 
due to the high uncertainty in the H$\beta$ emission. 
The reddening distributions for the E and S0 classes
 are shown in Fig.~\ref{ebv}. The median ${\rm E(B-V)}$ values for the two classes are 
 0.11 and 0.26 respectively. However, a Mann-Withney U test shows that 
 there is not a significant difference between the two distributions.

Once computed the ${\rm E(B-V)}$ values, 
we corrected the observed emission fluxes:

\begin{equation}
{\rm 
\displaystyle  F_{corr} = F_{obs} \times 10^{0.4 \times A},
\label{corr}
}
\end{equation}

where 

\begin{equation}
{\rm 
\displaystyle  A = \left( \frac{A_{\lambda}}{A_V} \right) \times 3.05 \times E(B-V)
}
\end{equation}

${\rm \left( \frac{A_{\lambda}}{A_V} \right)}$ is taken from the extinction curve, and  
$\lambda$ is the wavelength of the line of interest.

\subsection{Results}

The galaxies are classified in Table~4 according to the intensity of their nuclear emission lines.
We identify three categories: no emission galaxies (N), where no emission lines are 
detected within a 3 $\sigma$ error; weak emission line galaxies (W), with EW(H$\alpha$ $+$ [NII]6584)$<$3 \AA; 
and strong emission line galaxies (H), with EW(H$\alpha$ $+$ [NII]6584)$>$3 \AA.

Emission lines are detected in 58 out of 65 galaxies ($\sim$ 89\% of the sample), while 
strong emission is present in 57\% of the sample.
In the central r $\leq$r$_e$/16 region, we derive median EWs of $\approx$  0.51, 0.72, 1.4 and 1.8  
for H$\beta$, [OIII]$\lambda 5007$, H$\alpha$, and [NII]$\lambda$ 6584, respectively.
For the same lines, the median percentage errors amount to 
$\approx$ 31\%, 11\% , 6\% , and , 1\%.

 We investigated possible differences  in the emission intensity between the 
E and S0 subsamples, and between the fast (F) and slow (S) rotators subsamples.
The  distributions and the relative median values are shown in Fig~\ref{ew}.
An appreciable difference is observed only between F and S 
for the median [NII] value.
However, Mann-Whitney U tests provide P $\geq$ 0.05 (for a  two-tailed test) for all the distributions,
indicating that there is no significant difference between the E, S0, F and S subsamples.

 %--------------------------------------- Figure 6 --------------------------------------
%MACRO: /Users/annibali/science/ellittiche/paperIV/plots_xpaper/diagnostic_my.sm; command: vaiall
%DD are extintion corrected
\begin{figure*}[ht!]
   \centering
  \includegraphics[width=17cm]{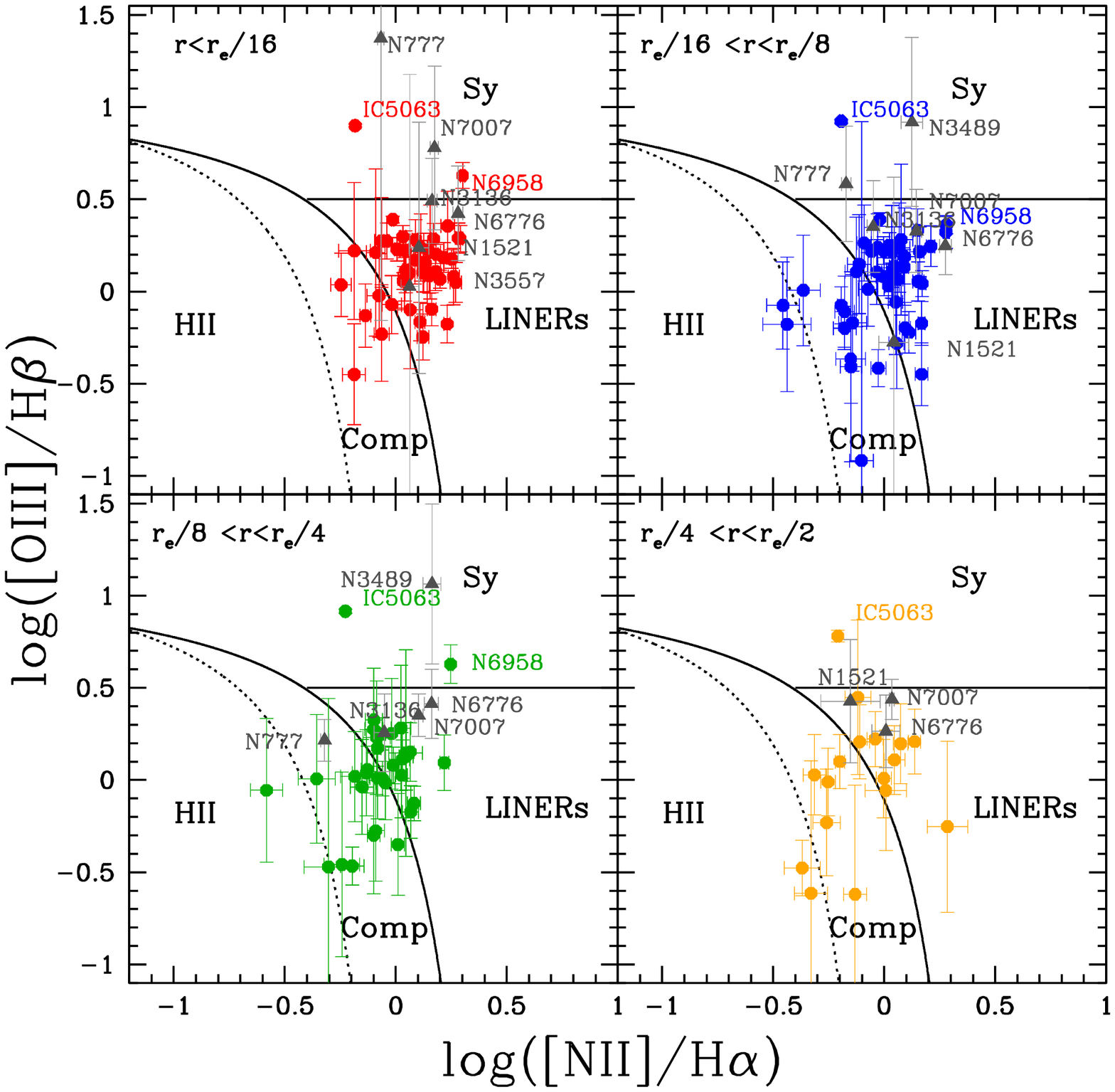}
  \caption{Extinction corrected BPT
  diagnostic diagram for the galaxy sample in annuli of increasing galacto-centric distance
 (r $\leq$r$_e$/16, r$_{e}$/16 $<$ r $\leq$r$_e$/8, r$_{e}$/8$<$ r $\leq$r$_e$/4, and r$_{e}$/4 $<$ r $\leq$r$_e$/2).
 Triangles are for galaxies with very large reddenings 
 (${\rm E(B-V)>1.5}$ at r $\leq$r$_e$/16).  
 The solid curve is the ``maximum starburst line'' of Kewley et al.~(\cite{Kewley01}), while
 the dashed line indicates the empirical division between pure star-forming galaxies and
 AGN-HII composite (or transition) objects from Kauffmann et al.~(\cite{Kauff03}).
 The horizontal line at ${\rm \log([OIII]/H\beta)=0.5}$ separates ``Seyfert'' and LINERs galaxies
 (Kewley et al.~\cite{kew06}). }
\label{bpt}
\end{figure*}
%--------------------------------- end Figure 6 ------------------------------------------------

To quantify the importance of the emission with respect to the underlying stellar continuum at 
increasing galacto-centric distance, we studied the behavior of the 
[NII]$\lambda 6584$ and of the H$\alpha$ lines in the 4 annuli defined in Section~2.
Notice that the use of adjacent annuli instead of ``apertures'' allows us to isolate the true 
emission contribution from different regions within the galaxy.
The [NII]$\lambda 6584$ line is relatively stronger, and thus 
more easily measured also in the external annuli. Furthermore, it is also very poorly affected 
by uncertainties in the underlying stellar population subtraction, and it has the lowest percentage error.
On the other hand the H$\alpha$ presents the advantage that, being a recombination line, directly reflects the ionizing photon flux, but its measure is much more affected than the [NII] by uncertainties in the underlying population subtraction.
The behaviors of the observed [NII]/([NII]$\leq$r$_e$/16) and H$\alpha$/(H$\alpha$$\leq$r$_e$/16) ratios 
with increasing galacto-centric distances are shown in Fig.~\ref{ewgrad}.
Negative values are consistent with no emission.
The plotted values were not corrected for reddening, but 
extinction could affect the EWs if the
nebular emission was more reddened than the continuum.
Indeed, Tran et al.~(\cite{tr01}) showed that the presence of emission lines in ETGs
tends to be associated with the presence of clumpy dust.

 Fig.~\ref{ewgrad} shows that, 
with some exceptions (NGC~3489, NGC~5898, NGC~7007, NGC~7079, NGC~7192, NGC~7377), the emission EWs tend to decrease from the center outwards.
The average decrease in the [NII] EW is $\approx$ 0.7, 0.5, and 0.4 of the central value 
at $\sim$0.1 r$_e$, $\sim$0.2 r$_e$, and $\sim$0.4 r$_e$, while the H$\alpha$ EW decreases more gently 
($\sim$ 0.8, 0.7, and 0.6  at the same radii). The steeper decrease of the [NII] compared 
to the H$\alpha$ indicates a progressive softening of the ionizing spectrum from the galaxy center to more
external regions. The decrease in the H$\alpha$ EW implies that the ionizing flux decrease more rapidly 
than the stellar continuum around $\sim$ 6500 \AA. However, given the uncertainties, we can consider 
significant only the decrease from the center to  $\sim$ 0.1 r$_e$. 
Deeper observations are needed to trace with higher confidence 
the EW behavior out to larger radii.

The galaxies NGC~7079, NGC~3489, NGC~7007, NGC~7192, NGC~7377, NGC~5898 deviate from 
the average decreasing trend, presenting flat or increasing emission EWs (although the H$\alpha$ 
of NGC~3489, NGC~5898 and  NGC~7079 drops in the most external annulus).
NGC~3489 and NGC~7007 have large reddenings, thus their increasing trend could be due to stronger extinction toward the galaxy center. 
Unfortunately, the errors in the derived $E(B-V)$ values are large, and it is difficult to establish 
from our data if indeed the reddening decreases with galacto-centric distance.
On the other hand, the other galaxies (NGC~7079, NGC~7192, NGC~7377, and NGC~5898) show very little or almost null extinction, thus the observed trends are likely to be intrinsic.

\section{Spectral classification}

\subsection{Diagnostic Diagrams}

%------------------------------------- Figure 7 -----------------------------------------
%MACRO: /Users/annibali/science/ellittiche/paperIV/plots_xpaper/n2ha.grad.sm; command: go
%DD are extintion corrected
\begin{figure}
   \centering
  \includegraphics[width=8.5cm]{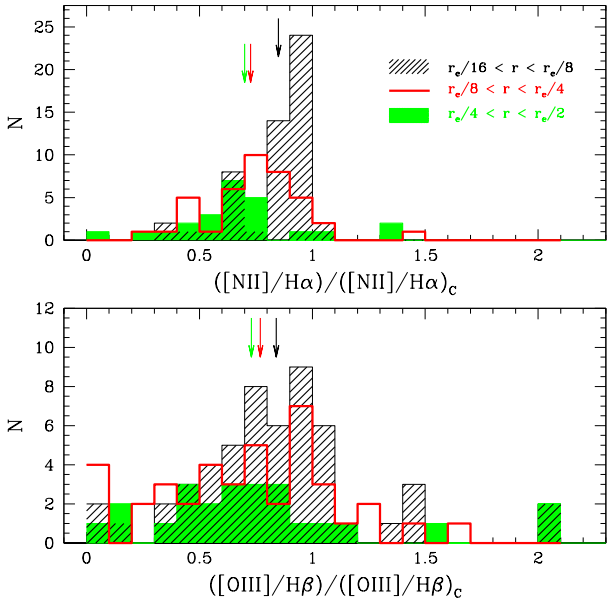}
  \caption{
Distribution of emission line ratios over the central r $\leq$r$_e$/16 value
for the galaxies in the four BPT diagrams of Fig.~\ref{bpt}.
The dashed, empty, and full histograms correspond respectively to the 
r$_{e}$/16 $<$ r $\leq$r$_e$/8, r$_{e}$/8$<$ r $\leq$r$_e$/4, and r$_{e}$/4 $<$ r $\leq$r$_e$/2) annuli.
Vertical arrows indicate the average values. From the galaxy center outwards,
the average [NII]/H$\alpha$ / [NII]/H$\alpha_{C}$ ratios drop to  
0.85, 0.73, and 0.69 in the three annuli, respectively.
For the same annuli, we derive average [OIII]/H$\beta$ / [OIII]/H$\beta_{C}$
values of 0.92, 0.77, and 0.64.}
\label{bptgrad}
\end{figure}
%-------------------------------------------- end Figure 7 -----------------------------

\addtocounter{table}{1}
\addtocounter{table}{1}

%------------------------------ Table 4 ----------------------------------
\begin{table}
\begin{minipage}[t]{\columnwidth}
\caption{Classification from optical lines in the nuclear (r $\leq$r$_e$/16) region}\label{class}
\centering
\renewcommand{\footnoterule}{} 
\tiny{
\begin{tabular}{lcc}
\hline \hline
Galaxy Id & Emission Type\footnote{N$=$no emission; W$=$weak emission 
(${\rm EW(H\alpha + [NII]6584)<}$3 \AA); H$=$high emission (${\rm EW(H\alpha + [NII]6584)>}$3 \AA)
 } & Activity Class\footnote{At larger radii, some LINERs are classified as Composites.}
 \\ 
\hline
 NGC~128   &  H  & LINER   \\
 NGC~777   &  W & Sy/ LINER   \\
 NGC~1052 &  H  & LINER  \\
 NGC~1209 &  H  & LINER   \\
 NGC~1297 &  H  & LINER   \\
 NGC~1366 &  N  &  \\
 NGC~1380 &  H   & LINER  \\
 NGC~1389 &  N  & \\
 NGC~1407 &  N  & \\
 NGC~1426 &   N & \\
 &  \\
 NGC~1453 &  H  &  LINER   \\
 NGC~1521 &  W &  LINER  \\
 NGC~1533 &  H  & LINER  \\
 NGC~1553 & W  &  LINER  \\
 NGC~1947 &  H  & LINER \\
 NGC~2749 &  H  & LINER  \\
 NGC~2911 &  H  & LINER  \\
 NGC~2962 &  H &  \\
 NGC~2974 &  H & LINER \\
 NGC~3136 & H & LINER/ Sy   \\
 & \\
 NGC~3258 &   H  & Comp  \\   
 NGC~3268 &   H  & LINER   \\ 
 NGC~3489 &   H  & Sy/ LINER    \\
 NGC~3557 &   W &  LINER    \\
 NGC~3607 &   H  &  LINER   \\
 NGC~3818 &   Traces &   \\
 NGC~3962 &   H  & LINER    \\
 NGC~4374 &   H  &  LINER   \\
 NGC~4552 &   W & Comp  \\
 NGC~4636 &   H &  LINER  \\
 &  \\   
 NGC~4696 & H & LINER  \\
 NGC~4697 & W & LINER   \\
 NGC~5011 &  W & LINER  \\
 NGC~5044 &  H  & LINER   \\
 NGC~5077 & H  & LINER  \\
 NGC~5090 & H  & LINER   \\
 NGC~5193 & W & Comp   \\
 NGC~5266 & H & LINER  \\
 NGC~5328 & W/Traces & Comp  \\
 NGC~5363 & H  & LINER  \\
  & \\  
 NGC~5638 & N &  \\ 
 NGC~5812 & N &  \\
 NGC~5813 & W &  LINER \\
 NGC~5831 &  N & \\
 NGC~5846 &  H & LINER  \\
 NGC~5898 &  W  & LINER  \\
 NGC~6721 & W/Traces & Comp  \\
 NGC~6758 &  H & LINER \\ 
 NGC~6776 & H &  LINER/ Sy  \\
 NGC~6868 & H  & LINER  \\
  & \\ 
NGC~6875 & W &    \\
NGC~6876 & W /Traces & Comp    \\
NGC~6958 &  H  & Sy/ LINER   \\
NGC~7007 &  W  & Sy/ LINER   \\
NGC~7079  &  W & LINER  \\
NGC~7097 &  H  & LINER   \\
NGC~7135 &  H  & LINER   \\
NGC~7192 & W & LINER    \\
NGC~7332 &  W/Traces &  \\
NGC~7377 &  W  &   LINER  \\
& \\
IC~1459    & H & LINER   \\ 
IC~2006    & W & Comp    \\
IC~3370    & H & LINER   \\
IC~4296    & H & LINER    \\
IC~5063    &  H & Sy \\
\hline
\end{tabular}
}
\end{minipage}
\end{table}
%------------ end Table 4 -------------------------------------

Galaxies were classified through the standard [OIII]$\lambda$5007/ H$\beta$ versus 
[NII]$\lambda$6584/ H$\alpha$ diagnostic diagram (Baldwin, Phillips and Terlevich~(\cite{Baldwin81}; hereafter BPT).
%, and then revised by Veilleux \& Osterbrock (\cite{Veil87}).
Using a combination of stellar population synthesis models and photoionization models,
Kewley et al.~(\cite{Kewley01}) identified a ``maximum starburst line'' in the BPT diagram.
Galaxies lying above this line are likely to be dominated by an AGN.
Kauffmann et al.~(\cite{Kauff03}) later revised the Kewley et al.~(\cite{Kewley01}) classification scheme 
introducing  an empirical line to divide
pure star-forming galaxies from AGN-HII Composite objects (also called ``Transition objects''),
with significant contribution from both AGN and star formation
(see also Stasi{\'n}ska et al.~\cite{sta06}).
The optical spectra of composite objects can be due to either 
1) a combination of star formation and a Seyfert nucleus, or 2) a combination of star formation 
and LINERs emission (Kewley et al.~\cite{kew06}).

From the distribution in the BPT diagram of the galaxies from the Sloan Digital Sky Survey, 
Kewley et al.~(\cite{kew06})  introduced an empirical horizontal line to separate ``Seyfert'' and LINERs.
Even though LINERs show  [NII]/ H$\alpha$ ratios larger than the star-forming 
galaxies, indicating a harder ionization continuum, 
it is still under debate if they are powered by active galactic nuclei or not (see Section~1).

%----------------------------- Figure 8 -------------------------------
\begin{figure*}
   \centering
  \includegraphics[width=12cm]{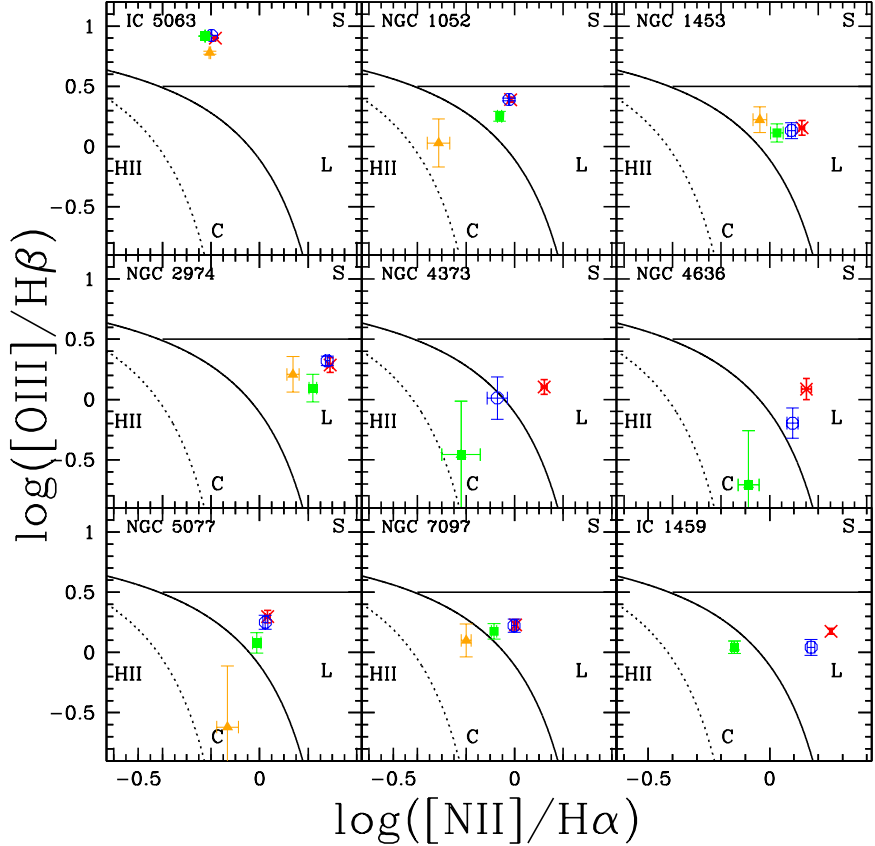}
  \caption{BPT diagrams  of some galaxies representing the general trends of the whole sample in the four annuli:
  r $\leq$r$_e$/16 (red cross), r$_{e}$/16 $<$ r $\leq$r$_e$/8 (blue open circle), 
  r$_{e}$/8$<$ r $\leq$r$_e$/4 (green full square), and r$_{e}$/4 $<$ r $\leq$r$_e$/2 (yellow full triangle). 
  Notice that while the error in  [OIII] / H$\beta$ is large, the error in [NII]/H$\alpha$ is significantly smaller.  
}
\label{ddex}
\end{figure*}
%-------------------- end Figure 8 ------------------------------- 

The BPT diagram for the R05$+$A06  sample is shown in Fig.~\ref{bpt}.
The [NII]$\lambda$6584/ H$\alpha$ line ratio was preferred to the
 [SII]($\lambda$6717 $+$ $\lambda$6731)/ H$\alpha$ and 
 [OI]$\lambda$6300/ H$\alpha$ ratios because
of the larger signal to noise of the [NII] line in our spectra.
Emission line ratios were corrected for extinction adopting the ${\rm E(B-V)}$ values
derived in Section~3.2 and reported in Col.~3 of Table~2, even though the BPT 
diagram is almost insensitive to reddening.
The galaxies with negative reddening, or reddening lower than the foreground value, where
corrected adopting the galactic foreground reddening from NED .

According to their nuclear (${\rm r<r_e/16}$) emission, the majority 
of our galaxies are classified as LINERs.
A few galaxies (NGC~3258, NGC~4552, NGC~5193, NGC~5328, NGC~6721, NGC~6876, IC~2006) fall in the region of ``Composites'', and possibly contain a combined contribution
from both star formation and AGN. 
IC~5063, NGC~777, NGC~3489, NGC~7007, and NGC~6958, fall in the Seyfert 
region.  Among them, only IC~5063 can be classified as a {\it bona fide} Seyfert galaxy.
The others all have very high uncertainties in the [OIII]/ H$\beta$ ratios, due to the low H$\beta$ 
emission (nuclear EW$<$0.3 \AA), and are consistent with a LINERs classification within the errors.
These galaxies also display very large reddenings (${\rm E(B-V)\gsim1}$).
 Vice versa, NGC~3136 and NGC~6776 are LINERs, but are consistent with being Seyferts within the errors.
Among the Seyferts, a spectral classification is present in the literature only for 
NGC~3489 (Ho et al.~\cite{ho97b}, Sarzi et al.~\cite{sar06}), NGC~777 (Ho et al.~\cite{ho97b}), and NGC~6958 (Saraiva et al. 2001). These studies classify the first two galaxies as Seyferts, and the last one as LINER.

The final spectral classification obtained from the nuclear (r $\leq$r$_e$/16) lines is given in Col.~3 of Table~4.

% ------------------------- Figure 9 ----------------------------

 \begin{figure*}
 \centering
  \includegraphics[width=16cm]{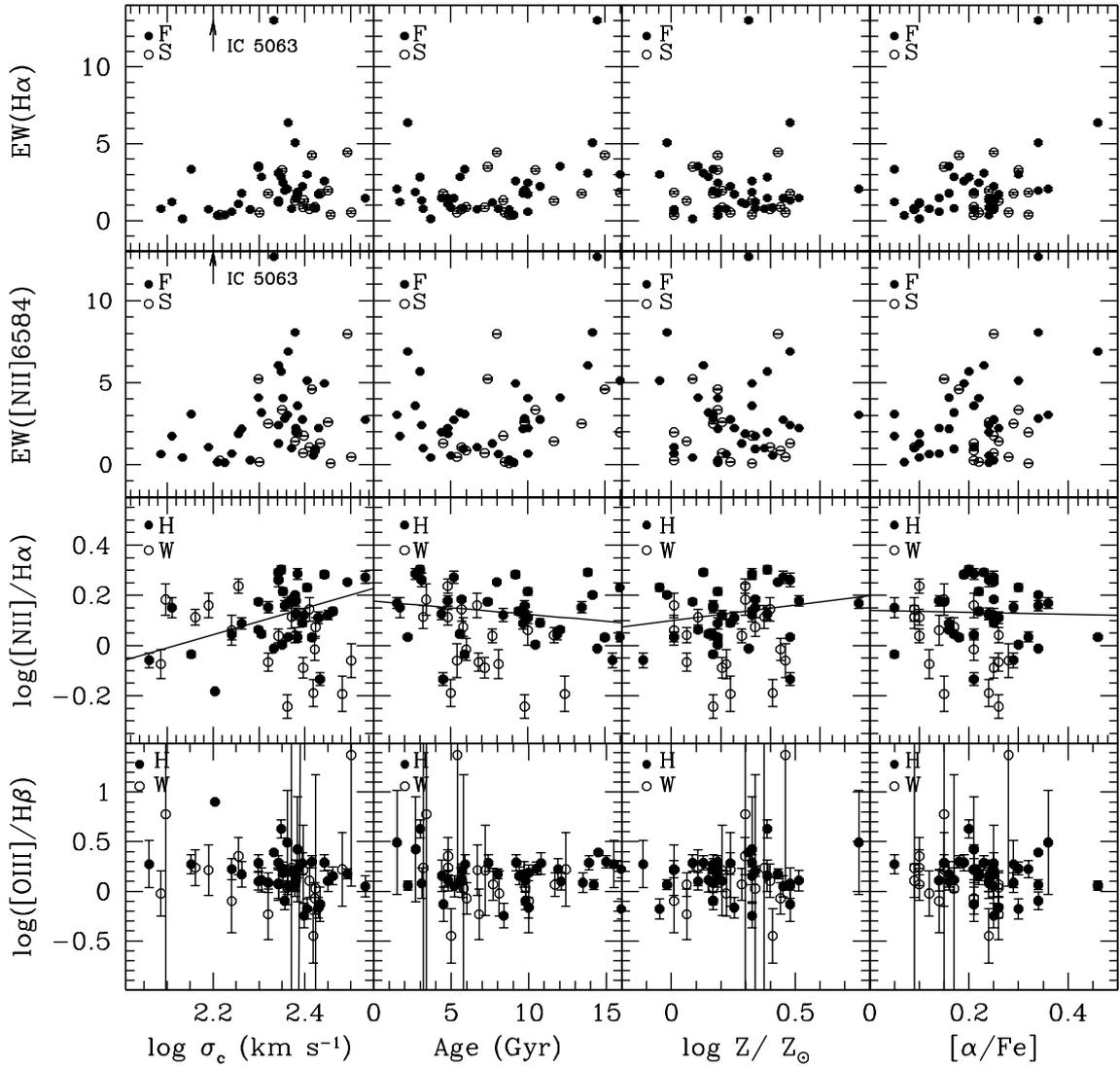}
  \caption{H$\alpha$ and  [NII] equivalent widths, [NII]/H$\alpha$ ratio, and [OIII]/H$\beta$ ratio versus 
  galaxy central velocity dispersion $\sigma_c$, age, metallicity Z, and [$\alpha$/Fe] ratio.
 In the two top rows, full dots are for fast rotators (F), while empty dots are for slow rotators (S).
 In the two bottom rows, full dots are for high emission galaxies (H), and empty dots
  are for weak emission galaxies (W), as defined in Section~3 and Table~4.
 For the third row panels, we show least square fits for the H subsample.}
\label{allcorr}
\end{figure*}
% ------------------------- end Figure 9 ----------------------------

% ------------------------- Figure 10 ----------------------------
\begin{figure*}
\centering
  \includegraphics[width=14cm]{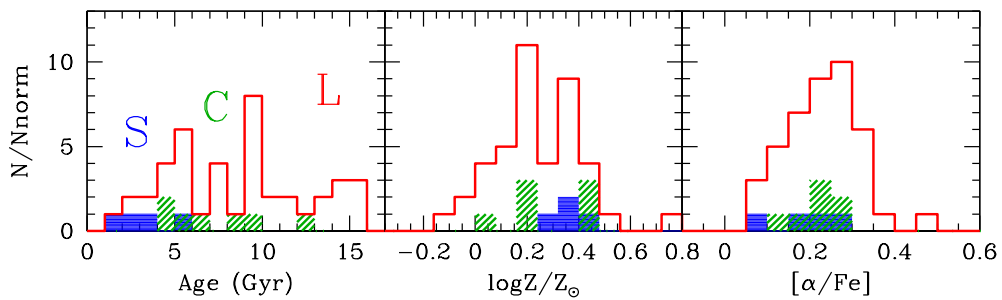}
  \caption{Age, metallicity, and [$\alpha$/Fe] distributions for Seyfert-like galaxies (S),
  LINERs (L), and ``Composites'' (C). 
  %The average ages for S, C and L are 3.8, 
 % 7.5, and 9 Gyr, respectively. For the same classes, the average metallicities 
 % are $\log Z/Z_{\odot}$  $\sim$ 0.22, 0.29, 0.4 dex. The average [$\alpha$/Fe] is always $\sim$ 0.2 .
 }
\label{scl}
\end{figure*}
% ------------------------- end Figure 10 ----------------------------

% ------------------------- Figure 11 ----------------------------
\begin{figure*}
\centering
  \includegraphics[width=14cm]{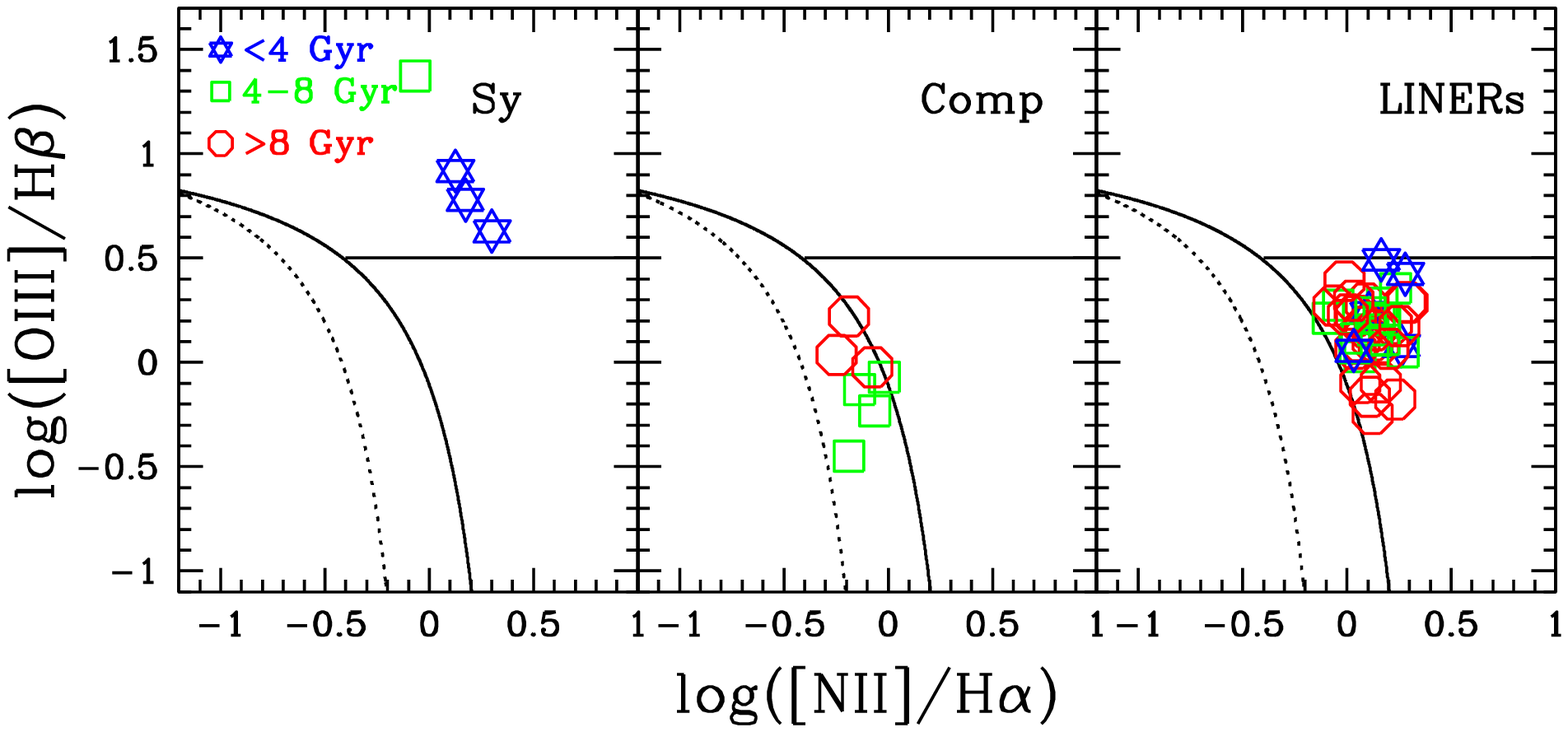}
  \caption{BPT diagnostic diagram for Seyfert galaxies (left panel),
  ``Composites'' (center), and LINERs (right panel).
 The data were color-coded according to the galaxy age 
 as derived in Paper~III: blue stars are for ages younger than 4 Gyr, 
 green squares for ages between 4 and 8 Gyr, and red circles 
 for ages older than 8 Gyr.
}
\label{3dd}
\end{figure*}
% ------------------------- end Figure 11 ----------------------------

 \subsection{Gradients}

We investigated how the distribution of the galaxies in the BPT diagram changes at annuli of increasing  galacto centric distance 
(r $\leq$r$_e$/16, r$_{e}$/16 $<$ r $\leq$r$_e$/8, r$_{e}$/8$<$ r $\leq$r$_e$/4, and r$_{e}$/4 $<$ r $\leq$r$_e$/2).
This emerges by comparing the four panels of Fig.~\ref{bpt}.
Only a few galaxies have sufficiently high S/N ratios that allows to measure the emission line ratios in the most external annuli.
 
The clear result is that, from the center outwards, the bulk of the galaxies 
moves left-down in the BPT diagram. An increasing number of galaxies pass from 
LINERs to ``Composite'' objects as the emission lines are measured at larger 
galacto-centric distances. Likely, this effect would not have been observed 
by considering apertures instead of annuli, 
since the central emission is almost always dominating.
The effect is better quantified in Fig.~\ref{bptgrad}, where we plot 
the distribution of the [NII]/H$\alpha$ and [OIII]/H$\beta$ ratios over the 
central r $\leq$r$_e$/16 values (C) for the different annuli.
The peak of the ([NII]/H$\alpha$)/([NII]/H$\alpha$)$_C$ distribution
moves from 1 to values $< 1$ at larger radii. The average ([NII]/H$\alpha$)/([NII]/H$\alpha$)$_C$
values are  $\approx$ 0.85, 0.73, and 0.69 at 
r$_{e}$/16 $<$ r $\leq$r$_e$/8, r$_{e}$/8$<$ r $\leq$r$_e$/4, and r$_{e}$/4 $<$ r $\leq$r$_e$/2, 
respectively. These correspond to horizontal shifts of $\approx$ 
$-0.07$, $-0.14$, and $-0.16$ dex  in the BPT diagram.
However, given the large errors in the emission line measures in the most external annulus,
we consider this result significant only out to r $\leq$r$_e$/4.
The ([OIII]/H$\beta$)/([OIII]/H$\beta$)$_C$ distribution is less peaked,
in part also because of the large errors in the derived [OIII]/H$\beta$ ratios.
The average values are  0.92, 0.77, and 0.64. for the three annuli,
corresponding to vertical shifts of  $\approx$ 
$-0.04$, $-0.11$, and $-0.19$ dex  in the BPT diagram.
However, given the intrinsic faintness and large errors of the [OIII] and H$\beta$ lines compared to [NII] and H$\alpha$, none of our conclusions will be based on the spatial behavior of the [OIII]/H$\beta$ ratio.

When the galaxies are inspected individually, we see that for the majority of 
them the [NII]/H$\alpha$ ratio decreases monotonically with galacto-centric distance. 
The only exceptions are:
IC~5063, NGC~1521, NGC~3136, NGC~3489, NGC~5813, NGC~5898, 
NGC~6721, NGC~6876, NGC~7135 and NGC~7192.
In all these galaxies the [NII]/H$\alpha$ ratio does not have a clear trend.
In NGC~3489, the ratio increases from the center outwards.
The [NII]/H$\alpha$ ratio decreases very steeply in IC~1459, IC~4296, NGC~3607, NGC~4374, NGC~4697, NGC~5090, and NGC~5266.
We show in Fig.~\ref{ddex} the individual BPT diagrams in different annuli for some galaxies of the sample.
The  [NII]/H$\alpha$ and  [OIII]/H$\beta$ ratios in the four annuli for the whole sample are shown in Appendix B. 

{\it Summarizing, the hardness of the ionizing continuum decreases with galacto-centric distance.}

\subsection{Central correlations with galaxy properties}

We investigated possible correlations between the emission line properties and the galaxy properties.
Our results are summarized in Fig.~\ref{allcorr}. 

In the top panels of Fig.~\ref{allcorr} we plotted the central ($r<r_e/16$)
H$\alpha$ and  [NII]$\lambda$6584 equivalent widths versus the central velocity dispersion $\sigma_c$, the age, the metallicity, and 
the $\alpha$/Fe enhancement. The H$\alpha$ and [NII] EWs quantify the strength of the emission with respect 
to the stellar continuum. However, the [NII] line is much less affected than the H$\alpha$ by uncertainties or systematics due to the stellar continuum subtraction.
The scatter in the H$\alpha$ and  and [NII] EW increases with $\sigma_c$. This still holds if the 
 [OIII] emission is considered. If we exclude the Seyfert galaxy IC 5063, strong emission 
is observed only for galaxies with $\sigma_c \gsim$ 200 km s$^{-1}$. The galaxies with the strongest emissions tend to be fast rotators. Also, all the low $\sigma_c$ galaxies are fast rotators.
Concerning the stellar population parameters, the trend is preserved only for the [NII] line versus 
the [$\alpha$/Fe] ratio (it is barely present in the H$\alpha$ vs [$\alpha$/Fe] plane).
We notice that this trend could be driven by the contamination from the 
[NI]5199 line to the Mgb index, implying larger derived [$\alpha$/Fe] values 
for galaxies with stronger emission (Goudfrooij \& Emsellem ~\cite{goem96}).
The lack  of a trend between the EW and metallicity is a bit surprising at the light of 
the well know metallicity-$\sigma$ relation for ETGs.

In the  third row panels of  Fig.~\ref{allcorr} we show the  central ($r<r_e/16$) [NII]$\lambda$5684/H$\alpha$ ratio, which is a measure of the hardness of the ionizing spectrum. In examining possible correlations between this emission line ratio and the galaxy properties, we separated the sample in high emission galaxies (H) and weak emission ones (W)
 (see Section~3.3 and classification in Table~4). In fact, the determination of the H$\alpha$ emission can be strongly affected by systematics in the underlying continuum subtraction in the W galaxies. To investigate the presence of correlations it is thus safer to consider only the H subsample.
The first panel in the third row shows that there is a positive correlation between 
 the central [NII]/H$\alpha$ and $\sigma_c$.
The Spearman correlation coefficient  $r_s$ is 0.31 (with 36 data points), 
implying that there is a 6\% probability that the variables are not correlated 
for a two-tailed test. However, if we restrict the analysis to the galaxies with 
EW([NII])$>3$, the probability that there is no correlation drops to 4\%. 
This result is in agreement with Phillips et al.(~\cite{ph86}), who derived a positive trend  between the 
[NII]/ H$\alpha$ ratio and the galaxy absolute magnitude.
On the other hand, Spearman tests show that there is no strong correlation between [NII]/ H$\alpha$ and the age, the metallicity, or the $\alpha$/Fe enhancement of the galaxy stellar population.

In the forth row panels of Fig.~\ref{allcorr} we show the central  
[OIII]/H$\beta$ ratio, which is related to the ionization 
parameter. No clear trend is observed with $\sigma_c$, age, metallicity, and [$\alpha$/Fe]. We only notice that 
the largest [OIII]/H$\beta$ values are associated to the youngest galaxy ages.

The relation between the nuclear emission spectral class (Seyfert, LINERs, composite), and the stellar population properties  is show in Figs.~\ref{scl}  and \ref{3dd}.
Seyfert  galaxies tend to have ages younger than $\sim$ 4 Gyr.
``Composite'' galaxies have ages older than 5 Gyr.
LINERs cover the whole age range from a few Gyrs to a Hubble time.
For the three classes, the average ages are 3.8, 7.5, and 9 Gyr, respectively.
The average metallicities are $\log Z/Z_{\odot}$ $\sim$ 0.39, 0.29, and 0.22, respectively. 
No significant difference is observed in the [$\alpha$/Fe] ratio.

It emerges that Seyferts tend to have luminosity-weighted ages significantly younger than LINERs,
likely because of recent star formation episodes.
The only caveat in this statement is that the only {\it bona fide} Seyfert in our sample (IC~5063),
does not have an age determination.

\section{Oxygen abundance}

If the ionizing source is known, one can derive the element abundance of the warm gas.
 In their nuclear (${\rm r<r_e/16}$) region,
the majority of the galaxies in our sample are classified as LINERs, a few are classified as 
``Composites'', and only IC~5063 is classified as a {\it bona fide} Seyfert galaxy.
We have also noticed a progressive increase of the fraction of ``Composites'' 
with respect to pure LINERs moving from the galaxy center 
to more external regions. 
As discussed in Section~1, the ionizing mechanism in LINERs/Composites is still far from being understood.
For this reason, we derived the oxygen metallicity for the R05$+$A06 sample 
in the two extreme assumptions of 1) photoionization by hot stars, and 2)  AGN excitation.
A calibration in the case of shock heating is not  present in the literature.

%------------------------------ Figure 12 -----------------------------------
 \begin{figure}
   \centering
  \includegraphics[width=8.5cm]{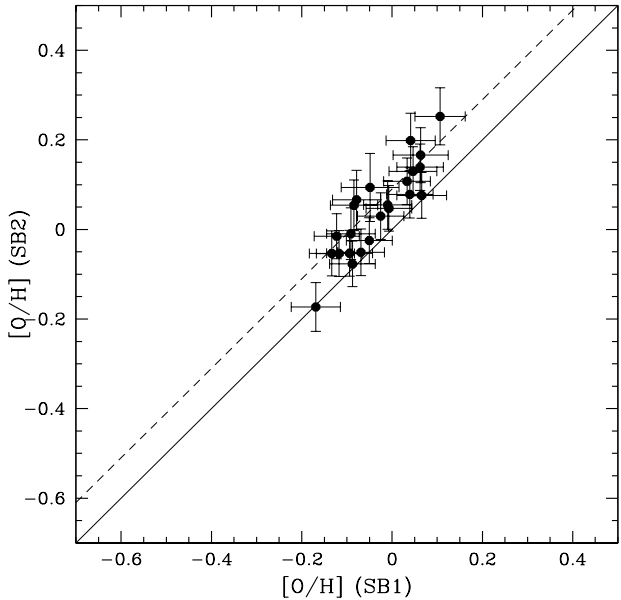}
  \caption{Oxygen abundance at r $\leq$r$_e$/16 obtained with the second calibration of 
  Storchi-Bergmann et al.~(\cite{Storchi98}) against the values obtained with the first
  calibration. For comparison, the one-to-one relation is drawn as a solid line,
  while the dashed line indicates a systematic shift of 0.09 dex.}
\label{cfr}
\end{figure}
%------------------------------ end Figure 12 -----------------------------------

%------------------------------ Figure 13 -----------------------------------
%macros: oxygen.sm, command distr
\begin{figure}
   \centering
  \includegraphics[width=8.5cm]{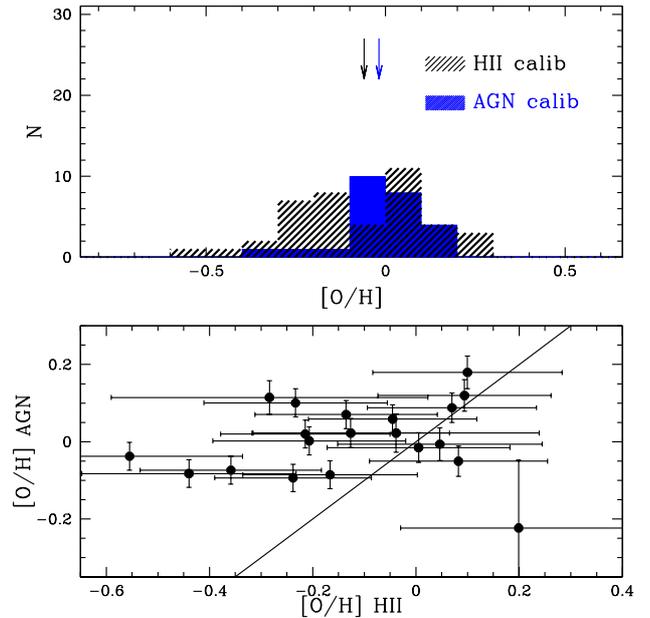}
  \caption{Top panel: oxygen abundance distributions   
  (${\rm [O/H]= \log(O/H) -\log(O/H)_{\odot}}$, where ${\rm 12 + \log(O/H)_{\odot} = 8.76}$ from 
  Caffau et al.~\cite{Caffau08}) derived for the R05$+$A06 sample  
  at r $\leq$r$_e$/16. The dashed histogram was obtained with the 
  Kobulnicky et al.~(\cite{Kobulnicky99}) calibration valid for HII regions, 
  while the full histogram was obtained with the Storchi-Bergmann et al.~(\cite{Storchi98}) 
  calibration for AGNs. Vertical arrows indicate the average [O/H] values obtained with the two methods.
Bottom panel: comparison of the metallicities obtained with the two methods 
for 20 galaxies in common. The solid line is the one-to-one relation.}
\label{ohdistr}
\end{figure}
%------------------------------ end Figure 13 -----------------------------------

%------------------------------ Figure 14 -----------------------------------
%sm oxygen.sm commnad gradi (only selected galaxies)
\begin{figure}
   \centering
  \includegraphics[width=8.5cm]{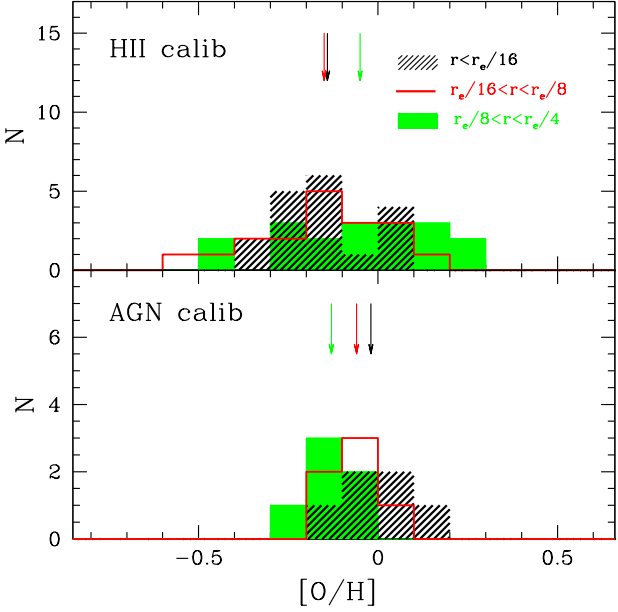}
  \caption{Oxygen abundance distributions in the three more internal annuli 
  (r $\leq$ r$_{e}$/16, r$_{e}$/16 $<$ r $\leq$r$_e$/8, r$_{e}$/8$<$ r $\leq$r$_e$/4)
  obtained with the Kob99 calibration (18 galaxies, top panel), and with the SB98 calibration
   (6 galaxies, bottom panel).
  Vertical arrows indicate the average values for the distributions.
 Notice that the metallicity increases from the galaxy center outwards 
 with the Kob99 calibration, while the opposite trend is obtained with 
 the SB98 calibration.
}
\label{ohgrad}
\end{figure}
%------------------------------ end Figure 14 -----------------------------------

%------------------------------ Figure 15 -----------------------------------
\begin{figure}
   \centering
  \includegraphics[width=8.5cm]{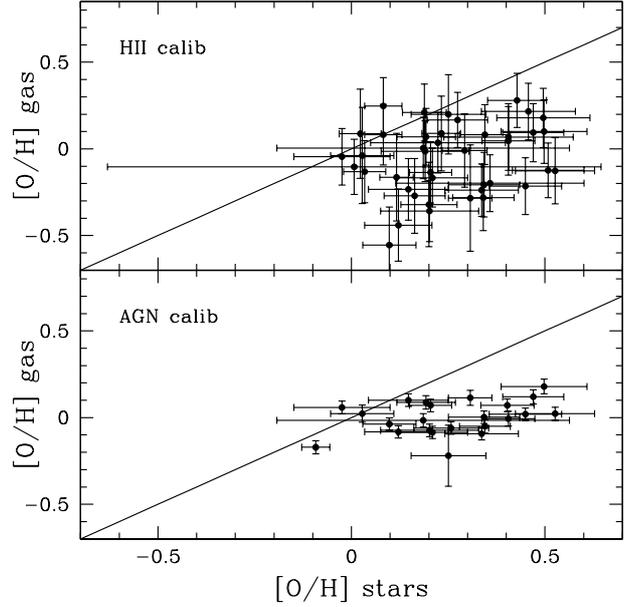}
  \caption{Nebular oxygen abundance versus star oxygen abundance.
  The latter was derived from the Z and [$\alpha$/Fe] values obtained in Paper~III.
 % To compute the stellar $[O/H]$ values, we assumed that oxygen traces the same abundances 
 % of the other $\alpha$-elements. The adopted solar oxygen abundance  is 
  %$[O/H]_{\odot}=8.76$, from Caffau et al.~(\cite{Caffau08}).
  In the top and bottom panels, the nebular metallicities were obtained with the 
  Kob99 and SB98 calibration, respectively.
  }
\label{stars}
\end{figure}
%------------------------------ end Figure 15 -----------------------------------

%------------------------------ Table 5 -----------------------------------

\begin{table}
\caption{Electron densities $n_e$ derived from [SII]6716/6731 assuming T=10,000 K} 
\label{ne}
\centering
\begin{tabular}{lcc}
\hline \hline
Galaxy &  [SII]6717/6731 & $n_e$ (cm$^{-3}$)\\
\hline
%      NGC~128  &    8.44  $\pm$     0.24    &      $-$    \\
     NGC~1052  &   1.11 $\pm$ 0.03    &   384.1   \\
%      NGC~777  &       $-$                     &    8.55 $\pm$      0.12  \\
     NGC~1209  & 1.19  $\pm$ 0.13  &    254.8  \\
     NGC~1297  &   1.08 $\pm$  0.10      &   423.6  \\
      NGC~1380  &   0.98 $\pm$  0.15    & 653.9  \\
  NGC~1453  &   1.28  $\pm$  0.09 &   141.5 \\
   %      NGC~1521  &       $-$                     &   8.68 $\pm$     0.06  \\
            NGC~1533  &  1.22  $\pm$ 0.16   &  212.0   \\
            NGC~1553  &  1.21  $\pm$ 0.30   &   229.3   \\
           NGC~1947  & 1.32    $\pm$ 0.04     &   95.0 \\
             NGC~2749  &  1.18  $\pm$ 0.08   &  267.7    \\
   %           NGC~2911  &   8.56  $\pm$     0.16    &  8.72  $\pm$    0.03  \\
      NGC~2974  &  1.10   $\pm$ 0.05  &  385.1 \\
     NGC~3136  &   1.60  $\pm$ 0.11 &   $<$26   \\
  %           NGC~3258  &     8.94  $\pm$      0.17    &       $-$                    \\
             NGC~3268  &  1.32 $\pm$ 0.13  &  103.4  \\
             NGC~3489  &  1.21 $\pm$ 0.14 &  223.1  \\
%     NGC~3557  &       $--$ $\pm$       5.442   &        $--$$\pm$       0.1697  \\
     NGC~3607  &  1.24  $\pm$ 0.20  & 185.4   \\
%     NGC~3818  &       $-$                     &     8.41 $\pm$      0.15  \\
     NGC~3962  &  1.27  $\pm$ 0.06  & 147.1     \\
      NGC~4374  &  1.45 $\pm$ 0.17  &  $<$26   \\
%           NGC~4552  &    8.98  $\pm$     0.16    &       $-$            \\
              NGC~4636  & 1.45 $\pm$ 0.15  &     $<$26        \\
            NGC~4696  &  1.31  $\pm$ 0.08 &  110.0  \\
%          NGC~4697  &   8.85  $\pm$     0.26    &       $-$         \\
%     NGC~5011  &       $--$ $\pm$       1.105   &        $--$$\pm$      0.04232 \\
    NGC~5044  &   1.51 $\pm$ 0.06  &    $<$26   \\
         NGC~5077  & 1.26 $\pm$ 0.06   & 158.7   \\
         NGC~5090  &  1.48 $\pm$ 0.20   &    $<$26   \\
         NGC~5193  & 1.57 $\pm$ 0.50  &       $<$26   \\
          NGC~5266  & 1.06 $\pm$ 0.04   & 476.7  \\
   %      NGC~5328  &    8.96  $\pm$     0.23    &   $-$   \\
          NGC~5363  &   1.27 $\pm$ 0.05   &  148.8  \\
                  NGC~5813  &  1.34 $\pm$ 0.22    &    83.4 \\
                 NGC~5846  & 1.10  $\pm$ 0.21  & 395.6 \\
           NGC~5898  & 1.55    $\pm$ 0.34 &    $<$26  \\
   %        NGC~6721  &    9.04  $\pm$     0.16    &   8.69  $\pm$    0.08  \
  NGC~6758  & 1.42   $\pm$ 0.17 &     $<$26  \\
 %      NGC~6776  &       $-$                     &     8.85 $\pm$     0.10  \\
          NGC~6868  &   1.43  $\pm$ 0.08  &  $<$26    \\
%               NGC~6876  &    8.97  $\pm$     0.16    &  8.69  $\pm$       0.07  \\
               NGC~6958  & 0.77 $\pm$ 0.03     &  1605.9   \\
     NGC~7007  &  1.53  $\pm$ 0.15 &    $<$26  \\
                    NGC~7079  & 0.95 $\pm$ 0.21   & 732.3  \\
                NGC~7097  &  1.23 $\pm$ 0.06  &  200.4 \\
    %              NGC~7135  &   8.64  $\pm$     0.16    &   8.52  $\pm$    0.03  \\
  %               NGC~7192  &   8.83  $\pm$     0.19    &  8.73  $\pm$    0.05   \\
              NGC~7377  &  1.43   $\pm$ 0.14  &  $<$26  \\
                       IC~1459   & 0.97 $\pm$ 0.04   & 674.2 \\
   
%              IC~2006   &  8.85   $\pm$    0.21     &      $-$      \\
              IC~3370   & 1.48    $\pm$ 0.05 &    $<$26  \\
              IC~4296   & 1.10  $\pm$ 0.15 & 400.7  \\
             IC5063   & 1.08    $\pm$ 0.01  &  433.3 \\ 
\hline
\end{tabular}
\end{table}
%----------------------- end Table 5-----------------------------------------------

%----------------------- Table 6-----------------------------------------------

\begin{table}
\caption{[O/H] values obtained the with Kobulnicky et al.~(\cite{Kobulnicky99}) calibration (Kob99),
and with the Storchi-Bergmann et al.~(\cite{Storchi98}) calibration (SB98).} \label{oh}
\centering
\begin{tabular}{lcc}
\hline \hline
Galaxy & ${\rm \log(O/H) + 12}$ (Kob99) & ${\rm \log(O/H) + 12}$ (SB98) \\
\hline
      NGC~128  &    8.44  $\pm$     0.24    &      $-$    \\
%      NGC~777  &       $-$                     &    $-$  \\
     NGC~1052  &   8.52  $\pm$     0.15    &   8.67  $\pm$    0.03   \\
     NGC~1209  &   8.63  $\pm$     0.19    &   8.78  $\pm$     0.04   \\
     NGC~1297  &       $-$                     &     8.59 $\pm$     0.04  \\
      NGC~1380  &  8.81  $\pm$     0.20    & 8.75  $\pm$    0.04   \\
  NGC~1453  &   8.55  $\pm$     0.19    &  8.76  $\pm$    0.04   \\
  %      NGC~1521  &       $-$                     &  $-$  \\
            NGC~1533  &   8.76  $\pm$     0.18    &  8.74  $\pm$    0.04   \\
            NGC~1553  &   8.48  $\pm$     0.31    &    8.87  $\pm$    0.04   \\
%     NGC~1947  &       $--$                    &        $--$$\pm$      0.03587  \\
             NGC~2749  &       $--$                     &   8.7 $\pm$     0.04   \\
            NGC~2911  &   8.56  $\pm$     0.16    &  $-$  \\
             NGC~2974  &   8.53  $\pm$     0.18    &    8.86  $\pm$    0.04  \\
%     NGC~3136  &       $--$ $\pm$       $--$   &        $--$$\pm$     $--$     \\
             NGC~3258  &     8.94  $\pm$      0.17    &       $-$                    \\
             NGC~3268  &   8.83  $\pm$     0.16    &   8.85  $\pm$    0.04  \\
%     NGC~3557  &       $--$ $\pm$       5.442   &        $--$$\pm$       0.1697  \\
     NGC~3607  &   8.86 $\pm$     0.18    &    8.94  $\pm$    0.04  \\
%     NGC~3818  &       $-$                     &     8.41 $\pm$      0.15  \\
     NGC~3962  &   8.62  $\pm$     0.18    &   8.83  $\pm$    0.04  \\
      NGC~4374  &   8.79  $\pm$     0.17    &       $-$          \\
           NGC~4552  &    8.98  $\pm$     0.16    &       $-$            \\
            NGC~4636  &  8.75  $\pm$     0.17    &       $-$           \\
            NGC~4696  &  8.71  $\pm$     0.16    &    8.82  $\pm$    0.04  \\
          NGC~4697  &   8.85  $\pm$     0.26    &       $-$         \\
%     NGC~5011  &       $--$ $\pm$       1.105   &        $--$$\pm$      0.04232 \\
    NGC~5044  &   8.66  $\pm$     0.16    &       $-$        \\
     NGC~5077  &   8.40  $\pm$     0.17    &  8.69  $\pm$    0.04   \\
     NGC~5090  &    8.93  $\pm$     0.16    &       $-$           \\
     NGC~5193  &    9.01  $\pm$     0.16    &       $-$          \\
      NGC~5266  &   8.20  $\pm$     0.22    &  8.72  $\pm$    0.04   \\
      NGC~5328  &    8.96  $\pm$     0.23    &   8.54 $\pm$   0.17   \\
          NGC~5363  &     8.32  $\pm$     0.21    &   8.68  $\pm$    0.03   \\
        NGC~5813  &   8.84  $\pm$     0.17    &       $-$      \\
          NGC~5846  &   8.84  $\pm$     0.17    &  8.71  $\pm$    0.04  \\
       NGC~5898  &  8.75  $\pm$     0.21    &       $-$     \\
          NGC~6721  &    8.88  $\pm$     0.16    &   $-$ \\
 NGC~6758  &   8.63  $\pm$     0.18    &       $-$      \\
      NGC~6776  &       $-$                     &    $-$  \\
          NGC~6868  &   8.48  $\pm$     0.19    &       $-$             \\
               NGC~6876  &    8.97  $\pm$     0.16    & $-$  \\
               NGC~6958  &       $-$           &    8.82 $\pm$     0.04  \\
%     NGC~7007  &       $--$ $\pm$       20.78   &        $--$$\pm$        1.165  \\
               NGC~7079  &  8.72  $\pm$     0.28    &   8.78  $\pm$    0.05  \\
            NGC~7097  &   8.59  $\pm$     0.17    &  8.67 $\pm$    0.03  \\
               NGC~7135  &   8.64  $\pm$     0.16    &   $-$  \\
               NGC~7192  &   8.83  $\pm$     0.19    &  $-$ \\
              NGC~7377  &   8.60  $\pm$     0.25    &       $-$      \\
         IC~1459   &  8.54   $\pm$      0.16  &   8.78  $\pm$     0.03 \\
              IC~2006   &  8.85   $\pm$    0.21     &      $-$      \\
              IC~3370   &   8.49   $\pm$    0.21     &      $-$    \\
              IC~4296   &    8.85   $\pm$    0.17     &   8.88   $\pm$   0.04  \\
             IC5063   &      $-$       &  8.69 $\pm$    0.03  \\ 
\hline
\end{tabular}
\end{table}
%----------------------- end Table 6-----------------------------------------------

%------------------------------ Figure 16 -----------------------------------
\begin{figure}
   \centering
  \includegraphics[width=8.5cm]{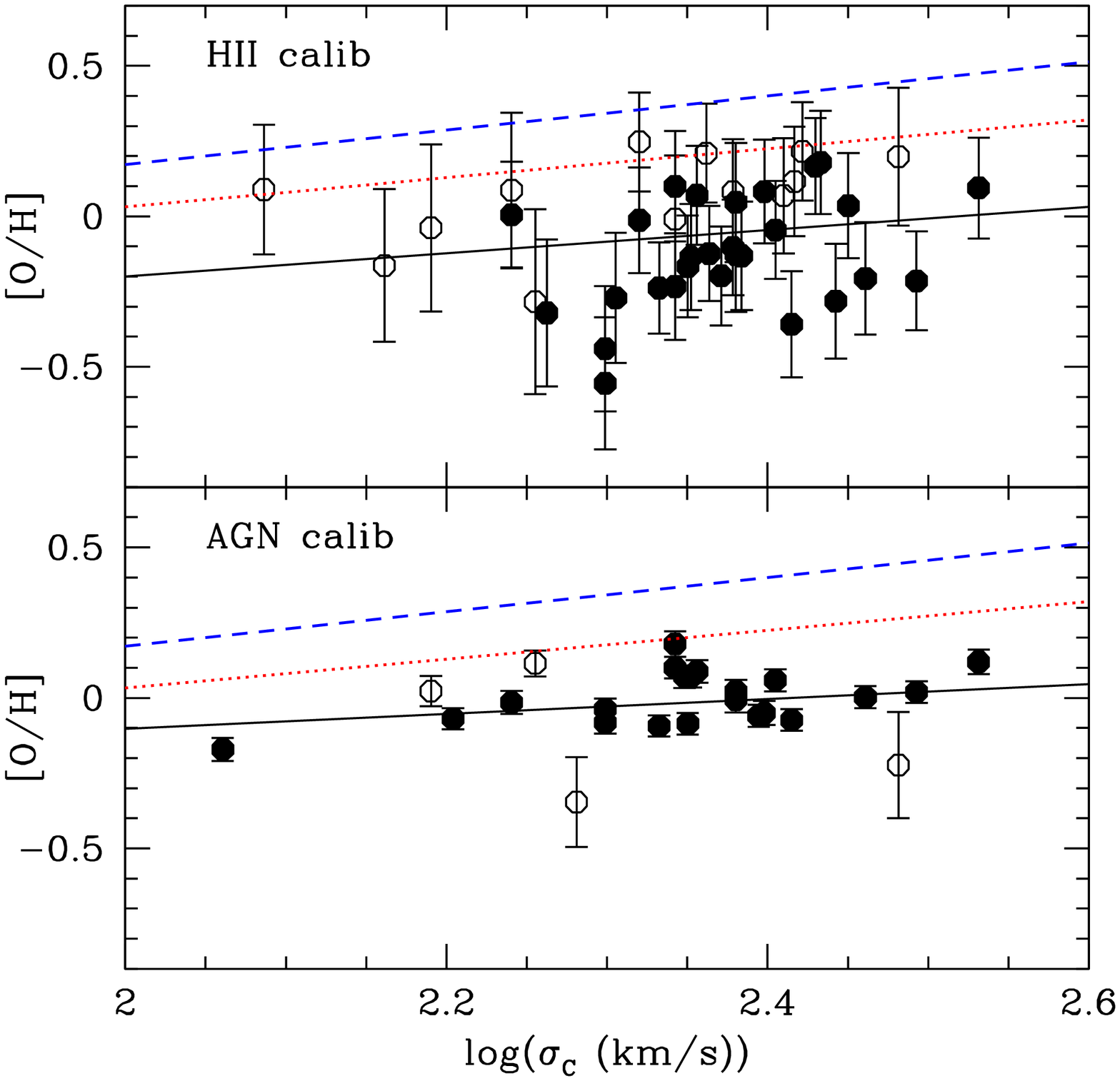}
  \caption{Oxygen abundances derived with the Kob99 (top) 
  and SB98 (bottom) calibration 
  at  r $\leq$r$_e$/16 
  versus the galaxy central stellar velocity dispersion. 
  Full symbols are the high emission galaxies (H), as defined in Table~4.
   The solid lines are the least-square fits to the data. 
  The dashed and dotted lines are the relations derived 
  by Thomas et al.~(\cite{tho05}) and A07 for the stars, 
   rescaled to account for the solar metallicity $Z_{\odot}=00156$ adopted in this paper.
 }
\label{oxysig1}
\end{figure}
%------------------------------ end Figure 16 -----------------------------------

\subsection{HII region calibration}

In HII region, the oxygen metallicity is determined through the standard 
${\rm R_{23} = ([OII]\lambda3727 + [OIII]\lambda\lambda4959, 5007)/H\beta}$ parametrization,
first introduced by Pagel et al.~(\cite{Pagel79}), and then revised by other authors 
(Edmund \& Pagel~\cite{Edmunds84}; McCall, Rybski \& Shields~\cite{McCall85}; 
Dopita \& Evans~\cite{Dopita86};  McGaugh~\cite{McGaugh91}; Zaritsky et al.~\cite{Zaritsky94}).
The main problem with this calibration is that the relationship between ${\rm R_{23}}$ and 
the oxygen abundance is double valued, requiring some a priori knowledge of a galaxy's
metallicity in order to determine its correct location in the upper or lower branch of the curve. 
A good discriminator is the [OIII]$\lambda$5007/ [NII]$\lambda$6584 ratio, which is usually 
less then $\approx$ 100 for galaxies with 12 + log(O/H) $>$ 8.3, on the metal rich branch
(Edmunds \& Pagel~\cite{Edmunds84}, Kobulnicky et al.~\cite{Kobulnicky99}). 
The various calibrations, based on different photoionization models and 
HII region data, show a dispersion of $\sim$ 0.2 dex in the metal rich branch. 
Kobulnicky et al.~(\cite{Kobulnicky99}) reports analytic expressions based on a set of photoionization models
that fit the empirical calibrations within less than 0.05 dex, and that include the effect of the ionization 
parameter  through ${\rm O_{32}=[OIII]\lambda \lambda4959,5007/[OII]\lambda 3727}$.

We computed the oxygen abundance for our sample adopting the calibration in  
Kobulnicky et al.~(\cite{Kobulnicky99}) (hereafter Kob99). 
 Because of the lack of star forming regions in ETGs, this calibration is valid if:  
1) the gas is photoionized by hot old stars, likely PAGB stars, and 2) the calibration 
is still valid even though PAGB stars have different spectral shapes than high-mass main sequence stars. 
The same approach was adopted by Athey \& Bregman~(\cite{ab09}) to determine
oxygen abundances in a sample of 7 ETGs.
We computed the ${\rm R_{23}}$ and ${\rm O_{32}}$ values using the emission line fluxes in 
Table~2, and correcting for reddening through the values in Col~3 of Table~2.
The derived oxygen metallicities at r $\leq$r$_e$/16 are given in Col.~2 of Table~6. 
The errors were computed by combining the intrinsic uncertainty in the 
calibration ($\sim$ 0.15 dex, see Kob99), 
with the uncertainty in the emission line fluxes.
We could not determine the oxygen abundance in 13 galaxies, 
namely IC~5063, NGC~777, NGC~1297, NGC~1521,
NGC~1947, NGC~2749 , NGC~3136, NGC~3489, NGC~3557, NGC~5011, 
NGC~6776, NGC~6958, and NGC~7007, since ${\rm R_{23}}$ is larger than 10, and 
falls outside the valid calibration range for HII regions.

There are 6 galaxies in common with the sample of 
Athey and Bregman~(\cite{ab09}): NGC~3489, NGC~3607, NGC~4374, NGC~4636, NGC~5044, 
and NGC~5846.
Excluding NGC~3489, we derived for the other galaxies
central $\log R_{23}$ values of 0.50 $\pm$ 0.15, 0.58 $\pm$ 0.11, 0.63 $\pm$ 0.1, 0.70 $\pm$ 0.05, 
0.51 $\pm$ 0.11.
For the same galaxies, they derive $0.93^{0.08}_{-0.11}$, $0.58^{0.09}_{-0.10}$,
$0.68^{0.09}_{-0.11}$, $0.75^{0.05}_{-0.06}$, $0.68^{0.08}_{-0.09}$.
With the exception of NGC~3607, our and their results are in agreement
within the errors. 
%Notice however that their metallicities were 
%computed assuming a slightly different solar metallicity 
%($12 + \log(O/H)_{\odot} = 8.7$), from Scott et al. (2009).

The oxygen metallicity distribution is shown in the top panel of Fig.~\ref{ohdistr} . 
The ${\rm [O/H] = \log (O/H) - \log(O/H)_{\odot}}$ values were computed adopting  
a solar abundance of ${\rm 12 + log(O/H) = 8.76}$ from Caffau et al.~(\cite{Caffau08}).
The average metallicity is ${\rm [O/H]=-0.06}$. Some values are as low as 
$\approx$ 0.25 times solar, while the largest derived metallicity is $\approx$ 
twice solar. 

\subsection{AGN calibration}

Storchi-Bergmann et al.~(\cite{Storchi98}) (hereafter SB98) derived oxygen metallicity calibrations
from a grid of models assuming photoionization by a typical AGN continuum
(the segmented power law of  Mathews \& Ferland~\cite{mat87}). 
The first calibration (hereafter SB1) is in terms of 
[OIII]$\lambda\lambda$4959,5007/ H$\beta$ and  [NII]$\lambda\lambda$6548,6584/ H$\alpha$ 
(Eq. (2) in SB98), while the second calibration (hereafter SB2) is in terms of 
[OII]$\lambda$3727/ [OIII]$\lambda$4959,5007 and  [NII]$\lambda\lambda$6548,6584/ H$\alpha$
 (Eq. (3) in SB98). The fitted calibrations are within $\sim$ 0.05 dex of the models.
 The models were computed assuming a gas density ${\rm n_g=300}$  cm$^{-3}$. 
 The correction due to deviations from this density is  
 ${\rm -0.1 \times \log (n_g/300)}$, valid for 100 cm${\rm ^{-3} \lesssim n_g \lesssim}$ 10,000 cm$^{-3}$.

We derived oxygen metallicities for the R05$+$A06 sample using both SB1 and SB2.
The electron densities n$_e$ were calculated from the [SII]$\lambda\lambda$6717, 31 line ratios
using the {\it temden} task within IRAF/STSDAS (Shaw \& Dufour~\cite{Shaw95}),  and assuming
an electronic temperature $T_e=$10,000 K.
This was possible only for some galaxies of our sample (see Table~5), 
because the quality of our spectra gets significantly worse beyond $\sim$ 6600 \AA, 
and the presence of telluric absorption lines around 6800 \AA \ prevents in many cases a reliable 
fit to the [SII] doublet. The derived electron densities  are given in Table~5.
As evident from the [SII] fluxes given in Table~2, both increasing and decreasing density profiles
are present within our sample.
The oxygen metallicities were then corrected for the density dependence assuming $n_g \approx n_e$.
 Because of the limited range of validity for the density correction, we discarded
 the galaxies with n$_e <$ 100 cm$^{-3}$.
 The [O/H] values obtained with SB1 and SB2 are compared in Fig.~\ref{cfr}.
 The metallicities obtained with SB2 are on average $\sim$ 0.09 dex larger than
 those obtained with SB1, in agreement with SB98 who derived a shift of $\sim$ 0.11 dex between the 
 two calibrations. The final metallicities are computed as the average between the two values.
The results are given in Col. 3 of Table~6. 

We compare the results obtained with the Kob99 and with the SB98 calibrations 
in Fig.~\ref{ohdistr}. The SB98 calibration provides an average oxygen 
abundance $\sim$ 0.04 dex higher than that obtained with Kob99.
Also, Kob99 provides a broader metallicity range than SB98.

\subsection{Abundance gradients}

 We studied the behavior of the oxygen metallicity as function of  galacto-centric distance.
 The [O/H] distributions obtained with Kob99 and SB98 in the three more internal annuli 
 (r $\leq$ r$_{e}$/16, r$_{e}$/16 $<$ r $\leq$r$_e$/8, r$_{e}$/8$<$ r $\leq$r$_e$/4) 
 are shown in Fig.~\ref{ohgrad}. The samples are not very large 
because of the increased difficulty in measuring the emission lines in the more external annuli. 
Emission lines are too faint to allow a reliable abundance measurement in 
the more external annulus (r$_{e}$/4$<$ r $\leq$r$_e$/2).
The results obtained with the Kob99 and SB98 calibrations are contradictory:
in the first case, the oxygen metallicity tends to increase from the galaxy center 
outwards. In the second case, the opposite behavior is observed. 
The latter trend is more consistent with the negative metallicity gradients 
of the stellar populations.

\subsection{Comparison with stellar metallicities}

To understand the origin of the warm gas in ETGs it is important  
to compare the nebular metallicities with those of the stellar populations.
We already derived the (total) stellar metallicities (Z) and the [$\alpha$/Fe] ratios 
for the R05$+$A06 sample through the Lick indices in Paper~III.
The comparison with the nebular metallicities is not straightforward, 
since the emission lines provide a direct measure of the oxygen abundance, while the Lick indices 
depend on a mixture of different elements. 
In this paper, we compute the oxygen stellar metallicities from 
the Z and  [$\alpha$/Fe] values derived in Paper~III, assuming 
 that O belongs to the {\it $\alpha$-enhanced} group 
(Ne, Na, Mg, Si, S, Ca, Ti, and also N).
The comparison between gas and star metallicities is shown in Fig.~\ref{stars}.
The clear result is that, irrespective of the adopted calibration (Kob99 or SB98), 
the gas metallicity tends to be lower than
the stellar metallicity. The effect is more severe for the galaxies with the largest stellar metallicities.
A Spearman test shows that in both cases the star and gas metallicities are poorly correlated.

We plot the central (r $\leq$r$_e$/16) nebular metallicities obtained with Kob99 and SB98
versus the galaxy stellar velocity dispersions 
in Fig.~\ref{oxysig1}. 
Linear square fits provide an increasing trend of ${\rm [O/H]}$ with $\sigma_c$ in both cases.
However, the Spearman correlation coefficients are low ($r_s = 0.25$ with $N=39$ degrees of freedom 
for Kob99, top panel, and $r_s = 0.08$ with $N=22$  degrees of freedom for SB98, bottom panel), 
 indicating that the correlations are weak.
 For comparison, we plotted in Fig.~\ref{oxysig1} the stellar metallicity-$\sigma_c$ relations derived by us in Paper~III (A07) and by Thomas et al.~(\cite{tho05}) for low density environment ETGs.
The A07 relation was appropriately re-scaled to account for the solar 
metallicity adopted in this paper \footnote{$Z_{\odot}=0.018$ in A07, while we adopt $Z_{\odot}=0.0156$ from 
Caffau et al.~(\cite{Caffau09}) in this paper. This implies a correction term of $\log(0.018/0.0156)=0.06$ dex
to be applied to the A07 relation.}.
In Fig.~\ref{oxysig1}, the fits provide ${\rm [O/H]_{gas}\sim -0.08}$ and 
${\rm [O/H]_{gas}\sim -0.03}$ at $\sigma_c=$ 200 km s$^{-1}$ for Kob99 and SB98, respectively.
For comparison, the A07 relation gives a stellar metallicity of $\sim$ 0.18 dex at the same $\sigma$, 
i.e. 0.26 dex and 0.21 dex larger than those obtained for the warm gas with  Kob99 and SB98.
The discrepancy is even larger if we consider the Thomas et al.~(\cite{tho05}) relation,
which provides a metallicity of 0.34 dex at $\sigma_c=$ 200 km s$^{-1}$,
i.e. $\sim$ 0.4 dex higher than the nebular metallicity.
We notice that the discrepancy between gas and stars persists (and is even more significant) 
if we consider only the high emission (H) galaxies (see classification in Table~4).

\section{Comparison with Models}

The ionizing source in LINERs is still not well understood.
Among the proposed mechanisms, we recall sub-Eddington accretion into a super massive black hole 
(e.g., Ho~\cite{ho99b}; Kewley et al.~\cite{kew06}, Ho~\cite{ho09a}), photoionization by old PAGB stars 
(e.g. Trinchieri \& di Serego Alighieri~\cite{tds91}; Binette et al.~\cite{bin94}; Stasi{\'n}ska et al.~\cite{sta08}), fast shocks (Koski \& Osterbrock~\cite{ko76}; Heckman~\cite{heck80}; Dopita \& Sutherland~\cite{dosu95}; Allen et al.~\cite{allen08}).

%------------------------------ Figure 17 -----------------------------------
%macro diagnostic_models.sm
\begin{figure}
  \includegraphics[width=9cm]{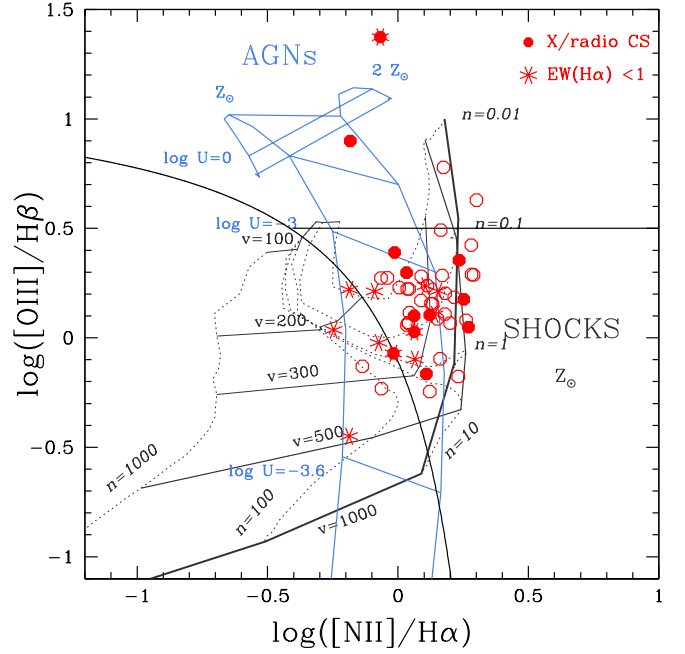}
  \caption{BPT diagram for our galaxies at r $\leq$r$_e$/16 with superimposed the 
  shock models of Allen et al.~\cite{allen08} and the dusty-AGN models of 
  Groves et al.~(\cite{gro04}). Full dots indicate objects where compact nuclear X-ray  and/or radio  
  sources have been detected (X/radio CS), consistently with the presence of an AGN, while stars are for 
 objects with equivalent width in H$\alpha < 1$ \AA \ , compatible with ionization from PAGBs.  
 Notice that three galaxies (NGC~777, NGC~4552, and NGC~3557) are X/radio CS and at the same 
 time have EW(H$\alpha$$<1$).
 The shock models have solar metallicity, densities  from $n=$ 0.01 cm$^{-3}$ to 
 1000 cm$^{-3}$, velocities from $v=$100 km $s^{-1}$ to 500 km $s^{-1}$, 
 and magnetic field $B=1$ $\mu$G.
 The plotted AGN models have solar and twice solar metallicities,  $n=$ 1000 cm$^{-3}$,
 and ionization parameters from $log U \sim -4$ to 0.
 The models were downloaded from the Brent Groves web page 
 http://www.strw.leidenuniv.nl/$\sim$brent/itera.html using the ITERA tool.
 }
\label{ddmod}
\end{figure}

%------------------------------ end Figure 17 -----------------------------------

%\begin{figure}
 % \includegraphics[width=9.5cm]{s2.hist.ps}
 % \caption{[SII](6717)/[SII](6731) ratio for the central r$<$r$_e$/16 region (top panel) and 
%  as a function of galacto-centric distance for a subsample where the [SII]
%  measurements were most reliable. The shaded blue histogram is for Seyfert galaxies.}
%\label{s2}
%\end{figure}

%------------------------------ Figure 18 -----------------------------------
\begin{figure*}
   \centering
  \includegraphics[width=14cm]{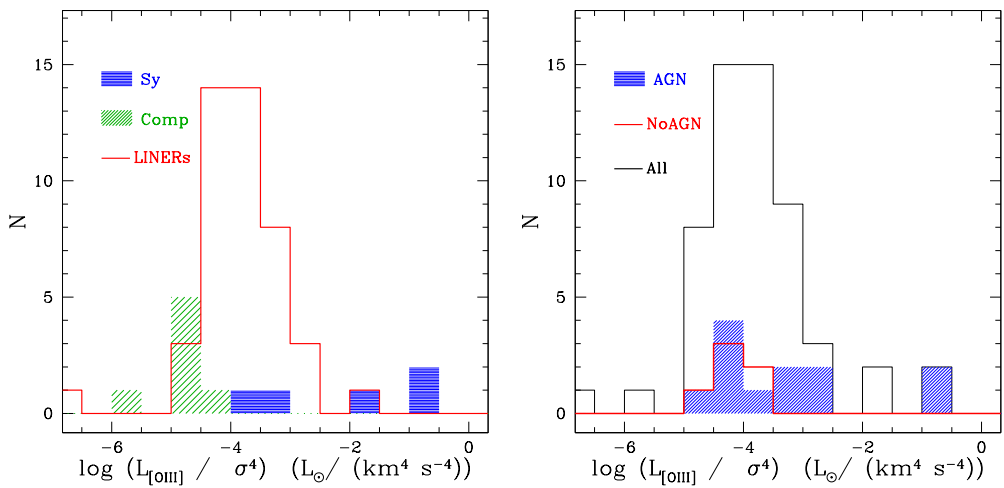}
  \caption{Distribution of  ${\rm L_{[OIII]}/\sigma^4}$ ratio for 
   Seyferts, Composites and LINERs (left panel), and 
   for the total sample and galaxies classified as AGN and non AGN 
   from X-ray/radio data (right panel).
   }
\label{edd}
\end{figure*}
%------------------------------ end Figure 18 -----------------------------------

The first hypothesis is strongly supported by our current knowledge of the demography of central 
BHs based on direct dynamical searches (Magorrian et al.~\cite{mago98}; Ho~\cite{ho99a}; Kormendy~\cite{kor04}),according to which massive BHs appear to be a generic component of galaxies with a bulge.
As shown by Ho et al.~(\cite{ho97c}), local stellar mass loss can supply the fuel necessary to accretion rates 
of $\dot{M} = 0.001-0.1 M_{\odot} yr^{-1}$. Thus, radiatively inefficient accretion flows onto a central BH 
(see Quataert~\cite{quat01} for a review) provide an attractive solution to explain the origin of LINERs.
Additional evidences for the presence of AGN activity come from X-ray and radio observations, 
which have revealed the presence of  compact nuclear X-ray and radio  
sources, sometimes with detected variability, in a significant fraction of LINERs 
(see e.g. Terashima~\cite{tera99}; Flohic et al.~\cite{flo06}; Filho et al.~\cite{filh06}; 
Gonzalez-Martin et al.~\cite{go09}; Pian et al.~\cite{pian10}).
On the other hand, some LINERs lack evidence of a central AGN, leaving space to  
alternative explanations for the observed emission lines. 
X-ray and radio studies are available in the literature for a few galaxies in our sample.
Compact nuclear X-ray and/or radio sources have been detected in 
IC~5063 (Koyama et al.~\cite{koy92}), NGC~777 (Ho \& Ulvestad~\cite{hu01}), NGC~1052, NGC~4374,   IC~1459 (Gonzalez-Martin et al.~\cite{go09}), NGC~5363 (Gonzalez-Martin et al.~\cite{go09},
Filho et al.~\cite{filh06}), NGC~1553 (Blanton et al.~\cite{bla01}), 
NGC~3557 (Balmaverde et al.~\cite{bal06}), NGC~5090 (Gr{\"u}tzbauch et al.~\cite{gru07}), 
IC~4296 (Pellegrini et al.~\cite{pel03}), NGC~5077, (Filho et al.~\cite{filh06}), 
and NGC~4552 (Nagar et al.~\cite{nag02}, Filho et al.~\cite{filh04}).
Other galaxies observed in X-ray and/or radio lack evidence for AGN activity:
NGC~3607, NGC~4636, NGC~5813, (Gonzalez-Martin et al.~\cite{go09}, Filho et al.~\cite{filh06}), 
NGC~4696, NGC~5846 (Gonzalez-Martin et al.~\cite{go09}),  
NGC~6876 (Machacek et al.~\cite{mac05}).
High quality X-ray imaging has revealed in NGC~4636 the signature of 
shocks, probably driven by energy deposited off-center by jets (Jones et al.~\cite{jo02}; 
Baldi et al.~\cite{bald09}). 
 
Shock heating was initially proposed as a viable excitation mechanism to solve the 
so-called ``temperature problem''  of Seyferts and LINERs, in which the electron temperatures observed 
were found to be systematically higher than predicted by photoionization models.
Indeed, Koski \& Osterbrock~(\cite{ko76}) and 
Fosbury et al.~(\cite{fos78}) argued that shocks are essential to explain the observed [OIII] line ratios in the 
prototypical LINER NGC~1052.
Dopita \& Sutherland~(\cite{dosu95}) and Allen et al.~(\cite{allen08}) presented extensive grids of 
high velocity shock models, and showed that the emission line ratios observed in LINERs can be modeled in 
terms of fast shocks (150 - 500 km s$^{-1}$) in a relatively gas poor environment.

The findings that the emission-line flux correlates with the host galaxy stellar luminosity 
within the emission-line region (Macchetto et al.~\cite{mac96}), and that the line flux distribution 
closely follows that of the stellar continuum (Sarzi et al.~\cite{sar06,sar09}), have been presented  
as evidences in support of the PAGB scenario. These results suggest in fact that the sources of ionizing photons are distributed  in the same way as the stellar population.
Recently, Sarzi et al.~(\cite{sar09}) tried to model the H$\beta$ EW spatial distribution under the assumption that the ionizing photons originate from an underlying old stellar population.
They showed that to obtain an almost constant EW it is necessary to assume that the gas radial profile 
decrease more gently than the ionizing radiation, or alternatively that the ionizing photons are more concentrated 
than the bulk of the stellar populations. Indeed, the presence of negative metallicity gradients in ETGs
(e. g., Annibali et al.~\cite{anni07})  could explain this effect, because in general the ionizing 
flux from old populations is larger for higher metallicities.
The Stasi{\'n}ska et al.~(\cite{sta08}) models are able to reproduce the emission line ratios 
of LINERs provided that the ionization parameter is sufficiently high (${\rm \log U > -4}$).
Binette et al.~(\cite{bin94}) and Cid Fernandes et al.~(\cite{cf09}) demonstrated that photoionization 
by PAGB stars can only explain LINERs with relatively weak emission lines.
Binette et al.~(\cite{bin94}) estimated that the H$\alpha$ equivalent width produced by 
a  8 Gyr old and a 13 Gyr old simple stellar populations 
amounts to  0.6 \AA \ and 1.7 \AA, respectively.
%Recent models (Stasinska et al 2008, Fernandes et al.(2009)) show that post -AGB and white dwarf stars in aging   %populations can well account for the ionizing photons in LINERs with weak emission lines (S/N $<$ 3 in H$\beta$ %and/or [OIII].) On the other hand, 
Stronger emission lines require far more ionizing photons than old stars can provide.

A first-order test on the validity of the PAGB scenario consists in a comparison 
of the observed and predicted EWs. Following the computations of Binette et al.~(\cite{bin94}), we 
assume that photoionization by old stars produce a typical emission of 1 \AA \ in H$\alpha$.
Comparing this value with our measured  {\underline{central}}  ($r<r_e/16$) EWs,  we obtain that 
PAGB photoionization alone can explain the observed nuclear emission 
in only 11 out of the 49 galaxies classified as LINERs/Composites in our sample (i.e. in $\approx$ 
 22\% of the LINERs/Comp sample).
 This implies that, for the majority of the cases, some mechanism other than PAGB stars must be at work 
 in the central regions of LINERs.
 On the other hand, as we progressively move toward larger radii, an increasing number of LINERs enter
 the EW(H$\alpha$)$\sim 1$  \AA \ {\it transition} region, where the emission strength is compatible with photoionization by PAGB stars alone. Thus, while our computations exclude that PAGB stars play 
 a significant role in the central galaxy regions, we can not rule out that they are 
 an important excitation source at larger radii.
Interestingly, Sarzi et al.~(\cite{sar09}) quote a 3'' radius for the maximum extent of central activity,
which for their sample galaxies corresponds on average to 0.11 r$_e$.

One of the arguments presented in support of the PAGB photoionization scenario is that 
the emission flux closely follows  the stellar continuum.
At odds with other results (e.g. Sarzi et al.~\cite{sar09}), we have shown in Section~3.3 that the 
EW (emission flux over continuum flux) decreases with radius. 
Is this incompatible with the PAGB scenario? Answering this question is not straightforward. 
First of all, we have shown that the strongest decrease in EW occurs from the nuclear to the second (r$_{e}$/16 $<$ r $\leq$r$_e$/8) annulus, being much gentler at larger radii. Indeed, our data lack the depth necessary to reliably trace the emission strength beyond r$_{e}/4$.
Second, neither a well established decreasing trend in EW would be {\it per se} incompatible with the PAGB scenario: in fact, because of the presence of age/metallicity gradients in the stellar population, the ratio of the ionizing flux over the optical continuum is not necessarily constant. 
A reliable radial prediction of the ionizing photon budget stems on a detailed analysis of the combined effect 
of age and metallicity on the PAGB population, which is beyond the aims of this paper.

\begin{figure}
  \includegraphics[width=9.5cm]{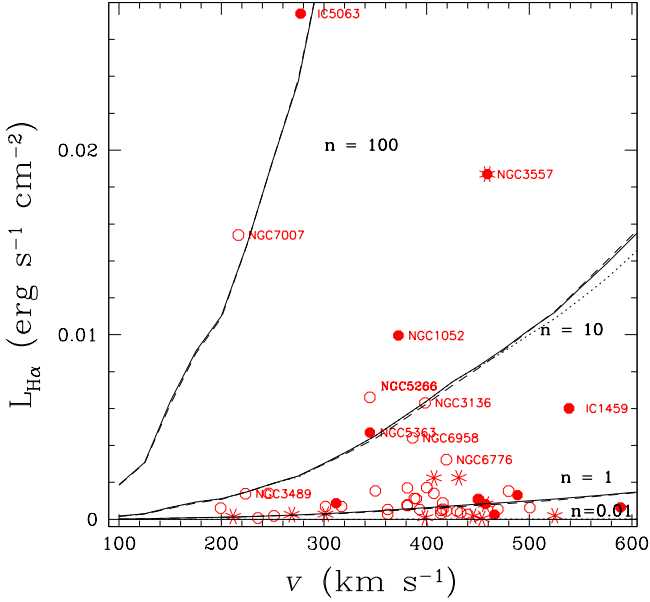}
  \caption{Shock models of Allen et al (2008) compared with our r $\leq$r$_e$/16 H$\alpha$ luminosities,
  corrected for reddening.
  The shock models are plotted for 4 density values ($n=$ 0.01, 1, 10, 100 cm$^{-3}$),
  and magnetic fields $B=0.1$ (dotted line), $B=1$ (solid line) and  $B=10$ $\mu$G (dashed line). 
  The velocity in the abscissa is the shock velocity for the models, and $\sqrt{3} \sigma_c$ for the galaxies.
  The different symbols used to plot the data are the same as in Fig.~\ref{ddmod}.
 We indicate the name of the galaxies with the largest ${\rm L_{H\alpha}}$ values. 
 NGC~777, once corrected for the high extinction (E(B$-$V)$=$1.6), has 
 ${\rm L_{H\alpha} \sim 0.06  \ erg s^{-1} cm^{-2}}$ and falls outside the plot boundaries.}
\label{shock}
\end{figure}

Since the nuclear emission requires a mechanism different than PAGB photoionization, 
we compare in Fig.~\ref{ddmod} our (r $\leq$r$_e$/16) data, corrected for reddening,  
with the dusty AGN models of Groves et al.~(\cite{gro04}), and with the fast shock models 
of Allen et al.~(\cite{allen08}).
Among the data points, we identify objects with compact nuclear 
X-ray and/or radio source detections, likely hosting an AGN, and objects with equivalent width in H$\alpha < 1$ \AA \ , compatible with sole ionization from PAGBs.  
We notice that three galaxies (NGC~777, NGC~4552, and NGC~3557) present evidence 
for AGN activity from X ray or radio observations, and at the same time have EW(H$\alpha$)$<1$.
Comparing dusty models with reddening-corrected data is not completely self-consistent;
however the effect of extinction in the BPT diagram is negligible.

Both the AGN and shock 
models were developed using the code MAPPINGS III (Dopita et al.~\cite{dop82}; Sutherland \& Dopita~\cite{sudo93}), implemented with dust and radiation pressure (Groves et al.~\cite{gro04}).
The AGN models were developed for different metallicities, 
and assuming a power law with slope $\alpha$ between  $-1.2$ and  $-2.0$ for the ionizing spectrum.
For the AGN models in Fig.~\ref{ddmod}, $\alpha=-1.4$, $Z= Z_{\odot}$ and  2 $Z_{\odot}$, $n_H$ =1000 cm$^{-3}$, and $\log U$ between $-4$ and 0. 
LINERs and Seyferts are both explainable with being AGNs, provided that the formers
have much lower ionization parameters than the latters.
According to these models, 
the galaxies in our sample should have gas metallicities much above solar. This is in disagreement 
with the oxygen metallicities derived in Section~5 with the Storchi-Bergmann et al.~(\cite{Storchi98}) calibration. The discrepancy arises from the use of a power law in the Groves et al.~(\cite{gro04}) models 
compared to the typical AGN continuum of Mathews \& Ferland~(\cite{mat87}) adopted by SB98.
SB98 noticed in fact that models obtained with power-law ionizing continua yield systematically larger 
(up to 0.5 dex) metallicities than models that adopt a typical AGN continuum.

Under the assumption that all the central LINERs activity is powered by a sub-Eddington accretion AGN, 
we can attempt a determination 
of the accretion rate from the ${\rm L_{[OIII]}/\sigma^4}$ ratio (where ${\rm L_{[OIII]}}$ is the extinction 
corrected [OIII] luminosity), proportional to  ${\rm L/L_{EDD}}$
(Heckman et al.~\cite{heck04}). 
The result is shown in Fig.~\ref{edd}. 
The accretion rate increases systematically along the sequence Composite $\rightarrow$ LINERs $\rightarrow$ Seyferts (left panel) in agreement with recent findings by Ho~(\cite{ho09a}). 
Also, galaxies with compact nuclear X-ray/ radio sources, likely hosting an AGN, tend to have
larger accretion rates than non -AGN galaxies (right panel).
Notice that since $L \propto \eta L_{EDD} \propto \eta \sigma^4$, 
and $\eta << 1$ in the sub-Eddington accretion regime, 
the AGN luminosity can vary significantly for a fixed $\sigma$
depending on the accretion $\eta$. This implies that 
it is  necessary to assume a maximum, sub-Eddington, accretion rate
to reproduce the observed trend between the emission strength and $\sigma$.

The Allen et al.~(\cite{allen08}) shock models are dust free, and were computed for different chemical abundances, pre-shock densities $n$ from 0.01 to 1000 cm$^{-3}$, velocities up to 1000 km s$^{-1}$, 
and magnetic parameters $B/n^{1/2}$ from 10$^{-4}$ to 10 $\mu$G cm$^{3/2}$, where B is the 
pre-shock transverse magnetic field.
In Fig.~\ref{ddmod}, we plotted the models with solar metallicity and  B$=1$.  
With these parameters, the bulk of LINERs in our sample is well reproduced by pre-shock densities $n=0.01 - 10$ cm$^{-3}$, and shock velocities $v=200 - 500$ km $s^{-1}$.
Using a higher magnetic field implies larger densities for our sample. For instance, for B$=$10 $\mu$G, our data 
are consistent with pre-shock densities as high as $n \approx100$ cm$^{-3}$. 
This range in B is typical of the ISM (e.g., Rand \& Kulkarni~\cite{rk89}).

Are these densities consistent with those derived from the observed [SII](6717)/[SII](6731) ratios?
As shown in Table~5, the majority of our galaxies have ratios $>1$, implying n$_e\lesssim$ 600 cm$^{-3}$ 
for an assumed T$=$10,000 K. A significant fraction of galaxies have ratios around 1.45, which 
is the saturation limit for n$\rightarrow$ 0. These values are consistent with pre-shock densities 
as low as n$<$10 cm$^{-3}$. Indeed, as the density of the material passing through the shock increases,
the density indicated by the [SII] ratio also increases. However,  the amount of compression 
depends critically upon the magnetic parameter (Dopita \& Sutherland~\cite{dosu95}).
In the Allen et al.~(\cite{allen08}) models, [SII](6717)/[SII](6731) $\sim$ 0.7 and $\sim$ 1 for a pre-shock density 
 n$=$10 cm$^{-3}$, and B$=$1 and 10 $\mu$G, respectively. 

In the fast shock scenario, where does the mechanical energy come from?
Dopita \& Sutherland~(\cite{dosu95}) distinguish between two possible types 
of gas dynamical flows: jet-driven and accretion-driven flows. 
Evidence for jets in LINERs comes from observations revealing the presence 
of low-luminosity AGNs associated with radio sources with a high brightness temperature 
and/or elongated radio morphology (e.g., Nagar et al.~\cite{nag01}). 
In accretion-driven flows, the shock is either the result of the collision of cool
clouds which have dropped out of a cooling flow into the potential of the galaxy 
(e.g. Crawford \& Fabian~\cite{crfa92}) or have been accreted from a companion galaxy.
Shocks may also occur in the outer accretion disk about the central object either trough the dissipation of turbulence, dynamical intsabilities, or spiral shocks. (e.g. Dopita et al.~\cite{dop97}).
We suppose that the mechanical energy could also come from turbulent motion of gas clouds, 
injected by stellar mass loss or accreted, within the potential well of the galaxy.
Fig.~\ref{ddmod} shows that the shock velocities required 
to reproduce the distribution of LINERs in the BPT diagram 
range between 200 and 500 km/s. Are the mentioned mechanisms able 
to account for such high shock velocities? 
It seems that jet driven flows or accretion into a central massive BH 
can well do the job (e.g., Nagar et al.~\cite{nag05}, Dopita et al.~\cite{dop97}). 
On the other hand, it is not straightforward to establish if, in absence of 
these sources, pure gravitational motion of clouds within the potential well of the galaxy
can produce collision at the required velocities.
Close to the center of a relaxed stellar system, the gas should quickly relax into a disk-like 
configuration and orbit at a velocity close to the circular one, even though not necessarily in the same 
direction of stars. Indeed, this is confirmed by observations, showing regular gaseous 
disk structures and coherent gas motions (see e.g. Zeilinger et al.~\cite{zei96}, Sarzi et al.~\cite{sar06}).
If most of the clouds rotate in the same direction, it would be then rather 
difficult for them to collide at the high speed required.
On the other hand, observations detect an additional gaseous component 
of high gas velocity dispersion, in the range 
150-250 km/s (Bertola et al.~\cite{ber95}; Zeilinger et al.~\cite{zei96}; Emsellem et al~\cite{Emsellem03}).
Sarzi et al.~(\cite{sar06}) found that $\sigma_{gas}$ is generally 
smaller than the stellar velocity dispersion $\sigma_{*}$, but in some cases  $\sigma_{gas}$ $\sim$  $\sigma_{*}$, either only in the central regions or over most of the field. 
For a sample of 345 galaxies, with Hubble type from E to  Sbc, Ho~\cite{ho09b} found 
that the gas dispersions strongly correlate with the stellar dispersions over the velocity range of 
$\sigma \approx$  30-350 km s$^{-1}$, such that $\sigma_{gas}$ /$\sigma_* \approx  0.6-1.4$, 
with an average value of 0.80. 
All these observations suggest that random motions are crucial for the dynamical support of the gas.
Projected velocity dispersions in the range 150-250 km/s  translate, under the assumption of isotropy,
into intrinsic 3D velocity dispersions in the range 260-430 km/s. \footnote{The total  3D velocity dispersion 
satisfies $\sigma^2 = \sigma_x^2 + \sigma_y^2 + \sigma_z^2 $, because of the Pythagoras theorem and 
basic statistics of the second moments. If the dispersion is isotropic, $\sigma_x =  \sigma_y = \sigma_z$ 
by definition, and $\sigma^2 = 3 \sigma_x^2$.} This is a crude estimate of the intrinsic velocity dispersions since 
studies indicate deviation from isotropy (e.g. Merrifield et al.~\cite{merri01}).
Given these values, collisions at a speed between 200 and  500 km/s can not be excluded.

We compare the shock models of Allen et al.~(\cite{allen08}) with our data in a luminosity vs velocity plane 
in Fig.~\ref{shock}. For the data, we adopted a velocity of  $\sqrt{3} \sigma_{c}$, which is a 
first-order approximation for the shock velocity under the assumption that the shocks are due to motions of gas clouds in the galaxy potential well.
The models are  plotted for four density values
($n=0.01, 1, 10, 100$ cm$^{-3}$), and for three values of the transverse magnetic field ($B=0.1, 1, 10$ $\mu$G).
The shock luminosity increases significantly with the shock velocity and with the gas density. The dependence on the magnetic parameter is almost null in H$\alpha$ (and modest in [NII]).
The observed nuclear H$\alpha$ luminosities were corrected for reddening through 
the values derived in Table~2.
Some galaxies have very large reddening, associated to large uncertainties, and display quite 
high luminosities once corrected for extinction (NGC~777, NGC~1521, NGC~7007, NGC~3557). 
Their ${\rm L_{H\alpha}}$ should be taken with caution. 
Fig. ~\ref{shock} shows that, for almost all the galaxies in our sample,  
the observed luminosities are well reproduced by models with 
pre-shock densities $n \lesssim 10$ cm$^{-3}$. 
This does not proof that shocks are the excitation mechanism in LINERs, but 
show that such a scenario is compatible with the data.
We notice that shocks could provide a natural explanation for the observed 
{\it half-cone-shaped} distribution in the line luminosity/EW vs $\sigma$ plane (see also Fig.\ref{allcorr}).
They could also explain the observed [NII]/H$\alpha$ vs $\sigma_c$ correlation 
(Fig.\ref{allcorr}).The shock models predict in fact an increase of the [NII]/H$\alpha$ line ratio with the shock velocity. This is because higher velocity shocks result in harder and more luminous ionizing spectra.
It is interesting that the weak emission galaxies in our sample, compatible with pure 
photoionization by PAGB stars at their center, are those deviating from the [NII]/H$\alpha$ vs $\sigma_c$ relation in Fig.\ref{allcorr}.
Phillips et al.~(\cite{ph86})  also found that the [NII]/H$\alpha$ ratio increases 
with the galaxy absolute magnitude in a sample of LINERs. They interpreted this behavior as due 
to differences in metallicities or to due to differing ionization conditions.

\section{Summary and Conclusions}

Optical emission lines are detected in 89\% of our sample.
This large fraction is somehow expected, since our sample was selected among 
a compilation of galaxies showing traces of ISM, and it is thus biased toward the presence 
of emission lines.
This fraction is higher than that derived by Phillips et al.~(\cite{ph86}) (55\%-60\%), Macchetto et al.~(\cite{mac96}); (72\% - 85\%),  Sarzi et al.~(\cite{sar06}) (75\%), 
Yan et al.~(\cite{yan06}) ( 52\%), and Serra et al.~(\cite{serra08}) (60\%), 
for other samples of early-type galaxies. If we consider only the galaxies in our sample with EW(H$\alpha$ $+$ [NII]6584)$>$3 \AA, the detection fraction drops to  57\%.

The incidence and the strength of emission in our sample is 
independent on the elliptical or lenticular morphological classes. 
These results are in agreement with Phillips et al.~(\cite{ph86}), while on the other hand Macchetto et al.~(\cite{mac96}) and Sarzi et al.~(\cite{sar06})  found a higher incidence of emission in lenticular than in elliptical galaxies.
Following the studies of Cappellari et al.~(\cite{capp07}) and Emsellem et al.~(\cite{Emsellem07}), we attempted a classification between fast and slow rotators in our sample. Fast and slow rotators account for $\sim$ 70\%  and  $\sim$ 30\% of the sample, respectively. This is interesting given the fact that the sample is composed of $\sim$ 70\% ellipticals and  $\sim$ 30\% lenticulars. Emission does not depend on the fast and slow rotator classification in our sample.

The EW of the lines tends to decrease from the center outwards. The [NII] line presents a steeper 
decrease than the H$\alpha$. However, given the uncertainties, we consider significant only
the decrease from the center to  $\sim$ 0.1 r$_e$. Deeper observations are needed to 
trace with higher confidence the EW out to larger radii. 
Some galaxies deviate from the general trend, 
either because they have a flatter emission profile (NGC~3489, NGC~7007, NGC~7377, NGC~5898), 
or because their emission is nuclear concentrated (IC~4296, NGC~4374, NGC~5090).
Previous narrowband imaging and integral field spectroscopy studies have revealed 
extended distribution for the ionized gas (Caldwell~\cite{cal84}; Goudfrooij et al.~\cite{gou94b};
Macchetto et al.~\cite{mac96}; Sarzi et al.~\cite{sar06,sar09}).

Extinction was derived from the observed H$\alpha$/H$\beta$ flux ratio, 
assuming an intrinsic value of $\approx$ 3.1 for AGN-like objects (Osterbrock~\cite{Osterbrock89}). 
This choice was driven by the fact that our galaxies show AGN-like [NII]/H$\alpha$ ratios. 
For the majority of the galaxies, the derived ${\rm E(B-V)}$ values are lower than 0.3.
Some galaxies display however very large reddenings, up to ${\rm E(B-V)\sim 1.5}$ or even more.
Since the observed continuum is incompatible with such large values, our result suggests 
that in these galaxies the dust has a patchy distribution. This is consistent with narrowband imaging studies revealing the presence of dust lanes and patches in early-type galaxies 
(e.g. Goudfrooij et al.~\cite{gou94b}).
 
We performed a spectral classification for our sample through the classical [OIII]/H$\beta$ versus [NII]/H$\alpha$ diagnostic diagram (BPT diagram). The classification  based on the central 
r$<$r$_{e}$/16 emission lines is provided in Table ~4. 
About 72\% of the galaxies present a  central LINERs emission activity.
Seyferts account for $\sim$ 9\% of the emission sample.
However, only IC~5063 can be classified as a {\it bona fide} Seyfert  galaxy.
The other 4 Seyferts (NGC~3489, NGC~777, NGC~7007, and NGC~6958),
have very large errors in [OIII]/H$\beta$ (because of the faint emission in H$\beta$), and the classification is less robust. We notice however that NGC~3489 and NGC~777 were classified as Seyferts also by other authors (Ho et al.~\cite{ho97b}, Sarzi et al.~\cite{sar06}).
7 galaxies ($\sim$ 12\% of the emission line sample) are  ``Composites'' (or Transition objects):
NGC~3258, NGC~4552, NGC~5193, NGC~5328, NGC~6721, NGC~6876, and IC~2006.
They have [NII]/H$\alpha$ ratios intermediate between HII regions and LINERs/Seyferts.
In our sample, these galaxies display the weakest emission lines. 
For the remaining 7\% of the emission line galaxies we can not provide a classification, either because the lines are too weak, or because the spectra are too noisy.

We derive spatial trends in the emission line ratios:
moving from the center toward annuli of increasing galacto-centric distance, we detect a clear decrease in the 
[NII]/H$\alpha$ ratio. The decrease in [OIII]/H$\beta$ is less clear.
It results that galaxies classified as LINERs from to their nuclear line ratios,
shift toward the region of ``Composites'' if one consider the emission from more external regions.
This could be interpreted either with a decrease in the nebular metallicity, or with 
a progressive ``softening'' of the ionizing spectrum.

We have investigated possible relations between the central line emission properties and the host galaxy properties.
The main results are:

\begin{enumerate}

\item Seyferts have young luminosity weighted ages ($\lesssim$ 5 Gyr), and are on average younger than LINERs.  LINERs span a wide age range, from a few Gyrs to a Hubble time. ``Composite'' have ages older than 5 Gyr, at variance with the idea that they may originate from a combined contribution of AGN and star formation.

\item Excluding the Seyferts, the spread in the emission equivalent width ([NII], H$\alpha$ or [OIII])
increases with the galaxy central velocity dispersion. Low-$\sigma$ galaxies have weak emission lines, while high-$\sigma$ galaxies show both weak and strong emission lines. Equivalent widths in [NII]$\lambda$6584 larger than $\sim$ 5 \AA \ are found only in galaxies with $\sigma_c >$ 200 km s$^{-1}$. 

\item If we consider only the high emission subsample (H) (see Table~4), where the emission line ratios are more reliable, a positive correlation exists between the [NII]/H$\alpha$ ratio and  $\sigma_c$.

\item Excluding the Seyferts, which have larger [OIII]/H$\beta$ and younger ages, 
no relation between the emission line ratios and the galaxy age, metallicity, or $\alpha$/Fe 
enhancement is found.

\end{enumerate}

\subsection{The AGN-starburst connection in Seyferts}
 
The first result is in agreement with previous findings that LINERs are older than Seyferts 
(e.g., Kewley et al.~\cite{kew06}), and supports the idea that the star formation and the AGN phenomena co-exist (Terlevich et al.~\cite{terl90}; Cid Fernandes \& Terlevich~\cite{cf95}; Heckman et al.~\cite{heck97}; Gonz{\'a}lez Delgado et al.~\cite{gd01}; Kauffmann et al.~\cite{Kauff03}; Cid Fernandes at al.~\cite{cf05}; Davies et al.~\cite{davies07}; Riffel et al.~\cite{riff09}). 
The young luminosity weighted ages in the Seyferts of our sample are likely the result of recent star formation  
episodes superimposed to a several Gyrs old stellar population (see Paper~III).
Davies et al.~(\cite{davies07}) provide evidence for a delay of 50-100 Myr between the onset of star formation and the subsequent fueling of the black hole. Given the short duration of the AGN phenomenon ($\lesssim$ 100 Myr), this suggests that star formation has occurred a few hundreds Myrs ago in the Seyfert galaxies of our sample.
Because of the short duration of the AGN phase, we also expect to observe galaxies with relatively young 
luminosity-weighted ages but no signatures of current Seyfert activity.
This is in agreement with the fact that LINERs span a wide luminosity-weighted age range, from a few Gyrs up to a Hubble time.

\subsection{The ionization mechanism in LINERs}

Several studies have demonstrated that a minimum level of ionizing photons are produced 
in any evolved stellar systems. Thus, if cold gas is present, some level of 
nebular emission is  always expected in early-type galaxies. 
However, the critical question is: can PAGB stars provide {\it all} the ionization observed in early-type galaxies? 
 In answering this question, we should first of all distinguish between the central LINERs activity, and 
the more extended LINERs-like emission. 
Indeed, through the BPT classification carried out in annuli of increasing galacto-centric distance, we have demonstrated that ETGs have both a central and an extended LINERs-like emission.
 When comparing the stronger  {\underline {nuclear}} r$<$r$_e$/16 emission with the results of the 
photoionization models from Binette et al.~(\cite{bin94}), we obtain 
that only 11 out of the 49 galaxies classified as LINERs/Composites in our sample (i.e. in $\approx$ 
 22\% of the LINERs/Comp sample) can be explained with photoionization by old stars alone.
Of these 11 galaxies, 6 are classified as LINERs and 5 as Composites.

For the other 78\% of LINERs/Composites, some mechanism different than photoionization by PAGB stars must be at work. We warn however that this fraction is very uncertain, since there is no precise description yet for the evolution and number of PAGB stars (e.g., Brown et al.~\cite{br08}).

On the other hand, we can not rule out the importance of hot old stars as a photoionization source at larger radii. Indeed, the emission strength progressively decreases 
toward larger galacto-centric distances, so that and an increasing number of LINERs become compatible with pure 
PAGB stars photoionization. 

{\it Summarizing, we can not exclude a scenario in which more than a source of ionizing photons are present,
with different roles at different radii. We can think of a transition region from the center,
where PAGB stars fail in accounting alone for the ionizing photon budget, to more extended regions, where they
may do well the job, in agreement with recent findings (Sarzi et al.~\cite{sar09}, Eracleous et al.~\cite{era10}). }

Studies of LINERs based on large sample, such as the SDSS, have been recently presented 
(Stasi{\'n}ska et al.~(\cite{sta08}, Cid Fernandes et al.~(\cite{cf09}). However, 
we warn against direct comparisons of our findings with the results from SDSS for at least two reasons:
 the 3''-wide aperture of the SDSS usually encompasses large Kpc-scale regions where diffuse emission
 contribute to the nebular fluxes; and our sample is biased toward the presence of emission lines.
%Here we just recall that Stasi{\'n}ska et al.~(\cite{sta08}) showed that only $\sim$ 1/4 of the LINERs in the SDSS can be explained with photoionization by old stars. Cid Fernandes et al.~(\cite{cf09}) showed that this fraction is higher if very weak emission galaxies (S/N in H$\beta$ $<$ 3) are also included.}

Once ascertained the limited role of PAGB stars in the nuclear LINERs region, 
we investigated other possible scenarios, more specifically  
photoionization by sub-Eddington accretion rate AGN and fast shocks.
Direct evidence for the presence of AGNs comes from the detection of compact nuclear X-ray/ radio sources. 
High resolution X-ray and/or radio data are available for 18 galaxies in our sample, 16 of which are classified as LINERs.
Among the 16 LINERs, evidence of an AGN is found in 10.  
Even if the statistics are small, this result suggests the presence of 
low accretion rate AGNs in $\sim$ 62\% of LINERs.
 However, the presence of an AGN does not exclude the coexistence of shock heating:
indeed, jet driven flows and dissipative accretion disks in active galaxies with 
strongly sub-Eddington accretion have ben proposed as possible sources of shocks (Dopita \& Sutherland~(\cite{dosu95}), Dopita et al.~\cite{dop97}), 
and bubbles likely produced by an AGN outburst 
have been observed in LINERs (e.g. Baldi et al.~\cite{bald09}).

Both AGN and shock models reproduce the central emission line ratios in our sample.
In the dusty AGN models of Groves et al.~(\cite{gro04}), LINERs require 
a ionization parameter ${\rm \log U < -3}$, significantly lower than for Seyferts.
This is consistent with a scenario in which LINERs are AGNs accreting at a much lower rate 
than Seyferts.
LINERs can also be modeled in terms of fast shocks 
(200-500 km s$^{-1}$)  in a relatively gas poor environment 
(pre-shock densities $n \approx 0.01 -100$ cm$^{-3}$, depending on the 
magnetic field), with gas metallicity $\sim$ solar. 
These low pre-shock densities are consistent with the observed [SII]6717/[SII]6731 ratios.
A critical point concerns the high shock velocities required to explain the 
emission line spectrum in LINERs. Jet driven flows or accretion into  a massive BH can well do the job
(Nagar et al.~\cite{nag05}, Dopita et al.~\cite{dop97}). 
We suppose that an alternative source of  
mechanical energy comes from turbulent motion of gas clouds within the potential well of the galaxy.
This scenario can not be excluded since several studies revealed  
a gaseous component of high velocity dispersion, in the range 
150- 250 Km/s, in the center of early-type galaxies (e.g. Bertola et al.~\cite{ber95}; Zeilinger et al.~\cite{zei96}, 
Sarzi et al.~\cite{sar06}), which translates into higher intrinsic 3D velocity dispersions.
This scenario is very appealing since it would naturally explain the 
observed trends between the emission strength, and the [NII]/H$\alpha$ ratio, with the 
galaxy stellar central velocity dispersion.

A solid result from our study is that the [NII]/H$\alpha$ ratio decreases with increasing galacto-centric distance. 
This trend may reflect metallicity gradients in the gas, similarly to what observed 
for the stars. However, this is difficult to fit in a scenario in which the gas has an external origin (see Section~7.3). If not due to metallicity variations, the [NII]/H$\alpha$ gradient should reflect variations in the properties of the ionization mechanism. 
In the shock scenario, this requires decreasing shock velocities  with increasing galacto-centric distances.
Unfortunately, we lack the spectral resolution necessary to determine the gas velocity dispersion and,
ultimately, to ascertain the presence of a radial gradient in the shock velocity.

A large fraction (40/65) of our sample was observed with {\it Spitzer}-IRS. 
In a forthcoming paper, we will use MIR ionic and molecular emission lines, and Polycyclic Aromatic Hydrocarbon features, to further investigate the powering mechanism in LINERs, and better disentangle
the contributions from AGN photoionization and shock heating.

\subsection{Origin of the gas}

 We derived the oxygen abundance using two calibrations present in the literature:
 the Kobulnicky et al.~(\cite{Kobulnicky99}) calibration, valid for HII regions, and the Storchi-Bergmann et al.~(\cite{Storchi98}) calibration, derived in the assumption of photoionization by a typical AGN continuum (Mathews \& Ferland~\cite{mat87}).
The first approach was also adopted by  Athey \& Bregman~(\cite{ab09}) to derive nebular metallicities in  7 early-type galaxies, and is probably reasonable in the case of photoionization by PAGB stars.
The second calibration is valid in the case of AGN excitation, and may be reasonable if LINERs are low-accretion rate AGNs.
Our result is the following: 
irrespective of the adopted calibration, the nebular oxygen metallicities are significantly lower than the stellar ones. The main caveat is that power-law ionizing spectra tend to provide abundances up to $\sim$0.5 dex larger compared to a typical AGN continuum.
A calibration in the case of shock heating is not available in the literature. However, we can see 
that shock models indicate nebular metallicities not higher than solar in our sample, in agreement with the results 
obtained with the Kobulnicky et al.~(\cite{Kobulnicky99}) and  Storchi-Bergmann et al.~(\cite{Storchi98}) calibrations.

Our result has an important implication for the origin of the gas, suggesting that 
it may have at least in part an external origin  
(e.g., from a cooling flow or from accretion from a companion galaxy).
This is in agreement with 
photometric and kinematical studies  in ETGs 
showing in many cases gas/star misalignment and gas/star angular momentum decoupling
(Bertola et al.~\cite{ber92}; van Dokkum \& Franx~\cite{vdf95}; Caon et al.~\cite{caon00}; Sarzi et al.~\cite{sar06}).
Indeed, kinematical and morphological peculiarities  are present in 50\% of our sample, suggesting 
that accretion is frequent. Accretion of fresh gas can feed the central supermassive BH, and activate 
the AGN emission. This is consistent with the fact that a major fraction of our galaxies studied in the X-ray 
 and/or radio show nuclear compact sources.

Finally, we can not exclude that the low oxygen abundance is unrelated to the gas origin, and 
reflects the fact that oxygen does not vary on lockstep with the other $\alpha$-elements.
In a survey of 27 Milky Way bulge giants, Fulbright, McWilliam, \& Rich~(\cite{fulbr06}) found that, 
although Mg appears to be enhanced at all [Fe/H], [O/Fe] declines with increasing [Fe/H] and is solar or 
mildly sub-solar at [Fe/H] $\ge 0$. Also, X -ray studies have derived 
subsolar [O/Fe] values, but supersolar [Mg/Fe] values, in the hot gas of elliptical galaxies
(e.g., Humphrey \& Buote~\cite{hb06}; Ji et al.~\cite{ji09}).
This may be the evidence of an overestimate of O yield by SNII models,
which do not consider  significant mass loss at the late stage of massive  progenitor stars 
(Ji et al.~\cite{ji09}).

\begin{acknowledgements}

We acknowledge a financial contribution from contract ASI-INAF I/016/07/0.
F. A. thanks R. van der Marel for useful suggestions.
We thank the anonymous referee for having carefully read the manuscript, and
for his/her useful comments and suggestions that helped to improve 
significantly the paper.

\end{acknowledgements}

%-----------------------------------------Table 2 ---------------------------------------
{\tiny{
\longtabL{2}{
\begin{landscape}
\begin{longtable}{cccccccccccccc}
\caption{Observed emission line fluxes in units of ${\rm 10^{-16} erg \ s^{-1} \  cm^{-2} \ arcsec^{-2}}$}\\
%\tiny
%see myrun_err/emissions.tab
%{\bf Table~1. Emission line intensities derived for the sample}
%\begin{tabular}{ccccccccccccc}
\hline \hline
Galaxy   &   Reg\footnote{Reg 1 $=$ r $\leq$r$_e$/16; Reg 2 $=$ r$_{e}$/16 $<$ r $\leq$r$_e$/8; Reg 3 $=$ r$_{e}$/8$<$ r $\leq$r$_e$/4; Reg 4 $=$ r$_{e}$/4 $<$ r $\leq$r$_e$/2.}
&  E(B-V) & [O II]\footnote{Negative emission values are consistent with no emission.}  & [Ne III] &  H$\beta$  & [O III] &  [O III]  & [O I] & [N II] &   H$\alpha$  & [N II] & [S II] & [S II ]    \\ 
& & & $\lambda3727$ & $\lambda3869$ & $\lambda4861$ & $\lambda4959$ & $\lambda5007$ & $\lambda6300$ & 
$\lambda6548$ & $\lambda6563$ & $\lambda6584$ & $\lambda6717$ &  $\lambda6731$ \\ 
%& & & $erg s^{-1} cm^{-2} arcsec^{-2}$ & & & & & & & & & &  \\
\hline
NGC128     & 1    &    0.26 $\pm$     0.29 &     6.77 $\pm$     0.55 &     0.40 $\pm$     0.55 &     1.76 $\pm$     0.50 &     1.35 $\pm$     0.20 &     2.71 $\pm$     0.25 &     0.99 $\pm$     0.06 &     2.96 $\pm$     0.16 &     7.13 $\pm$     0.34 &     8.72 $\pm$     0.10 &     2.70 $\pm$     0.23 &     1.51 $\pm$     0.18 \\ 
NGC128     & 2    &    0.03 $\pm$     0.28 &     5.87 $\pm$     0.49 &     0.07 $\pm$     0.44 &     1.68 $\pm$     0.45 &     0.85 $\pm$     0.15 &     1.96 $\pm$     0.19 &     0.51 $\pm$     0.09 &     2.29 $\pm$     0.17 &     5.26 $\pm$     0.30 &     6.02 $\pm$     0.04 &     2.16 $\pm$     0.19 &     1.64 $\pm$     0.18 \\ 
NGC128     & 3    &    0.15 $\pm$     0.41 &     2.96 $\pm$     0.34 &    -0.06 $\pm$     0.31 &     0.74 $\pm$     0.30 &     0.48 $\pm$     0.08 &     1.02 $\pm$     0.11 &    -0.67 $\pm$     0.06 &     1.09 $\pm$     0.11 &     2.68 $\pm$     0.20 &     2.99 $\pm$     0.04 &     1.07 $\pm$     0.12 &     0.98 $\pm$     0.12 \\ 
NGC128     & 4    &    0.03 $\pm$     0.46 &     0.62 $\pm$     0.15 &    -0.09 $\pm$     0.15 &     0.39 $\pm$     0.17 &     0.22 $\pm$     0.05 &     0.35 $\pm$     0.06 &     0.57 $\pm$     0.03 &     0.46 $\pm$     0.05 &     0.76 $\pm$     0.10 &    -0.12 $\pm$     0.01 &     0.36 $\pm$     0.07 &     0.39 $\pm$     0.07 \\ 
&    &    &    &    &    &    &    &    &    &    &    &    &    \\
NGC777 \footnote{The continuum was fitted with 2 SSPs}    & 1    &    2.81 $\pm$     3.53 &     0.89 $\pm$     0.16 &     0.13 $\pm$     0.23 &     0.03 $\pm$     0.12 &    -0.03 $\pm$     0.00 &     1.14 $\pm$     0.07 &     0.81 $\pm$     0.02 &     1.13 $\pm$     0.12 &     1.71 $\pm$     0.05 &     1.51 $\pm$     0.02 &     7.44 $\pm$     0.13 &   $-$ \\ 
NGC777 $^5$    & 2    &    0.52 $\pm$     0.71 &     0.46 $\pm$     0.10 &    -0.21 $\pm$     0.15 &     0.16 $\pm$     0.12 &    -0.09 $\pm$     0.05 &     0.67 $\pm$     0.08 &     0.15 $\pm$     0.03 &     0.65 $\pm$     0.09 &     0.85 $\pm$     0.04 &     0.58 $\pm$     0.01 &     3.82 $\pm$     0.09 &   $-$ \\ 
NGC777$^5$     & 3    &    0.05 $\pm$     0.24 &     0.20 $\pm$     0.04 &    -0.14 $\pm$     0.04 &     0.14 $\pm$     0.03 &    -0.00 $\pm$     0.00 &     0.23 $\pm$     0.02 &     0.16 $\pm$     0.02 &     0.09 $\pm$     0.01 &     0.35 $\pm$     0.01 &     0.17 $\pm$     0.00 &     1.74 $\pm$     0.01 &   $-$ \\ 
NGC777$^5$     & 4    &    0.05 $\pm$     0.94 &     0.13 $\pm$     0.04 &    -0.01 $\pm$     0.05 &    -0.04 $\pm$     0.04 &     0.02 $\pm$     0.01 &     0.06 $\pm$     0.02 &     0.13 $\pm$     0.01 &     0.04 $\pm$     0.01 &     0.07 $\pm$     0.02 &     0.00 $\pm$     0.00 &     0.43 $\pm$     0.01 &   $-$ \\ 
&    &    &    &    &    &    &    &    &    &    &    &    &    \\
NGC1052    & 1    &    0.22 $\pm$     0.04 &    88.41 $\pm$     0.87 &    10.79 $\pm$     1.11 &    29.21 $\pm$     1.18 &    15.98 $\pm$     0.55 &    73.78 $\pm$     0.64 &    53.41 $\pm$     0.45 &    49.57 $\pm$     0.94 &   112.80 $\pm$     0.94 &   109.76 $\pm$     0.55 &    67.68 $\pm$     1.16 &    60.98 $\pm$     0.99 \\ 
NGC1052    & 2    &    0.38 $\pm$     0.09 &    30.98 $\pm$     0.76 &     2.71 $\pm$     0.54 &     7.68 $\pm$     0.65 &     8.54 $\pm$     0.56 &    19.70 $\pm$     0.39 &    13.29 $\pm$     0.20 &    15.98 $\pm$     0.42 &    34.57 $\pm$     0.53 &    32.99 $\pm$     0.18 &    23.96 $\pm$     0.44 &    17.44 $\pm$     0.32 \\ 
NGC1052    & 3    &    0.24 $\pm$     0.20 &     6.71 $\pm$     0.32 &     0.53 $\pm$     0.28 &     1.37 $\pm$     0.27 &     0.74 $\pm$     0.09 &     2.52 $\pm$     0.13 &     1.59 $\pm$     0.07 &     1.98 $\pm$     0.13 &     5.38 $\pm$     0.18 &     4.67 $\pm$     0.04 &     3.62 $\pm$     0.15 &     2.62 $\pm$     0.12 \\ 
NGC1052    & 4    &    0.03 $\pm$     0.45 &     0.60 $\pm$     0.14 &     0.04 $\pm$     0.19 &     0.29 $\pm$     0.13 &     0.07 $\pm$     0.06 &     0.31 $\pm$     0.08 &     0.00 $\pm$     0.01 &     0.34 $\pm$     0.04 &     0.71 $\pm$     0.08 &     0.35 $\pm$     0.01 &     0.25 $\pm$     0.05 &     0.21 $\pm$     0.03 \\ 
&    &    &    &    &    &    &    &    &    &    &    &    &    \\
NGC1209    & 1    &    0.04 $\pm$     0.23 &    12.74 $\pm$     0.96 &    -0.39 $\pm$     0.96 &     3.74 $\pm$     0.85 &     2.04 $\pm$     0.30 &     4.82 $\pm$     0.37 &     1.10 $\pm$     0.17 &     5.46 $\pm$     0.28 &     9.39 $\pm$     0.46 &    14.21 $\pm$     0.09 &     4.69 $\pm$     0.36 &     3.94 $\pm$     0.31 \\ 
NGC1209    & 2    &    0.04 $\pm$     0.24 &     7.80 $\pm$     0.71 &    -0.57 $\pm$     0.84 &     3.10 $\pm$     0.71 &     1.72 $\pm$     0.22 &     3.52 $\pm$     0.29 &     0.91 $\pm$     0.13 &     3.43 $\pm$     0.19 &     6.34 $\pm$     0.36 &     9.02 $\pm$     0.11 &     2.41 $\pm$     0.25 &     2.36 $\pm$     0.22 \\ 
NGC1209    & 3    &    0.04 $\pm$     0.25 &     2.83 $\pm$     0.47 &    -0.63 $\pm$     0.63 &     1.79 $\pm$     0.43 &     0.80 $\pm$     0.22 &     1.21 $\pm$     0.28 &     0.53 $\pm$     0.04 &     1.08 $\pm$     0.11 &     2.63 $\pm$     0.21 &     3.07 $\pm$     0.05 &     0.73 $\pm$     0.14 &     0.75 $\pm$     0.11 \\ 
NGC1209    & 4    &    0.04 $\pm$     0.52 &     0.14 $\pm$     0.18 &    -0.13 $\pm$     0.26 &     0.35 $\pm$     0.17 &     0.31 $\pm$     0.10 &     0.31 $\pm$     0.18 &     0.13 $\pm$     0.03 &     0.26 $\pm$     0.04 &     0.42 $\pm$     0.09 &     0.43 $\pm$     0.02 &    -0.04 $\pm$     0.05 &    -0.00 $\pm$     0.04 \\ 
&    &    &    &    &    &    &    &    &    &    &    &    &    \\
NGC1297    & 1    &    0.52 $\pm$     0.52 &     3.74 $\pm$     0.43 &     0.16 $\pm$     0.51 &     0.67 $\pm$     0.34 &     0.64 $\pm$     0.19 &     1.35 $\pm$     0.24 &     0.64 $\pm$     0.06 &     1.38 $\pm$     0.10 &     3.49 $\pm$     0.23 &     3.04 $\pm$     0.04 &     2.07 $\pm$     0.15 &     1.91 $\pm$     0.11 \\ 
NGC1297    & 2    &    0.06 $\pm$     0.44 &     1.30 $\pm$     0.17 &    -0.01 $\pm$     0.25 &     0.40 $\pm$     0.17 &     0.34 $\pm$     0.10 &     0.74 $\pm$     0.15 &     0.41 $\pm$     0.04 &     0.59 $\pm$     0.04 &     1.33 $\pm$     0.10 &     1.08 $\pm$     0.02 &     0.77 $\pm$     0.06 &     0.64 $\pm$     0.04 \\ 
NGC1297    & 3    &    0.03 $\pm$     0.50 &     0.42 $\pm$     0.08 &     0.07 $\pm$     0.10 &     0.18 $\pm$     0.09 &     0.19 $\pm$     0.05 &     0.19 $\pm$     0.05 &     0.04 $\pm$     0.02 &     0.19 $\pm$     0.02 &     0.36 $\pm$     0.05 &     0.24 $\pm$     0.01 &     0.14 $\pm$     0.03 &     0.11 $\pm$     0.01 \\ 
NGC1297    & 4    &    1.00 $\pm$     1.83 &     0.08 $\pm$     0.02 &     0.03 $\pm$     0.04 &     0.03 $\pm$     0.05 &     0.03 $\pm$     0.02 &     0.07 $\pm$     0.02 &     0.09 $\pm$     0.01 &     0.14 $\pm$     0.01 &     0.22 $\pm$     0.02 &    -0.03 $\pm$     0.00 &     0.18 $\pm$     0.01 &     0.04 $\pm$     0.01 \\ 
&    &    &    &    &    &    &    &    &    &    &    &    &    \\
NGC1380    & 1    &    0.25 $\pm$     0.35 &     3.20 $\pm$     0.84 &    -0.58 $\pm$     0.86 &     2.31 $\pm$     0.80 &     1.23 $\pm$     0.25 &     3.41 $\pm$     0.39 &     1.85 $\pm$     0.16 &     5.32 $\pm$     0.31 &     9.21 $\pm$     0.47 &    12.32 $\pm$     0.11 &     3.49 $\pm$     0.37 &     3.55 $\pm$     0.40 \\ 
NGC1380    & 2    &    0.51 $\pm$     0.47 &     2.86 $\pm$     0.70 &     0.00 $\pm$     0.73 &     1.45 $\pm$     0.68 &     0.76 $\pm$     0.20 &     2.13 $\pm$     0.29 &     0.94 $\pm$     0.15 &     3.09 $\pm$     0.22 &     7.44 $\pm$     0.37 &     8.29 $\pm$     0.10 &     2.73 $\pm$     0.29 &     1.98 $\pm$     0.25 \\ 
NGC1380    & 3    &    0.44 $\pm$     0.49 &     1.52 $\pm$     0.45 &    -0.10 $\pm$     0.50 &     0.91 $\pm$     0.44 &    -0.02 $\pm$     0.06 &     1.02 $\pm$     0.22 &     0.59 $\pm$     0.11 &     1.69 $\pm$     0.14 &     4.38 $\pm$     0.26 &     4.66 $\pm$     0.05 &     1.80 $\pm$     0.18 &     1.16 $\pm$     0.16 \\ 
NGC1380    & 4    &    0.02 $\pm$     0.44 &     0.47 $\pm$     0.23 &     0.01 $\pm$     0.32 &     0.61 $\pm$     0.26 &     0.03 $\pm$     0.10 &     0.62 $\pm$     0.15 &     0.01 $\pm$     0.02 &     0.93 $\pm$     0.07 &     1.64 $\pm$     0.13 &     1.63 $\pm$     0.03 &     0.55 $\pm$     0.09 &     0.26 $\pm$     0.07 \\ 
&    &    &    &    &    &    &    &    &    &    &    &    &    \\
NGC1453    & 1    &    0.11 $\pm$     0.19 &    10.06 $\pm$     0.51 &     0.08 $\pm$     0.42 &     2.72 $\pm$     0.52 &     1.72 $\pm$     0.24 &     3.93 $\pm$     0.27 &     1.42 $\pm$     0.12 &     4.37 $\pm$     0.23 &     9.21 $\pm$     0.31 &    12.56 $\pm$     0.05 &     6.28 $\pm$     0.29 &     4.89 $\pm$     0.26 \\ 
NGC1453    & 2    &    0.11 $\pm$     0.21 &     5.37 $\pm$     0.39 &    -0.14 $\pm$     0.38 &     1.88 $\pm$     0.38 &     1.35 $\pm$     0.13 &     2.59 $\pm$     0.17 &     0.59 $\pm$     0.07 &     2.45 $\pm$     0.14 &     4.96 $\pm$     0.23 &     6.10 $\pm$     0.04 &     2.65 $\pm$     0.18 &     2.40 $\pm$     0.20 \\ 
NGC1453    & 3    &    0.11 $\pm$     0.23 &     1.39 $\pm$     0.16 &    -0.17 $\pm$     0.21 &     0.75 $\pm$     0.17 &     0.54 $\pm$     0.06 &     0.98 $\pm$     0.09 &     0.09 $\pm$     0.02 &     0.58 $\pm$     0.06 &     1.34 $\pm$     0.10 &     1.43 $\pm$     0.02 &     0.40 $\pm$     0.06 &     0.59 $\pm$     0.07 \\ 
NGC1453    & 4    &    0.11 $\pm$     0.29 &     0.43 $\pm$     0.06 &    -0.03 $\pm$     0.09 &     0.21 $\pm$     0.06 &     0.09 $\pm$     0.03 &     0.36 $\pm$     0.07 &    -0.00 $\pm$     0.02 &     0.15 $\pm$     0.02 &     0.44 $\pm$     0.03 &     0.40 $\pm$     0.01 &     0.14 $\pm$     0.02 &     0.08 $\pm$     0.02 \\ 
&    &    &    &    &    &    &    &    &    &    &    &    &    \\
NGC1521$^5$    & 1    &    1.22 $\pm$     1.51 &     0.80 $\pm$     0.40 &     0.04 $\pm$     0.63 &     0.24 $\pm$     0.36 &     0.51 $\pm$     0.12 &     0.49 $\pm$     0.23 &    -0.00 $\pm$     0.03 &     1.16 $\pm$     0.14 &     2.54 $\pm$     0.25 &     3.27 $\pm$     0.06 &     0.56 $\pm$     0.22 &     0.46 $\pm$     0.19 \\ 
NGC1521$^5$   & 2    &    0.30 $\pm$     1.20 &     0.54 $\pm$     0.37 &     0.05 $\pm$     0.59 &     0.30 $\pm$     0.36 &     0.07 $\pm$     0.18 &     0.17 $\pm$     0.28 &    -0.01 $\pm$     0.02 &     0.33 $\pm$     0.09 &     1.26 $\pm$     0.19 &     1.40 $\pm$     0.05 &     0.48 $\pm$     0.09 &     0.28 $\pm$     0.07 \\ 
NGC1521$^5$    & 3    &    0.04 $\pm$     8.66 &     0.32 $\pm$     0.17 &    -0.29 $\pm$     0.27 &    -0.02 $\pm$     0.18 &    -0.01 $\pm$     0.10 &     0.14 $\pm$     0.15 &     0.00 $\pm$     0.01 &     0.42 $\pm$     0.04 &     0.29 $\pm$     0.09 &     0.52 $\pm$     0.03 &     0.10 $\pm$     0.04 &     0.09 $\pm$     0.04 \\ 
NGC1521$^5$   & 4    &    0.04 $\pm$     0.77 &     0.04 $\pm$     0.05 &    -0.06 $\pm$     0.08 &     0.08 $\pm$     0.06 &     0.08 $\pm$     0.04 &     0.22 $\pm$     0.06 &     0.03 $\pm$     0.01 &    -0.01 $\pm$     0.01 &     0.14 $\pm$     0.04 &     0.10 $\pm$     0.01 &     0.01 $\pm$     0.01 &     0.01 $\pm$     0.01 \\ 
&    &    &    &    &    &    &    &    &    &    &    &    &    \\
NGC1533    & 1    &    0.01 $\pm$     0.24 &     9.45 $\pm$     1.24 &     0.43 $\pm$     0.76 &     4.77 $\pm$     1.13 &     2.71 $\pm$     0.40 &     7.99 $\pm$     0.47 &     3.88 $\pm$     0.19 &     6.02 $\pm$     0.32 &    12.56 $\pm$     0.63 &    13.90 $\pm$     0.09 &     4.66 $\pm$     0.42 &     3.82 $\pm$     0.38 \\ 
NGC1533    & 2    &    0.01 $\pm$     0.23 &     4.99 $\pm$     0.70 &    -0.42 $\pm$     0.80 &     2.94 $\pm$     0.65 &     1.31 $\pm$     0.23 &     3.54 $\pm$     0.30 &     1.23 $\pm$     0.13 &     2.06 $\pm$     0.17 &     4.71 $\pm$     0.36 &     4.70 $\pm$     0.08 &     1.62 $\pm$     0.24 &     1.34 $\pm$     0.20 \\ 
NGC1533    & 3    &    0.01 $\pm$     0.30 &     2.54 $\pm$     0.32 &     0.01 $\pm$     0.39 &     1.01 $\pm$     0.28 &     0.83 $\pm$     0.12 &     1.90 $\pm$     0.18 &     0.29 $\pm$     0.06 &     0.56 $\pm$     0.07 &     1.72 $\pm$     0.15 &     1.37 $\pm$     0.03 &     0.88 $\pm$     0.10 &     0.65 $\pm$     0.08 \\ 
NGC1533    & 4    &    0.01 $\pm$     0.45 &     0.62 $\pm$     0.11 &    -0.01 $\pm$     0.14 &     0.18 $\pm$     0.08 &     0.17 $\pm$     0.05 &     0.30 $\pm$     0.04 &     0.01 $\pm$     0.01 &     0.16 $\pm$     0.02 &     0.50 $\pm$     0.05 &     0.39 $\pm$     0.01 &     0.33 $\pm$     0.03 &     0.17 $\pm$     0.02 \\ 
&    &    &    &    &    &    &    &    &    &    &    &    &    \\
%NGC1553    & 1    &    0.26 $\pm$     0.43 &     6.83 $\pm$     1.12 &    -1.38 $\pm$     1.26 &     2.26 $\pm$     0.96 &     2.74 $\pm$     0.40 &     5.26 $\pm$     0.46 &     3.26 $\pm$     0.22 &     6.16 $\pm$     0.36 &     9.02 $\pm$     0.62 &    15.55 $\pm$     0.19 &     2.84 $\pm$     0.48 &     2.35 $\pm$     0.43 \\ 
%NGC1553    & 2    &    0.04 $\pm$     0.55 &     1.82 $\pm$     0.43 &    -0.76 $\pm$     0.62 &     0.74 $\pm$     0.39 &     0.79 $\pm$     0.15 &     1.29 $\pm$     0.20 &     0.44 $\pm$     0.13 &     1.20 $\pm$     0.14 &     2.37 $\pm$     0.25 &     2.77 $\pm$     0.05 &     0.57 $\pm$     0.20 &     0.62 $\pm$     0.18 \\ 
%NGC1553    & 3    &    0.07 $\pm$     0.69 &     0.51 $\pm$     0.20 &    -0.28 $\pm$     0.30 &     0.26 $\pm$     0.18 &     0.36 $\pm$     0.04 &     0.45 $\pm$     0.09 &     0.07 $\pm$     0.02 &     0.09 $\pm$     0.05 &     0.87 $\pm$     0.12 &     0.72 $\pm$     0.03 &     0.17 $\pm$     0.08 &     0.21 $\pm$     0.07 \\ 
%NGC1553    & 4    &    0.19 $\pm$     1.53 &     0.01 $\pm$     0.06 &    -0.23 $\pm$     0.10 &     0.04 $\pm$     0.06 &     0.07 $\pm$     0.02 &     0.04 $\pm$     0.02 &     0.12 $\pm$     0.02 &     0.06 $\pm$     0.02 &     0.16 $\pm$     0.04 &     0.11 $\pm$     0.01 &     0.02 $\pm$     0.04 &     0.01 $\pm$     0.02 \\ 
&    &    &    &    &    &    &    &    &    &    &    &    &    \\
\hline
\end{longtable}
\end{landscape}}
}
}
%------------------------------ end Table 2 ----------------------------------------------

%\newpage
%  --------------------------- Table 3 ----------------------------------------------

{\tiny{
\longtabL{3}{
\begin{landscape}
\begin{longtable}{ccccccccccccc}
\caption{Observed emission line equivalent widths.}\\
%\tiny
%see myrun_err/emissions.tab
%{\bf Table~1. Emission line intensities derived for the sample}
%\begin{tabular}{ccccccccccccc}
\hline \hline
Galaxy   &   Reg\footnote{Reg 1 $=$ r $\leq$r$_e$/16; Reg 2 $=$ r$_{e}$/16 $<$ r $\leq$r$_e$/8; Reg 3 $=$ r$_{e}$/8$<$ r $\leq$r$_e$/4; Reg 4 $=$ r$_{e}$/4 $<$ r $\leq$r$_e$/2.}
&  [O II]\footnote{Negative emission values are consistent with no emission.} & [Ne III] &  H$\beta$  & [O III] &  [O III]  & [O I] & [N II] &   H$\alpha$  & [N II] & [S II] & [S II ]    \\ 
& & $\lambda3727$ & $\lambda3869$ & $\lambda4861$ & $\lambda4959$ & $\lambda5007$ & $\lambda6300$ & 
$\lambda6548$ & $\lambda6563$ & $\lambda6584$ & $\lambda6717$ &  $\lambda6731$ \\ 
\hline
&&&&&&&&&&&& \\
NGC128     & 1    &    7.25 $\pm$     0.59 &     0.59 $\pm$     0.81 &     0.57 $\pm$     0.16 &     0.43 $\pm$     0.06 &     0.87 $\pm$     0.08 &     0.27 $\pm$     0.02 &     0.74 $\pm$     0.04 &     1.78 $\pm$     0.09 &     2.18 $\pm$     0.03 &     0.73 $\pm$     0.06 &     0.41 $\pm$     0.05 \\ 
NGC128     & 2    &    7.13 $\pm$     0.59 &     0.13 $\pm$     0.75 &     0.64 $\pm$     0.17 &     0.32 $\pm$     0.05 &     0.73 $\pm$     0.07 &     0.17 $\pm$     0.03 &     0.69 $\pm$     0.05 &     1.58 $\pm$     0.09 &     1.81 $\pm$     0.01 &     0.71 $\pm$     0.06 &     0.54 $\pm$     0.06 \\ 
NGC128     & 3    &    5.00 $\pm$     0.58 &    -0.15 $\pm$     0.74 &     0.40 $\pm$     0.16 &     0.25 $\pm$     0.04 &     0.54 $\pm$     0.06 &    -0.31 $\pm$     0.03 &     0.48 $\pm$     0.05 &     1.18 $\pm$     0.09 &     1.32 $\pm$     0.02 &     0.51 $\pm$     0.06 &     0.47 $\pm$     0.06 \\ 
NGC128     & 4    &    1.92 $\pm$     0.48 &    -0.41 $\pm$     0.68 &     0.39 $\pm$     0.17 &     0.22 $\pm$     0.05 &     0.34 $\pm$     0.06 &     0.51 $\pm$     0.03 &     0.38 $\pm$     0.04 &     0.62 $\pm$     0.08 &    -0.10 $\pm$     0.01 &     0.32 $\pm$     0.06 &     0.34 $\pm$     0.06 \\ 
&    &    &    &    &    &    &    &    &    &    &    &    \\
NGC777\footnote{The continuum was fitted with 2 SSPs}     & 1    &    1.36 $\pm$     0.24 &     0.27 $\pm$     0.50 &     0.01 $\pm$     0.04 &    -0.01 $\pm$     0.00 &     0.42 $\pm$     0.03 &     0.27 $\pm$     0.01 &     0.36 $\pm$     0.04 &     0.54 $\pm$     0.02 &     0.47 $\pm$     0.01 &     2.53 $\pm$     0.05 &    $-$ \\ 
NGC777$^8$     & 2    &    1.16 $\pm$     0.25 &    -0.71 $\pm$     0.50 &     0.11 $\pm$     0.08 &    -0.06 $\pm$     0.03 &     0.47 $\pm$     0.06 &     0.09 $\pm$     0.02 &     0.37 $\pm$     0.05 &     0.49 $\pm$     0.02 &     0.33 $\pm$     0.01 &     2.38 $\pm$     0.06 &    $-$  \\ 
NGC777$^8$      & 3    &    1.10 $\pm$     0.20 &    -1.04 $\pm$     0.28 &     0.23 $\pm$     0.05 &    -0.00 $\pm$     0.00 &     0.39 $\pm$     0.04 &     0.24 $\pm$     0.03 &     0.12 $\pm$     0.02 &     0.49 $\pm$     0.02 &     0.24 $\pm$     0.01 &     2.64 $\pm$     0.02 &    $-$  \\ 
NGC777$^8$      & 4    &    2.34 $\pm$     0.74 &    -0.31 $\pm$     1.10 &    -0.22 $\pm$     0.20 &     0.09 $\pm$     0.05 &     0.33 $\pm$     0.12 &     0.66 $\pm$     0.03 &     0.17 $\pm$     0.04 &     0.33 $\pm$     0.08 &     0.02 $\pm$     0.01 &     2.09 $\pm$     0.07 &    $-$  \\ 
&    &    &    &    &    &    &    &    &    &    &    &    \\
NGC1052    & 1    &   56.64 $\pm$     0.56 &     9.73 $\pm$     1.00 &     4.44 $\pm$     0.18 &     2.32 $\pm$     0.08 &    10.71 $\pm$     0.09 &     6.84 $\pm$     0.06 &     5.73 $\pm$     0.11 &    13.03 $\pm$     0.11 &    12.68 $\pm$     0.06 &     8.54 $\pm$     0.15 &     7.69 $\pm$     0.12 \\ 
NGC1052    & 2    &   31.55 $\pm$     0.78 &     3.89 $\pm$     0.77 &     2.03 $\pm$     0.17 &     2.23 $\pm$     0.14 &     5.14 $\pm$     0.10 &     2.96 $\pm$     0.04 &     3.23 $\pm$     0.09 &     6.98 $\pm$     0.11 &     6.66 $\pm$     0.04 &     5.23 $\pm$     0.10 &     3.81 $\pm$     0.07 \\ 
NGC1052    & 3    &   13.17 $\pm$     0.64 &     1.47 $\pm$     0.77 &     0.81 $\pm$     0.16 &     0.43 $\pm$     0.05 &     1.48 $\pm$     0.08 &     0.83 $\pm$     0.04 &     0.94 $\pm$     0.06 &     2.56 $\pm$     0.09 &     2.22 $\pm$     0.02 &     1.85 $\pm$     0.08 &     1.34 $\pm$     0.06 \\ 
NGC1052    & 4    &    2.69 $\pm$     0.63 &     0.24 $\pm$     1.14 &     0.43 $\pm$     0.19 &     0.10 $\pm$     0.08 &     0.46 $\pm$     0.11 &     0.01 $\pm$     0.01 &     0.42 $\pm$     0.05 &     0.87 $\pm$     0.09 &     0.43 $\pm$     0.02 &     0.33 $\pm$     0.07 &     0.28 $\pm$     0.05 \\ 
&    &    &    &    &    &    &    &    &    &    &    &    \\
NGC1209    & 1    &    6.97 $\pm$     0.53 &    -0.30 $\pm$     0.73 &     0.67 $\pm$     0.15 &     0.37 $\pm$     0.05 &     0.87 $\pm$     0.07 &     0.19 $\pm$     0.03 &     0.85 $\pm$     0.04 &     1.47 $\pm$     0.07 &     2.22 $\pm$     0.01 &     0.80 $\pm$     0.06 &     0.67 $\pm$     0.05 \\ 
NGC1209    & 2    &    5.42 $\pm$     0.50 &    -0.55 $\pm$     0.82 &     0.73 $\pm$     0.17 &     0.41 $\pm$     0.05 &     0.84 $\pm$     0.07 &     0.21 $\pm$     0.03 &     0.73 $\pm$     0.04 &     1.35 $\pm$     0.08 &     1.91 $\pm$     0.02 &     0.55 $\pm$     0.06 &     0.54 $\pm$     0.05 \\ 
NGC1209    & 3    &    3.49 $\pm$     0.58 &    -1.06 $\pm$     1.07 &     0.78 $\pm$     0.19 &     0.35 $\pm$     0.10 &     0.53 $\pm$     0.12 &     0.23 $\pm$     0.02 &     0.43 $\pm$     0.04 &     1.05 $\pm$     0.08 &     1.22 $\pm$     0.02 &     0.31 $\pm$     0.06 &     0.32 $\pm$     0.05 \\ 
NGC1209    & 4    &    0.43 $\pm$     0.55 &    -0.55 $\pm$     1.08 &     0.39 $\pm$     0.19 &     0.35 $\pm$     0.11 &     0.35 $\pm$     0.20 &     0.15 $\pm$     0.04 &     0.27 $\pm$     0.04 &     0.43 $\pm$     0.09 &     0.44 $\pm$     0.02 &    -0.04 $\pm$     0.06 &    -0.00 $\pm$     0.05 \\ 
&    &    &    &    &    &    &    &    &    &    &    &    \\
NGC1297    & 1    &    7.88 $\pm$     0.90 &     0.47 $\pm$     1.46 &     0.40 $\pm$     0.21 &     0.38 $\pm$     0.11 &     0.81 $\pm$     0.14 &     0.33 $\pm$     0.03 &     0.66 $\pm$     0.05 &     1.67 $\pm$     0.11 &     1.46 $\pm$     0.02 &     1.07 $\pm$     0.08 &     0.99 $\pm$     0.06 \\ 
NGC1297    & 2    &    4.66 $\pm$     0.59 &    -0.04 $\pm$     1.17 &     0.45 $\pm$     0.19 &     0.37 $\pm$     0.11 &     0.82 $\pm$     0.17 &     0.38 $\pm$     0.04 &     0.51 $\pm$     0.03 &     1.15 $\pm$     0.08 &     0.93 $\pm$     0.02 &     0.71 $\pm$     0.06 &     0.59 $\pm$     0.04 \\ 
NGC1297    & 3    &    3.22 $\pm$     0.64 &     0.63 $\pm$     0.99 &     0.45 $\pm$     0.22 &     0.48 $\pm$     0.12 &     0.47 $\pm$     0.13 &     0.09 $\pm$     0.04 &     0.38 $\pm$     0.04 &     0.72 $\pm$     0.10 &     0.47 $\pm$     0.02 &     0.29 $\pm$     0.06 &     0.25 $\pm$     0.03 \\ 
NGC1297    & 4    &    1.43 $\pm$     0.31 &     0.75 $\pm$     0.82 &     0.17 $\pm$     0.31 &     0.19 $\pm$     0.10 &     0.47 $\pm$     0.10 &     0.54 $\pm$     0.03 &     0.71 $\pm$     0.05 &     1.17 $\pm$     0.12 &    -0.17 $\pm$     0.02 &     1.02 $\pm$     0.08 &     0.22 $\pm$     0.05 \\ 
&    &    &    &    &    &    &    &    &    &    &    &    \\
NGC1380    & 1    &    1.88 $\pm$     0.50 &    -0.46 $\pm$     0.69 &     0.42 $\pm$     0.15 &     0.23 $\pm$     0.05 &     0.64 $\pm$     0.07 &     0.32 $\pm$     0.03 &     0.86 $\pm$     0.05 &     1.48 $\pm$     0.08 &     1.98 $\pm$     0.02 &     0.60 $\pm$     0.06 &     0.61 $\pm$     0.07 \\ 
NGC1380    & 2    &    1.95 $\pm$     0.47 &     0.00 $\pm$     0.67 &     0.32 $\pm$     0.15 &     0.18 $\pm$     0.05 &     0.49 $\pm$     0.07 &     0.21 $\pm$     0.03 &     0.62 $\pm$     0.04 &     1.50 $\pm$     0.07 &     1.67 $\pm$     0.02 &     0.59 $\pm$     0.06 &     0.43 $\pm$     0.05 \\ 
NGC1380    & 3    &    1.43 $\pm$     0.42 &    -0.13 $\pm$     0.64 &     0.30 $\pm$     0.14 &    -0.01 $\pm$     0.02 &     0.34 $\pm$     0.07 &     0.19 $\pm$     0.04 &     0.50 $\pm$     0.04 &     1.30 $\pm$     0.08 &     1.38 $\pm$     0.01 &     0.58 $\pm$     0.06 &     0.37 $\pm$     0.05 \\ 
NGC1380    & 4    &    0.79 $\pm$     0.38 &     0.02 $\pm$     0.71 &     0.36 $\pm$     0.15 &     0.02 $\pm$     0.06 &     0.37 $\pm$     0.09 &     0.01 $\pm$     0.01 &     0.50 $\pm$     0.04 &     0.88 $\pm$     0.07 &     0.88 $\pm$     0.02 &     0.32 $\pm$     0.05 &     0.15 $\pm$     0.04 \\ 
&    &    &    &    &    &    &    &    &    &    &    &    \\
NGC1453    & 1    &   10.44 $\pm$     0.53 &     0.12 $\pm$     0.61 &     0.82 $\pm$     0.16 &     0.52 $\pm$     0.07 &     1.20 $\pm$     0.08 &     0.36 $\pm$     0.03 &     1.02 $\pm$     0.05 &     2.14 $\pm$     0.07 &     2.93 $\pm$     0.01 &     1.57 $\pm$     0.07 &     1.22 $\pm$     0.06 \\ 
NGC1453    & 2    &    7.87 $\pm$     0.57 &    -0.28 $\pm$     0.77 &     0.82 $\pm$     0.17 &     0.58 $\pm$     0.06 &     1.12 $\pm$     0.07 &     0.22 $\pm$     0.03 &     0.84 $\pm$     0.05 &     1.70 $\pm$     0.08 &     2.09 $\pm$     0.01 &     0.97 $\pm$     0.07 &     0.88 $\pm$     0.07 \\ 
NGC1453    & 3    &    4.15 $\pm$     0.49 &    -0.70 $\pm$     0.88 &     0.69 $\pm$     0.15 &     0.51 $\pm$     0.06 &     0.91 $\pm$     0.09 &     0.08 $\pm$     0.01 &     0.44 $\pm$     0.04 &     1.01 $\pm$     0.07 &     1.08 $\pm$     0.02 &     0.33 $\pm$     0.05 &     0.48 $\pm$     0.06 \\ 
NGC1453    & 4    &    3.90 $\pm$     0.51 &    -0.33 $\pm$     1.07 &     0.62 $\pm$     0.17 &     0.26 $\pm$     0.10 &     1.05 $\pm$     0.21 &    -0.00 $\pm$     0.04 &     0.35 $\pm$     0.05 &     1.05 $\pm$     0.08 &     0.96 $\pm$     0.02 &     0.36 $\pm$     0.06 &     0.20 $\pm$     0.04 \\ 
&    &    &    &    &    &    &    &    &    &    &    &    \\
NGC1521$^8$     & 1    &    0.82 $\pm$     0.41 &     0.06 $\pm$     0.90 &     0.09 $\pm$     0.13 &     0.19 $\pm$     0.04 &     0.18 $\pm$     0.08 &    -0.00 $\pm$     0.01 &     0.36 $\pm$     0.04 &     0.78 $\pm$     0.08 &     1.00 $\pm$     0.02 &     0.18 $\pm$     0.07 &     0.15 $\pm$     0.06 \\ 
NGC1521$^8$     & 2    &    0.76 $\pm$     0.51 &     0.09 $\pm$     1.14 &     0.15 $\pm$     0.18 &     0.03 $\pm$     0.09 &     0.08 $\pm$     0.14 &    -0.00 $\pm$     0.01 &     0.14 $\pm$     0.04 &     0.55 $\pm$     0.08 &     0.61 $\pm$     0.02 &     0.23 $\pm$     0.04 &     0.13 $\pm$     0.03 \\ 
NGC1521$^8$     & 3    &    0.88 $\pm$     0.48 &    -1.06 $\pm$     0.99 &    -0.02 $\pm$     0.18 &    -0.01 $\pm$     0.10 &     0.14 $\pm$     0.16 &     0.00 $\pm$     0.01 &     0.38 $\pm$     0.04 &     0.27 $\pm$     0.08 &     0.47 $\pm$     0.02 &     0.10 $\pm$     0.04 &     0.09 $\pm$     0.04 \\ 
NGC1521$^8$     & 4    &    0.33 $\pm$     0.42 &    -0.61 $\pm$     0.82 &     0.25 $\pm$     0.18 &     0.23 $\pm$     0.11 &     0.67 $\pm$     0.19 &     0.09 $\pm$     0.02 &    -0.02 $\pm$     0.02 &     0.39 $\pm$     0.11 &     0.27 $\pm$     0.03 &     0.03 $\pm$     0.04 &     0.02 $\pm$     0.02 \\ 
&    &    &    &    &    &    &    &    &    &    &    &    \\
NGC1533    & 1    &    4.51 $\pm$     0.59 &     0.29 $\pm$     0.52 &     0.75 $\pm$     0.18 &     0.43 $\pm$     0.06 &     1.26 $\pm$     0.07 &     0.55 $\pm$     0.03 &     0.77 $\pm$     0.04 &     1.60 $\pm$     0.08 &     1.77 $\pm$     0.01 &     0.64 $\pm$     0.06 &     0.53 $\pm$     0.05 \\ 
NGC1533    & 2    &    3.80 $\pm$     0.53 &    -0.45 $\pm$     0.88 &     0.77 $\pm$     0.17 &     0.35 $\pm$     0.06 &     0.94 $\pm$     0.08 &     0.31 $\pm$     0.03 &     0.46 $\pm$     0.04 &     1.06 $\pm$     0.08 &     1.06 $\pm$     0.02 &     0.39 $\pm$     0.06 &     0.33 $\pm$     0.05 \\ 
NGC1533    & 3    &    4.38 $\pm$     0.56 &     0.03 $\pm$     0.96 &     0.64 $\pm$     0.18 &     0.53 $\pm$     0.08 &     1.22 $\pm$     0.11 &     0.17 $\pm$     0.04 &     0.31 $\pm$     0.04 &     0.94 $\pm$     0.08 &     0.75 $\pm$     0.02 &     0.52 $\pm$     0.06 &     0.38 $\pm$     0.05 \\ 
NGC1533    & 4    &    3.53 $\pm$     0.64 &    -0.11 $\pm$     1.15 &     0.40 $\pm$     0.17 &     0.36 $\pm$     0.12 &     0.64 $\pm$     0.08 &     0.02 $\pm$     0.02 &     0.30 $\pm$     0.04 &     0.94 $\pm$     0.09 &     0.73 $\pm$     0.02 &     0.68 $\pm$     0.06 &     0.35 $\pm$     0.04 \\ 
&    &    &    &    &    &    &    &    &    &    &    &    \\
%NGC1553    & 1    &    3.04 $\pm$     0.50 &    -0.85 $\pm$     0.78 &     0.32 $\pm$     0.14 &     0.39 $\pm$     0.06 &     0.75 $\pm$     0.07 &     0.43 $\pm$     0.03 &     0.74 $\pm$     0.04 &     1.09 $\pm$     0.07 &     1.88 $\pm$     0.02 &     0.37 $\pm$     0.06 &     0.31 $\pm$     0.06 \\ 
%NGC1553    & 2    &    1.66 $\pm$     0.39 &    -0.95 $\pm$     0.78 &     0.22 $\pm$     0.12 &     0.24 $\pm$     0.05 &     0.40 $\pm$     0.06 &     0.13 $\pm$     0.04 &     0.33 $\pm$     0.04 &     0.65 $\pm$     0.07 &     0.76 $\pm$     0.01 &     0.17 $\pm$     0.06 &     0.19 $\pm$     0.05 \\ 
%NGC1553    & 3    &    1.07 $\pm$     0.41 &    -0.78 $\pm$     0.83 &     0.19 $\pm$     0.13 &     0.26 $\pm$     0.03 &     0.33 $\pm$     0.06 &     0.05 $\pm$     0.02 &     0.06 $\pm$     0.04 &     0.57 $\pm$     0.08 &     0.47 $\pm$     0.02 &     0.12 $\pm$     0.06 &     0.14 $\pm$     0.05 \\ 
%NGC1553    & 4    &    0.06 $\pm$     0.29 &    -1.58 $\pm$     0.70 &     0.08 $\pm$     0.12 &     0.13 $\pm$     0.03 &     0.08 $\pm$     0.04 &     0.23 $\pm$     0.03 &     0.10 $\pm$     0.03 &     0.27 $\pm$     0.08 &     0.18 $\pm$     0.01 &     0.03 $\pm$     0.07 &     0.02 $\pm$     0.04 \\ 
&    &    &    &    &    &    &    &    &    &    &    &    \\
\hline
\end{longtable}
\end{landscape}}
}}
%------------------------------------- end Table 3 -----------------------------------

\begin{appendix} %First online appendix

\section{Short description of morphological and kinematic  
characteristics and peculiarities of
the galaxy sample (Paper~I+Paper~II)}

We collect in Table~A1 the morphological and kinematic characteristics, 
as well as the peculiarities, of the complete R05$+$A06 sample. 
The morphological and kinematic properties in Table~A1 are
derived from the literature, and were in part already described in
the Appendix (provided as on-line material) of Paper~I and Paper~II.

We list in Cols.~3 and 4 the galaxy ellipticity, $\epsilon$, and the   
(V/$\sigma$)$_{scaled}$. The latter is computed from slit spectroscopy measurements
 through the relation provided by Cappellari et al.~(\cite{capp07}),
(V/$\sigma$)$_{scaled}=$0.57$\times$(V/$\sigma)_{slit}$. It should 
be a reasonable approximation of the (V/$\sigma$)$_{e}$,
computed on a luminosity weighted spectrum within $r_e$.
For the eight galaxies in common with the {\tt SAURON}
sample, i.e. NGC~2974, NGC~3489, NGC~4374, NGC~4552, NGC~5813, 
NGC~5831, NGC~5846 and NGC~7332, the $\epsilon$ and  V/$\sigma$
values are those derived through integral field spectroscopy.
Cappellari et al.~(\cite{capp07}) noticed that there is 
a good agreement between $\epsilon$ values derived in the literature 
with different methods.

Emsellem et al.~(\cite{Emsellem07}) defined a new parameter,
$\lambda_R$=$\langle R|V|\rangle/\langle R \sqrt{V^2 + \sigma^2}\rangle$,
where $R$ is the galacto-centric distance, and  $V$ and $\sigma$ are 
luminosity weighted averages over the two-dimensional kinematic field.
This parameter is a proxy to quantify the observed projected stellar angular momentum per unit mass,
and can be used to classify ETGs into slow or fast rotators.
In the $\epsilon$ vs. (V/$\sigma$)$_e$ plane, Cappellari et al.~(\cite{capp07}) 
identified a region (0.0$ \lesssim   \epsilon \lesssim$0.3, 0.0$\lesssim (V/\sigma)_e \lesssim$0.13)
where  slow- rotators are generally found. 
Lacking the 2D V and $\sigma$ information for most of our galaxies, 
we used (V/$\sigma$)$_{scaled}$ to identify in the $\epsilon$ vs. (V/$\sigma$)$_e$ plane 
the fast and slow rotators.
The derived classification is provided in  Col.~5 of Table~A1 (F$=$fast rotator; S$=$slow rotator).
We obtain that $\approx$68\% (36/53) of our sample is fast rotator 
(to be compared with  75\% fast rotators in the {\tt SAURON} sample). 
Less than 1/3 of our galaxies are slow rotators 
(vs 25\% in the {\tt SAURON} sample).
%Considering ETGs for which a spatially resolved kinematics is available
%and using the above recipes, our class of fast rotators contains $ 
%\approx$68\%  (36/53) of our ETGs
% (75\% in the {\tt SAURON} sample). Less than 1/3 of our galaxies are  
%slow rotators (17/53) vs. 1/4
% in the {\tt SAURON} sample.

In Col.~6 and 7 of Table~A1 we provide the kinematic and 
morphological peculiarities, respectively.
A large fraction, approximately 50\% of our ETGs,
displays kinematic peculiarities, like counter-rotation, 
or peculiar faint features, like shells, tails,
or decoupled gas distributions with respect to the stars.

%----------------------------------------------Table  
\begin{table*}
\tiny{
\begin{tabular}{llcccll}
%& & & & \\
\multicolumn{7}{c}{\bf Table A1. kinematic and morphological overview}  
\\
\hline\hline
%& & & &  &\\
\multicolumn{1}{c}{ident}
& \multicolumn{1}{l}{RSA}
& \multicolumn{1}{c}{$\epsilon$}
& \multicolumn{1}{c}{(V/$\sigma$)$_{scaled}$}
& \multicolumn{1}{c}{Rot.}
& \multicolumn{1}{l}{Gas vs. stars kinematic}
& \multicolumn{1}{l}{Gas vs. stars morphological} \\
\multicolumn{1}{c}{}
& \multicolumn{1}{c}{}
& \multicolumn{1}{c}{}
& \multicolumn{1}{c}{}
& \multicolumn{1}{c}{class}
& \multicolumn{1}{l}{peculiarities}
& \multicolumn{1}{l}{ peculiarities}\\
\hline
  & & & & & & \\
  NGC~128  & S02(8) pec      & 0.67 & 0.44  &  F   &  CR g-s (1)  &  g- 
d and g-maj t$\approx$23$^\circ$ (5)\\
  NGC~777  & E1             & 0.21 & 0.06  &  S   &               &  \\
  NGC~1052 & E3/S0         & 0.28 & 0.27 &  F   &  CR g-g (1,5)   & \\
  NGC~1209 & E6             & 0.52 & 0.50  &  F   &               & X- 
like struct.; NW linear feature (2) \\
  NGC~1297 & S02/3(0)    & 0.13 &      &     &  &\\
  NGC~1366 & E7/S01(7)    & 0.56 &       &     &  &\\
  NGC~1380 & S03(7)/Sa     & 0.41 & 0.59  & F    &  &\\
  NGC~1389 & S01(5)/SB01  & 0.37 &      &     &  &\\
  NGC~1407 & E0/S01(0)      & 0.07 & 0.11  &  S    & rotat.  min. axis  
(5) &\\
  NGC~1426 & E4             & 0.34 & 0.40  & F    &  &\\
  & & & &  & &\\
  NGC~1453 & E0             & 0.17 &       &     &  & g-d and g-maj t  
(5) \\
  NGC~1521 & E3             & 0.35 & 0.54   &  F   &  & \\
  NGC~1533 & SB02(2)/SBa   & 0.19 &      &     &  & \\
  NGC~1553 & S01/2(5)pec    & 0.38 & 0.68  & F    &  & shells (3) \\
  NGC~1947 & S03(0) pec     & 0.11 & 0.23  & F     &rotat.  min. axis  
(5)  & dust-lane  min. axis (5)\\
  NGC~2749 & E3             & 0.07 & 0.21 &  F   & rotat.  min. axis  
(5) & \\
  NGC~2911 & S0p or S03(2)  & 0.32 &    &     &  & \\
  NGC~2962 & RSB02/Sa       & 0.37 &  0.51  &  F   &  & \\
  NGC~2974 & E4             & 0.37 & 0.70 & F    & & shells (2); g-d  
and g-maj t$\approx$20$^\circ$ (5) \\
  NGC~3136 & E4             & 0.24 &  0.67  &  F   & CR s-s (5) & \\
  & & & & & &\\
  NGC~3258 & E1             & 0.13 & 0.08 & S    & CR g-s (1)  &\\
  NGC~3268 & E2            & 0.24 & 0.53    & F    &  &\\
  NGC~3489 & S03/Sa     & 0.37 & 0.29 & F    & gas rot. min. axis (5) & 
\\
  NGC~3557 & E3             & 0.21 &  0.30 & F    &  & SW fan; asym.  
outer isophotes (2)\\
  NGC~3607 & S03(3)       & 0.11 & 0.28  &  F   &  &\\
  NGC~3818 & E5             & 0.36 & 0.27 &  F   &  &\\
  NGC~3962 & E1             & 0.22 & 0.23 &  F   &  &gas disk+outer  
arc-like struct. (5)\\
  NGC~4374 & E1              & 0.13 & 0.03 &  S   &  &\\
  NGC~4552 & S01(0)      & 0.06 & 0.05 &  S   &  & shells (3) \\
  NGC~4636 & E0/S01(6)     & 0.24 & 0.07 &  S  & gas irr. motion (5)  & 
\\
  & & & & &\\
  NGC~4696 &(E3)           & 0.34 &  0.18  & F    & &  Faint outer  
shells (2) \\
  NGC~4697 & E6             & 0.32 &  0.39  &F     &  & Non spherical  
inner isophotes (2) \\
  NGC~5011 & E2             & 0.15 & 0.11  &S     & &\\
  NGC~5044 & E0             & 0.11 & 0.13  &F     & CR s-s; gas  
irr.motion (5)  & gas fil. shape (5)\\
  NGC~5077 & S01/2(4)   & 0.15 & 0.06  &   S  &  CR s-s (5);  g-d and  
g-maj t$\approx$90$^\circ$ (5) &\\
  NGC~5090 & E2             & 0.15 & 0.17  & F    & & \\
  NGC~5193 & S01(0)       & 0.07 &    &     &  &\\
  NGC~5266 & S03(5) pec    & 0.31 & 0.12  & S &  stellar rot. min.  
axis (5)   & dust-lane and gas along min. axis (5) \\
  NGC~5328 & E4             & 0.31 &    &     & & \\
  NGC~5363 & [S03(5)]    & 0.34 & 0.39   & F    &  &  dust -lane along  
min. axis (5)\\
   & & & & & &\\
  NGC~5638 & E1              & 0.11 &  0.21   &  F   & & \\
  NGC~5812 & E0              & 0.13 & 0.11    & S    & & Tidal tail  
(2) \\
  NGC~5813 & E1              & 0.15 & 0.14    &  S   & CR s-s; gas  
irr. motion (5) & gas fil. shape (5)  \\
  NGC~5831 & E4             & 0.15 & 0.08    &  S   &  CR s-s (6,7) & \\
  NGC~5846 & S01(0)      & 0.07 & 0.03   & S    & gas irr. motion (5)   
& Faint outer shells (2) \\
  NGC~5898 & S02/3(0)   & 0.07 &  0.21 & F    &  CR g-s (1) &  Three  
spiral arm-like tidal tails (2) \\
  NGC~6721 & E1              & 0.15 &  0.26   &  F   & & \\
  NGC~6758 & E2 (merger)    & 0.22 & 0.12    & S    & CR g-s (5) & \\
  NGC~6776 & E1 pec         & 0.17 &  0.24   & F    & &  shell or  
loops NE (3,4); gas fil.  shape (5) \\
  NGC~6868 & E3/S02/3(3)    & 0.19 & 0.14 & F    & CR g-s (5)  & \\
   & & & & & &\\
NGC~6875 & S0/a(merger)    & 0.41  &    &     & & \\
NGC~6876 & E3              & 0.13  &    &     &  & \\
NGC~6958 & R?S01(3)   & 0.15  & 0.20   & F    & & Shells (2,3)\\
NGC~7007 & S02/3/a       & 0.42  &    &     & CR g-s (1) & g-d and g- 
maj t$\approx$30$^\circ$ (5) \\
NGC~7079  & SBa            & 0.32  & 0.96 &  F& CR g-s + s-s (1) &  \\
NGC~7097 & E4              & 0.29  & 0.10   & S    & CR s-s (1) & \\
NGC~7135 & S01 pec       & 0.31  &0.20    & F    &  & jet-like debris  
and shells (3) \\
NGC~7192 & S02(0)         & 0.15  &  0.04  & S    & CR s-s (5) &   
Shell (2) \\
NGC~7332 & S02/3(8)      & 0.42  & 0.32   & F & CR g-g (1) & \\
NGC~7377 & S02/3/Sa pec   & 0.19  &    &     & & \\
& & & & & &\\
IC~1459    & E4            & 0.28  &  0.10  &  S   & CR g-g (1) &  
shells (2) \\
IC~2006    & E1            & 0.15  &  0.16  &  F   & CR g-s (1) &  \\
IC~3370    & E2 pec    & 0.21  & 0.28  &  F   &  &  X-like struct.;  
broad N fan (2); polar ring ? (5) \\
IC~4296    & E0           & 0.17  & 0.13   &  F   &  CR s-s (5)  & \\
IC~5063    & S03(3)pec/Sa  & 0.28  & 0.26  & F    &  &\\
& & & & & &\\
\hline\hline
\end{tabular}}
\label{tab1app}

Cols.~4 and 5 list respectively (V/$\sigma)_{scaled}$=0.57(V/$\sigma)_{slit}$, and the rotation class
(F=fast rotator; S=slow rotator). 
Legenda: {\bf CR g-s}: counter rotation gas vs. stars;  
{\bf CR s-s}: counter rotation stars vs. stars;
{\bf CR g-g}: counter rotation gas vs. gas; {\bf stars rotat.  min.  
axis}: stars rotate along the
galaxy minor axis; {\bf g-d and g-maj t}: gas disk and galaxy major  
axis are tilted by the
reported angle, if provided in the literature.
References: (1) Corsini \& Bertola~\cite{Corsini98}; (2) Tal et al.~\cite{Tal09};
(3) Malin \& Carter~\cite{MC83};  (4) Pierfederici \& Rampazzo~\cite{Pierfederici04};
(5) the description of the kinematic and morphological peculiarities  
of the galaxies and full references are reported in the on-line notes of Paper~I and  
Paper~II; (6)  Davies et al.~\cite{Davies83}; (7) Emsellem et al.~\cite{Emsellem04}.
\end{table*}
%-------------------------end Table  

\end{appendix}

\Online

\begin{appendix} %Second online appendix

\section{Emission line ratios for the total sample in annuli of increasing galacto-centric distance.}

\begin{figure*}
\centering
\includegraphics[width=16.4cm,clip]{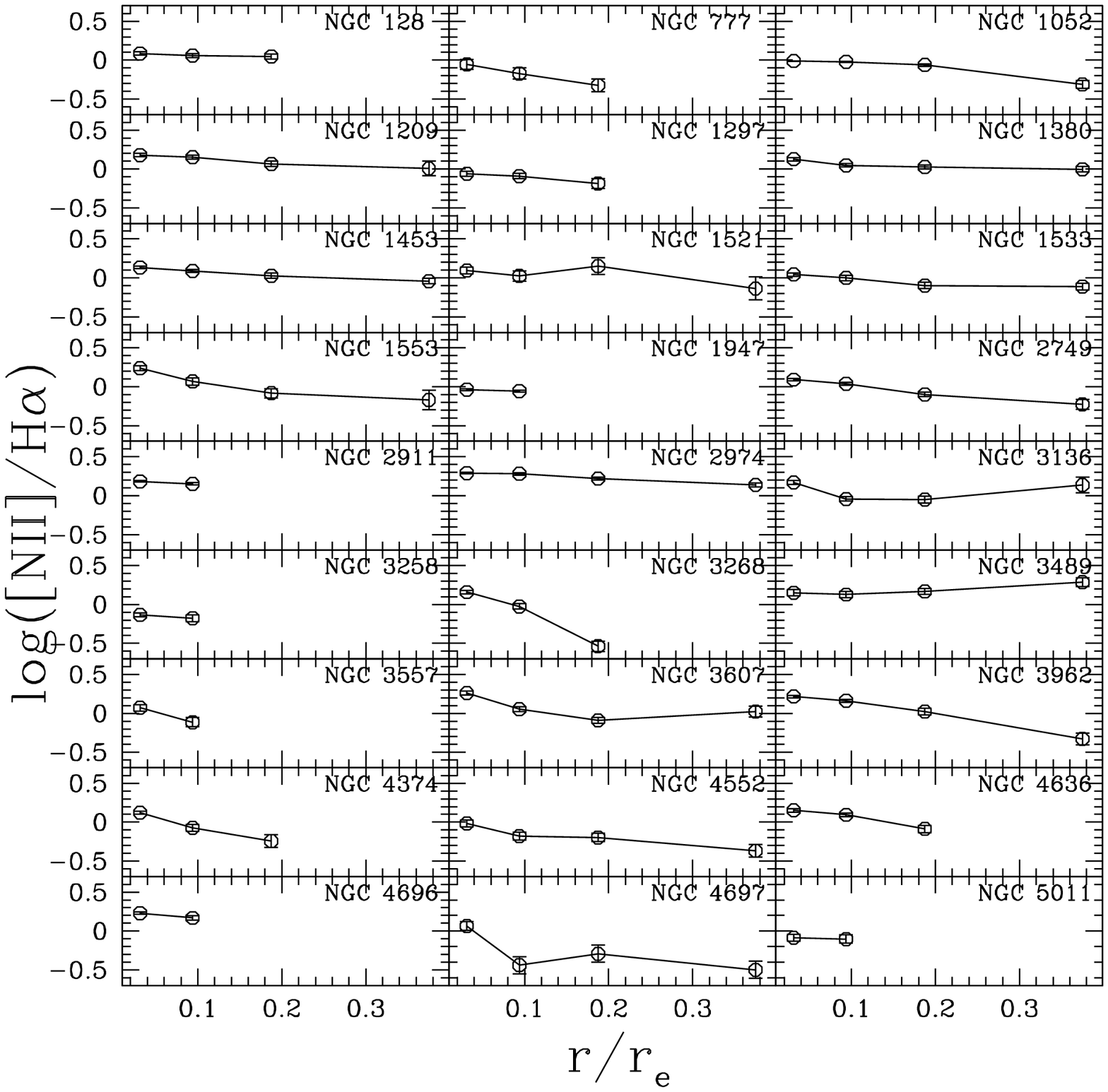}
\caption{${\rm \log [NII]/H\alpha}$ ratio for the sample in the annuli 
r $\leq$r$_e$/16, r$_{e}$/16 $<$ r $\leq$r$_e$/8, r$_{e}$/8$<$ r $\leq$r$_e$/4, and r$_{e}$/4 $<$ r $\leq$r$_e$/2.}
\label{}
\end{figure*}

\begin{figure*}
\centering
\includegraphics[width=16.4cm,clip]{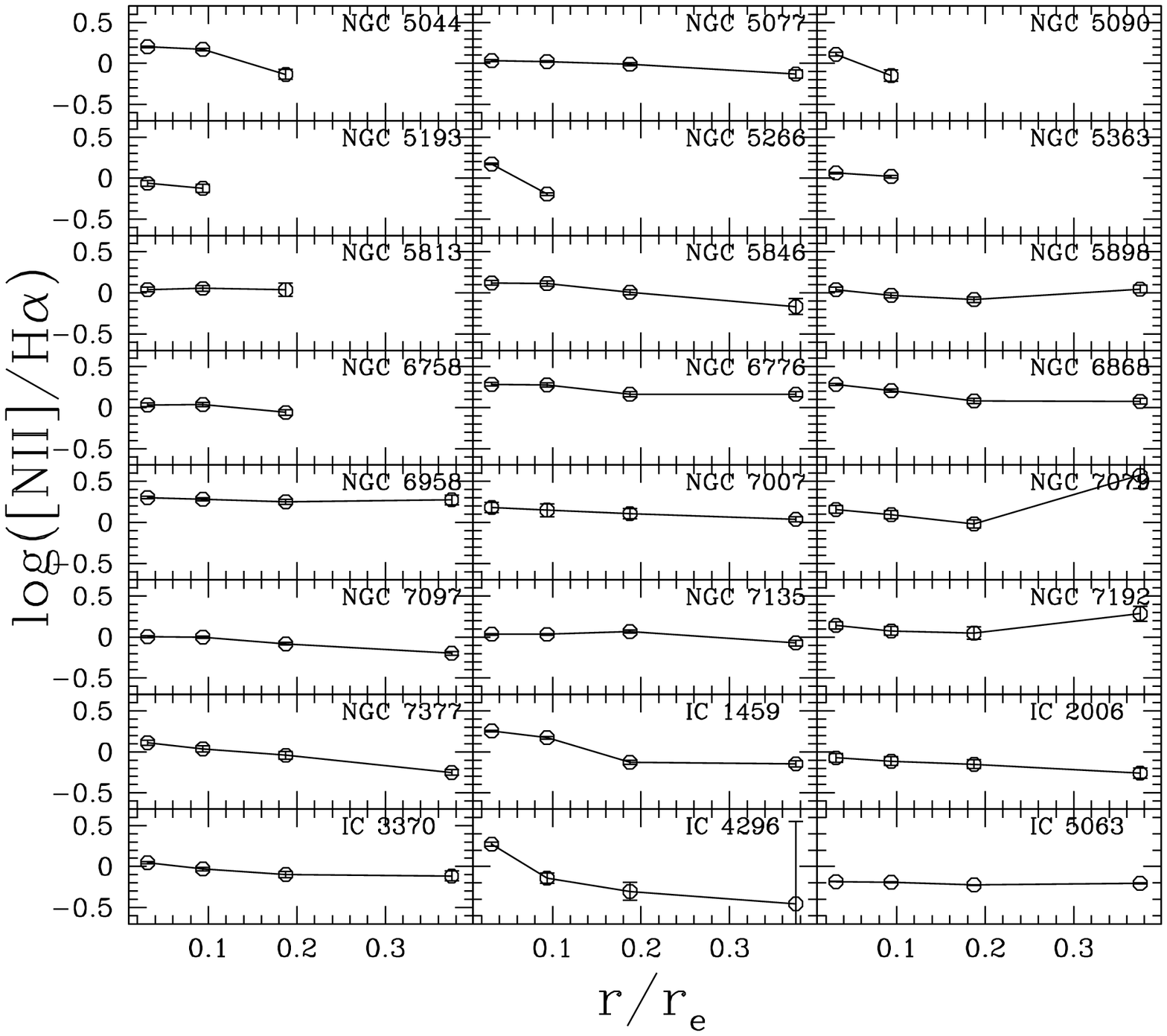}
\caption{${\rm \log [NII]/H\alpha}$ ratio for the sample in the annuli 
r $\leq$r$_e$/16, r$_{e}$/16 $<$ r $\leq$r$_e$/8, r$_{e}$/8$<$ r $\leq$r$_e$/4, and r$_{e}$/4 $<$ r $\leq$r$_e$/2.}
\label{}
\end{figure*}

\begin{figure*}
\centering
\includegraphics[width=16.4cm,clip]{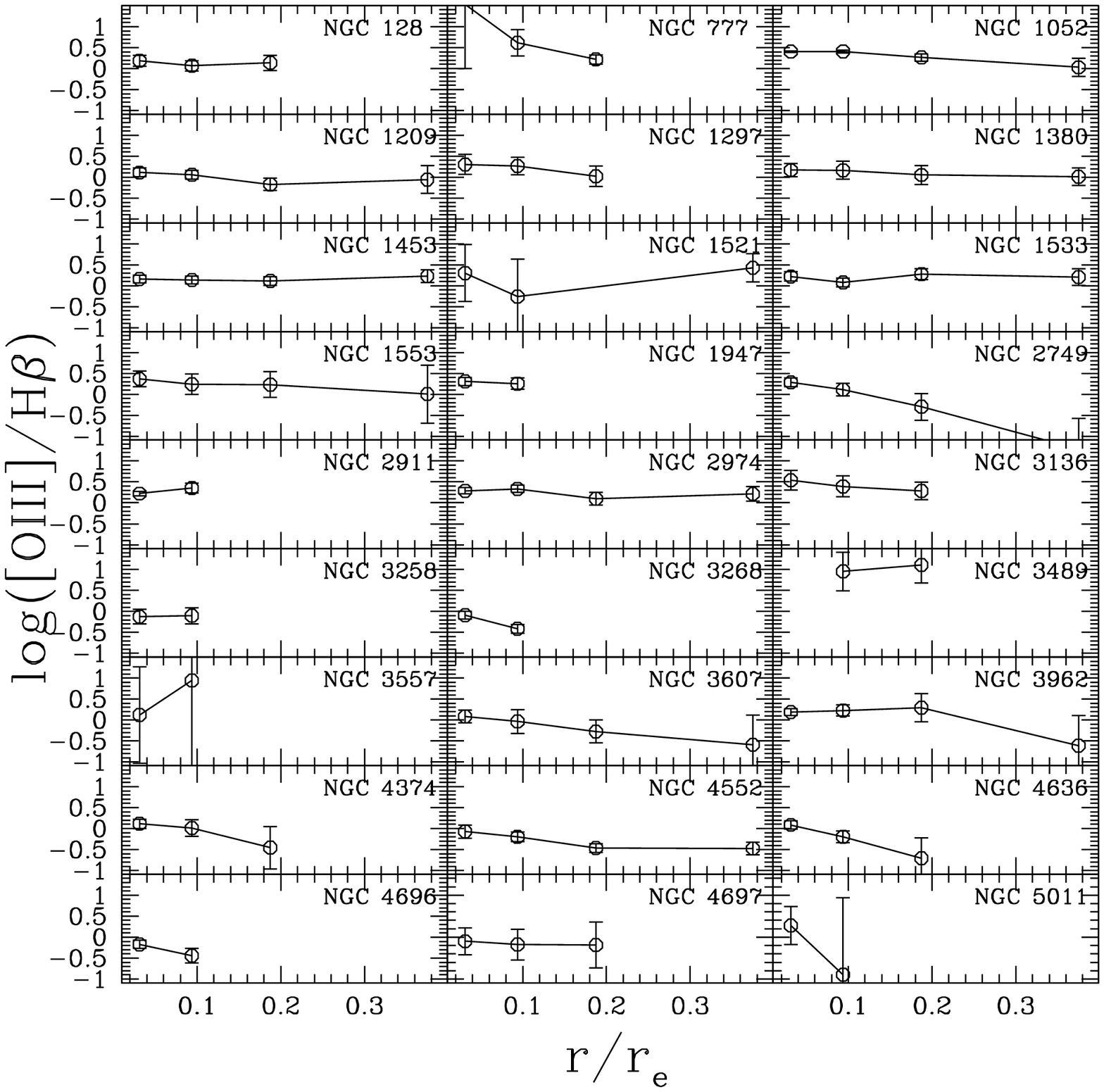}
\caption{${\rm \log [OIII]/H\beta}$ ratio for the sample in the annuli 
r $\leq$r$_e$/16, r$_{e}$/16 $<$ r $\leq$r$_e$/8, r$_{e}$/8$<$ r $\leq$r$_e$/4, and r$_{e}$/4 $<$ r $\leq$r$_e$/2.}
\label{}
\end{figure*}

\begin{figure*}
\centering
\includegraphics[width=16.4cm,clip]{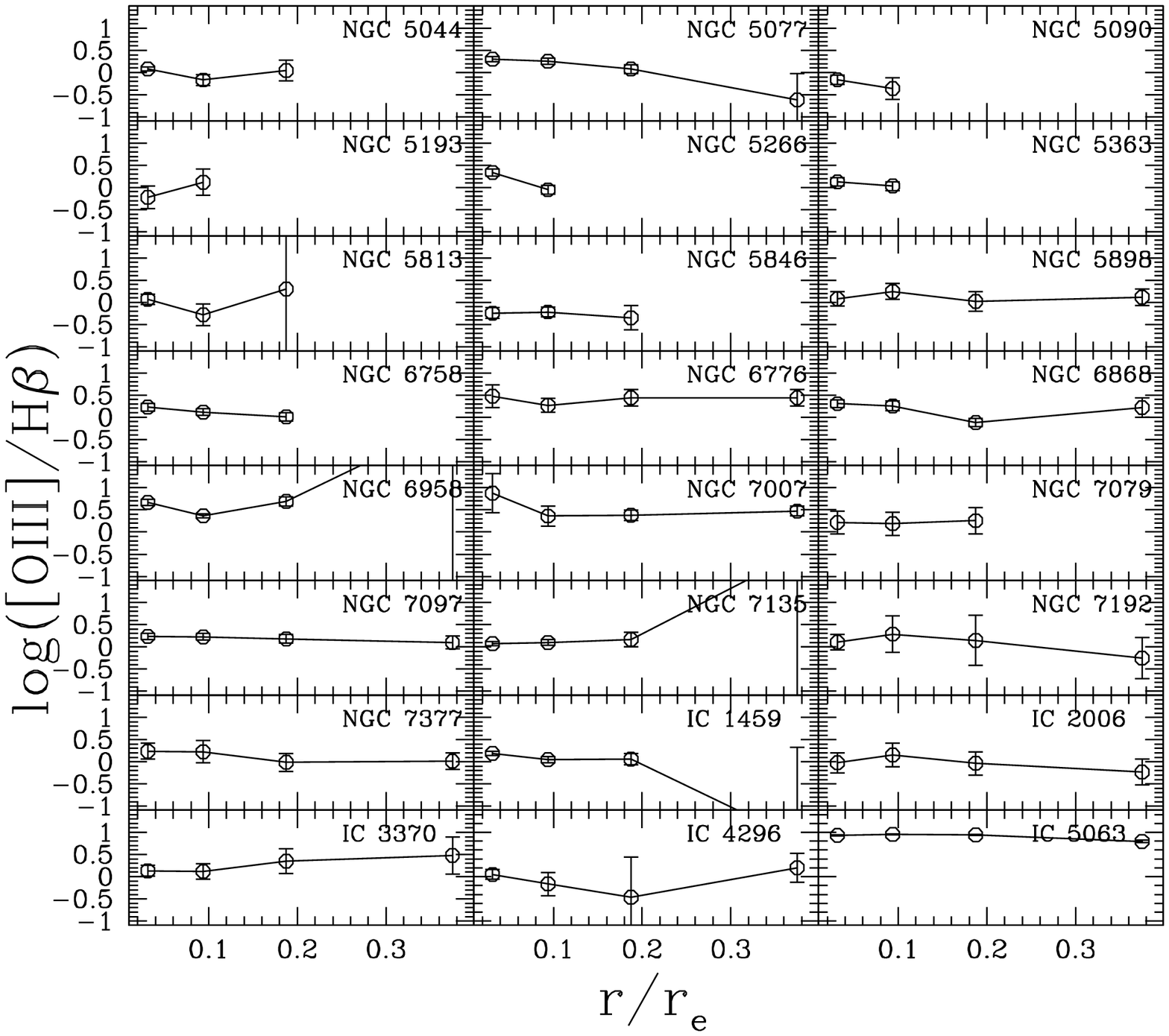}
\caption{${\rm \log [OIII]/H\beta}$ ratio for the sample in the annuli 
r $\leq$r$_e$/16, r$_{e}$/16 $<$ r $\leq$r$_e$/8, r$_{e}$/8$<$ r $\leq$r$_e$/4, and r$_{e}$/4 $<$ r $\leq$r$_e$/2.}
\label{}
\end{figure*}

\end{appendix}

\end{document}